\documentclass[%
 reprint,
nofootinbib,
 amsmath,amssymb,
 aps,
floatfix,
]{revtex4-2}

\usepackage{aasmacros}
\usepackage{graphicx}
\usepackage{dcolumn}
\usepackage{multirow}
\usepackage{bm}
\usepackage{newtxtext,newtxmath}
\usepackage{subcaption}
\usepackage[T1]{fontenc}
\usepackage{makecell}
\usepackage{animate}
\usepackage{mathtools}
\usepackage{url}
\usepackage[hidelinks,colorlinks]{hyperref}
\usepackage{xcolor}
\hypersetup{
    colorlinks=true,
    linkcolor={blue},
    citecolor={blue},
    urlcolor={blue}
}

\def\hlinewd#1{%
\noalign{\ifnum0=`}\fi\hrule \@height #1 %
\futurelet\reserved@a\@xhline} 
\usepackage{graphicx}	
\usepackage{amsmath}	

\begin{document}

\preprint{APS/123-QED}

\title{Feeding the Void: Co-evolution of the SIDM-seeded Black Hole and the dark halo after core-collapse}

\author{Zhichao Carton Zeng$^{1}$}\email{E-mail: zczeng@tamu.edu}
\author{Annika H. G. Peter$^{2,3,4}$}
\author{Louis E. Strigari$^{1}$}
\author{Andrew Benson$^{5}$}
\author{Xiaolong Du$^{5,6}$}
\author{Charlie Mace$^{2,3}$}

\affiliation{$^{1}$Department of Physics and Astronomy, Mitchell Institute for Fundamental Physics and Astronomy, Texas A\&M University, College Station, Texas 77843, USA \\
$^{2}$Department of Physics, The Ohio State University, 191 W. Woodruff Ave., Columbus OH 43210, USA \\
$^{3}$Center for Cosmology and Astroparticle Physics, The Ohio State University, 191 W. Woodruff Ave., Columbus OH 43210, USA\\
$^{4}$Department of Astronomy, The Ohio State University, 140 W. 18th Ave., Columbus OH 43210, USA\\
$^{5}$ Carnegie Observatories, 813 Santa Barbara Street, Pasadena CA 91101, USA\\
$^{6}$ Department of Physics and Astronomy, University of California, Los Angeles, CA 90095, USA\\
}

\date{\today}

\begin{abstract}
While the gravothermal collapse of self-interacting dark matter (SIDM) halos provides a compelling mechanism for seeding supermassive black holes (BHs), the post-collapse macroscopic co-evolution of the BH and its host halo remains largely unexplored. In this paper, we introduce a novel N-body framework to dynamically model the dark matter accretion onto the central BH. By bracketing the accretion rate with two complementary phenomenological schemes---a `multi-round core-collapse' scenario and a `continuous accretion' scenario---we establish bounds on the BH mass growth trajectory. We find that the BH mass reaches $5\%-12\%$ of the total halo mass within $\lesssim2$ Gyr post-collapse. Concurrently, this sustained mass depletion drives a `gravothermal supply chain' that forces outer dark matter inward, severely depleting the inner halo density ($\lesssim 2r_s$). This unique structural imprint qualitatively matches the recently detected anomaly in the strong lens JVAS B1938+666 (detection V), where the best-fit lens model consists of an unresolved central point mass and a shallow, uniform-surface-density extended disk, offering a distinct observational signature for SIDM-seeded BHs.
\end{abstract}

\maketitle


\section{Introduction}

Self-interacting dark matter (SIDM), an alternative paradigm that allows dark matter particles to have collisions, has attracted significant recent interest. It provides a compelling framework for resolving various small-scale observational anomalies that appear incompatible with standard Cold Dark Matter (CDM) predictions.  These anomalies include the unexpectedly large structural diversity in dwarf galaxy rotation curves and mass-size relations \cite{kdn14, oman15, bullock17, santos20, hayashi20, li20, sales22, nadler23}, ultra-concentrated subhalo perturbers in strong lensing systems \cite{minor21b, vegetti23, enzi24, tajalli25, vegetti26}, excess galaxy-galaxy lensing signals induced by low-mass substructures \cite{meneghetti20, meneghetti23, dutra24},  and dynamical perturbations to stellar streams by dense, dark substructures \cite{bonaca18, xyzhang24, hbyu25, nibauer25}. SIDM naturally addresses these discrepancies because the central density distribution of SIDM halos undergoes a distinct two-phase evolutionary pathway. Initially, frequent scatterings among SIDM particles drive rapid thermalization in the inner halo within a short time ($\lesssim1$ Gyr), forming a nearly constant density core that is shallower than its CDM counterpart. Subsequently, on a much longer timescale ($\sim \mathcal{O}(10^2)$ times the core-formation timescale), a temperature gradient between the hot core and the cooler outer halo drives a slow but relentless outward heat flux. Due to the negative heat capacity of self-gravitating systems, this energy loss forces the central dark matter to contract and heat up. This rising central temperature further steepens the temperature gradient, creating a positive feedback loop.
This self-accelerating cycle eventually drives the inner halo into a gravothermal runaway state known as `core-collapse' or `gravothermal catastrophe', during which the central core becomes increasingly denser, smaller, and hotter. This unique core-collapse mechanism can naturally produce a central density much higher than that of CDM halos, thus alleviating the aforementioned small-scale anomalies.

Despite this well-established evolutionary track, the ultimate physical fate of this violent runaway core-collapse process remains a critical open question. At the computational limits of both traditional N-body simulations \cite{koda11, zzc22, zzc23, zzc24, palubski24, fischer24, turner21, mace24, mace25, engelhardt26} and gravothermal fluid frameworks \cite{balberg02, essig19, nishikawa20, o22, sq22, ymzhong23}, it is widely agreed that the central density of the dark halo does increase dramatically during core-collapse. However, owing to the fundamental limitations of these macroscopic numerical techniques, they cannot directly resolve the ultimate final state; naive mathematical extrapolations merely point toward a singularity-like state.  Drawing upon early studies of relativistic star clusters, \cite{balberg02b} qualitatively proposed that a radial relativistic instability would inevitably trigger the formation of a central seed black hole (BH) when the core's 1D velocity dispersion approaches the velocity threshold of $\sim c/3$. \cite{wxfeng22} later quantified this dynamical instability scenario by solving the general relativistic equilibrium equations. They demonstrated that as the collapsing SIDM particles become increasingly relativistic, the actual adiabatic index $\gamma$ of the SIDM gas steadily decreases from $5/3$. Simultaneously, the critical adiabatic index $\gamma_{cr}$ rises above the Newtonian limit of $4/3$ because, in general relativity, the intense thermal pressure not only counters gravity but also acts as an additional gravitational source term. 
\cite{wxfeng22} showed that when the 3D velocity dispersion of the SIDM particles reaches $\sim0.57c$ (equivalent to a 1D dispersion of $\sim0.33c$), Chandrasekhar’s instability criterion $\gamma<\gamma_{cr}$ \cite{chandra64} is met. At this juncture, hydrostatic pressure can no longer balance self-gravity, triggering a dynamical collapse that inevitably births a seed black hole.

A lingering question that remains insufficiently studied is the mass of the SIDM-seeded BH. Specifically, the determination of this mass can be decoupled into two consecutive phases: (a) what is the initial microscopic seed mass formed directly by the relativistic instability, and (b) how does this initial seed grow, and what is its macroscopic mass at the end of the rapid core-collapse phase? Addressing (a), \cite{sophia23} developed a scaling framework across the late, unorganized stages of core-collapse, while \cite{gu26} went beyond the hydrostatic equilibrium approximation by directly incorporating general relativistic effects into the gravothermal equations and numerically evolving the system until an apparent horizon forms. Both methods independently agree on an initial BH seed mass of merely $\mathcal{O}(10^{-8})$ of the halo mass. To address (b), \cite{wxfeng21} provided an estimate of $\mathcal{O}(10^{-2})$ halo mass based on the phenomenological observation of late-time gravothermal fluid evolution. They found that within a certain enclosed mass, the inner density increases, physically resembling a dense central core rapidly contracting and effectively decoupling from the rest of the dark halo. \cite{wxfeng21} showed that this characteristic enclosed mass coincides with the short-mean-free-path (SMFP) region of the halo. Consequently, this SMFP mass has been widely adopted as the macroscopic prediction for the SIDM-seeded BH mass. However, the exact link between the physically rigorous microscopic seed (a) and the phenomenological macroscopic SMFP estimate (b) has only recently begun to be explored. For instance, recent efforts by \cite{wxfeng25} have attempted to bridge the gap from $\mathcal{O}(10^{-8})$ to $\mathcal{O}(10^{-2})$ halo mass by modeling the combined accretion of baryonic gas and dark matter. Complementing this, very recent work by \cite{meng26} moved beyond the traditional quasi-static gravothermal approximations. By employing advanced computational fluid dynamics (CFD) techniques to explicitly resolve the extreme non-equilibrium hydrodynamics, they demonstrate that purely dark matter-driven accretion onto a central BH seed can be exceptionally rapid during the first few Myr, before locally saturating due to the regulation by SIDM heat transport.

Even if the BH successfully grows to the SMFP mass scale, another question highly relevant to observational scales remains unanswered: does this SMFP estimate ($\mathcal{O}(10^{-2})$ halo mass) reflect the `ultimate' BH mass after $\mathcal{O}(1)$ Gyrs of evolution? This question is particularly pressing given the recent, unexpected discoveries of the JWST Little Red Dots (LRDs) \cite{lrd, kokorev24, greene24, taylor25}, which point to an unusual abundance of supermassive black holes at high redshifts $z\gtrsim5$. Exploring the post-SMFP evolution, a recent proof-of-concept study \cite{fz25} modeled the uncapped Eddington accretion of gas on top of the SMFP BH mass. Their statistical models demonstrated that if such long-term gas accretion is efficient, the abundance of LRDs could indeed be naturally explained by SIDM-seeded BHs. 

However, one crucial aspect remains largely unexplored in the macroscopic BH-halo co-evolution: does the surrounding SIDM continue to feed the BH post-collapse, and if so, how does this sustained accretion back-react on the macroscopic structure of the host halo, potentially leaving unique observational imprints?  Addressing this question directly, however, is obstructed by a mismatch of length and time scales. The physical scales of accretion, the Schwarzschild radius and the Bondi radius of the seed BH, lie many orders of magnitude below what N-body simulations can resolve, even though N-body remains the most faithful tool available for following the gravothermal collapse and evolution of SIDM halos. The temporal scales are similarly mismatched: DM accretion onto the BH proceeds on $\sim$Myr timescales \cite{meng26}, far shorter than both the snapshot cadence of a halo-scale simulation and the Gyr-scale timescales that govern the halo's evolution. We therefore need a phenomenological compromise that circumvents the unresolved accretion scales rather than resolving them directly. 

In this work, we tackle this problem by introducing a novel N-body framework to dynamically model the post-core-collapse co-evolution of the SIDM halo and the central BH.  We first propose a `multi-round core-collapse' scenario: after converting the inner DM region into a central BH at the onset of primary core-collapse, the simulation is resumed. We find that the halo repeatedly undergoes gravothermal rebuilding and reaches the core-collapse density threshold again. By iteratively repeating this cycle---terminating the simulation, updating the BH mass with newly condensed DM, and resuming the simulation---we establish a BH mass growth trajectory that serves as a lower bound, since active DM accretion is restricted solely to discrete core-collapse events. As a complementary approach, we simulate a `continuous-accretion' scenario where the BH accretes individual DM particles on-the-fly as they cross the designated central boundary. Since this continuous depletion bypasses the gravothermal bottleneck and inherently overestimates the accretion rate, it serves as an upper bound on the BH mass. We systematically test and provide recommended numerical recipes for two key parameters governing both methods: $\tilde{\rho}_{cc}$, the predefined density threshold for core-collapse, and $Kn_{\rm thres}$, the Knudsen number threshold defining the central region for DM-to-BH conversion. Using the best-converged parameter values, our framework brackets the final BH mass to $5\%-12\%$ of the total halo mass after $\lesssim2$ Gyr of post-collapse evolution.  

Concurrently, regarding the structural evolution of the halo, the conversion of central DM into the BH creates a profound pressure deficit. This forces the outer DM to migrate inward to replenish the central void. This process establishes a `gravothermal supply chain': the extended halo continuously feeds the inner core, which in turn feeds the BH. Consequently, the DM density within the inner $\lesssim 2r_s$ region is severely depleted, dropping by $\gtrsim1$ dex compared to its CDM counterpart, falling even significantly below the density of a standard SIDM core. Remarkably, this physical picture qualitatively agrees with the recent strong-lensing anomaly JVAS B1938+666 (detection V) reported by \cite{vegetti26}, where the best-fit lens model for the dark subhalo perturber consists of an unresolved central point mass surrounded by a shallow, uniform-surface-density extended disk.

This work is organized as follows: In Sec. \ref{sec:method}, we detail our two phenomenological methods for modeling the post-core-collapse evolution of the SIDM halo-BH system using N-body simulations. In Sec. \ref{sec:primary}, we present the primary core-collapse simulation from the initial NFW halo to the onset of gravothermal runaway, establishing the initial conditions for our subsequent BH seeding and accretion models. In Sec. \ref{sec:bh-mass}, we present our results on the BH mass growth trajectories, including extensive convergence tests over key numerical parameters. We also provide a comprehensive comparison and review of recent literature regarding SIDM-seeded BH formation scenarios. In Sec. \ref{sec:halo}, we investigate the detailed structural and kinematic evolution of the host dark halo during this post-collapse phase, with a particular focus on the density profiles and velocity anisotropy. Finally, in Sec. \ref{sec:summary}, we summarize our main findings, discuss the limitations of the current framework, and outline potential pathways for future work.

\section{Methods}\label{sec:method}

In this section, we present our numerical framework for modeling the gravothermal collapse of an SIDM halo and the subsequent mass growth of the central BH. Our methodology consists of two main components: first, a primary DM-only N-body simulation to track the halo's evolution up to the onset of core collapse; and second, the implementation of two distinct simulation-based accretion schemes, `multi-round core-collapse' and `black hole continuous accretion', which are specifically designed to bracket the lower and upper bounds of the BH's mass growth trajectory.

We first perform a DM-only simulation of an isolated SIDM halo until core collapse. Following the setup in \cite{zzc24}, we model a typical bright dwarf halo initialized with a spherically symmetric Navarro-Frenk-White (NFW) profile using the N-body initial condition generator \texttt{SpherIC}. The halo has an initial virial mass of $M_{200c}=3.02\times10^{10}M_\odot$, a virial radius of $r_{200c}=37.32$ kpc, and a scale radius of $r_s=5.12$ kpc. Following previous works \cite{mace24, palubski24, fischer24} on SIDM numerical artifacts and their recommended choice of N-body parameters for the convergence against these artifacts, we choose a DM particle mass of $m_p=5\times10^4M_\odot$, such that the number of DM particles in this target halo is $N_p=8.5\times10^5$ (including the exponentially truncated profile beyond $r_{200c}$). The corresponding softening length of DM particles is thus set to be $0.055\ \rm kpc$, following the recipe in \cite{vdb18}; and the softening length of the black hole particle is set to a much smaller value of $0.001\ \rm kpc$ in subsequent simulations, to mock a macroscopic black hole in reality. We note that our resolution configuration has been validated for SIDM core-collapse convergence in previous studies with similar halo masses and cross sections (e.g., \cite{mace24}), which demonstrated that the primary core-collapse phase is well resolved with $N_p \sim 10^5$--$10^6$ particles. Nevertheless, we caution that the per-particle Knudsen-number estimation and the early BH seed mass may still be subject to resolution-dependent systematics at the current particle number. Because this paper serves as an initial demonstrative study focusing on the physics and mechanisms of post-core-collapse evolution, we restrict our analysis to this single idealized halo.  

The simulation is carried out in an isolated environment using \texttt{Arepo} \cite{Springel10}, equipped with an established SIDM module \cite{mv12, mv14, mv19}. To ensure that the gravothermal catastrophe occurs within a Hubble time, we adopt a constant, velocity-independent SIDM cross section of 100 $\rm cm^2/g$ throughout this work. While incorporating a velocity-dependent cross section is a more realistic and widely favored approach, since it allows models to preserve desirable diversity features in low-mass systems \cite{tulin17, zzc23, zzc24, sq22, mace26} while simultaneously evading stringent cluster-scale constraints \cite{rocha13, Peter13, robertson17, kim17,  robertson19, McDaniel21, andrade22, cross23, fischer23}, our constant cross-section choice simplifies the numerical modeling. Because our analysis focuses only on the evolution of a single, isolated halo in this demonstrative study, this simplification is physically justified. In this context, the constant value effectively acts as a velocity-averaged proxy evaluated at the characteristic velocity scale of the halo, representing a discrete slice of a broader velocity-dependent SIDM model.

Following the methodology from our previous works \cite{zzc22,zzc23,zzc24}, we track the progress of core collapse by evaluating the halo's central density, $\rho_{\rm cen50}$, at every timestep, defined as the mean density of the 50 densest particles. We define a core-collapse threshold, $\tilde{\rho}_{cc}$, as the ratio between the instantaneous central density $\rho_{\rm cen50}(t)$ and its initial value $\rho_{\rm cen50}(t=0)$. Once this ratio reaches a pre-assigned value, the N-body simulation is terminated to allow for follow-up analysis and treatment of the black hole seed. Terminating the N-body simulation at this stage is physically and computationally necessary. First, as the core density increases drastically, the timestep required to accurately resolve SIDM scattering probabilities becomes prohibitively small. Second, as the central region enters the short-mean-free-path (SMFP) regime, the discrete particle noise inherent to the N-body approach makes the results physically questionable. 

In our previous studies focusing on SIDM subhalos \cite{zzc22,zzc23,zzc24}, we utilized fixed collapse thresholds of $\tilde{\rho}_{cc}$ being 5, 10 or 100. Because gravothermal collapse is a runaway process, the exact threshold has a negligible effect on the overall core-collapse timescale. Instead, it primarily served as a safety margin to ensure the collapsing core was dense enough to survive from subhalo-host interactions \cite{zzc22, zzc23, zzc24}.  In the current work, however, our primary objective is to explore the formation of a central massive black hole via core collapse and its subsequent mass growth through accretion. Here, the threshold parameter $\tilde{\rho}_{cc}$ becomes a key physical variable, as the density at termination directly dictates the mass of the initial black hole seed in our model. Consequently, we systematically compare three threshold choices throughout this work: $\tilde{\rho}_{cc}=[10, 30, 100]$.  While a larger $\tilde{\rho}_{cc}$ pushes the halo deeper into the core-collapse stage (at the expense of significantly higher computational cost), this manual threshold choice should not physically alter the final black hole mass after a prolonged period of accretion.  Therefore, a robust numerical prescription must demonstrate convergence over $\tilde{\rho}_{cc}$, ensuring that the long-term evolution of the black hole mass, $M_{\rm BH}(t)$, is independent of the exact termination threshold for core-collapse. 

We refer to the initial phase of evolving the dark-matter-only NFW halo until the onset of core-collapse as the `primary core-collapse simulation'. Upon terminating this primary simulation, we convert the DM particles within the collapsed inner region into a single massive black hole (BH) particle, representing the initial BH seed, while leaving the remaining DM particles unaffected. Defining this ``collapsed inner region'' is a critical physical variable, as it strictly determines the initial BH seed mass and influences its subsequent evolution. To quantify this dependency, we employ the Knudsen number, $Kn$, to determine the conversion boundary. The Knudsen number, $Kn\equiv\lambda/H$, is defined as the ratio of the self-interaction mean free path, $\lambda=1/n\sigma$ (where $n$ is the particle number density and $\sigma$ the interaction cross section), to the gravitational scale height, $H=\sqrt{\sigma_v^2/4\pi G\rho}$ with $\sigma_v$ denoting the 1D velocity dispersion and $\rho$ the DM mass density. The condition $Kn<1$ marks the short-mean-free-path (SMFP) regime in the halo center, where heat conduction transitions to being dominated by local fluid-like collisions. The mass enclosed within this SMFP core is conventionally used as an estimator for the final black hole mass by SIDM core collapse \cite{balberg02b, wxfeng21}. In this work, we test a series of $Kn$ thresholds [0.05, 0.1, 0.3, 1.0, 2.0] to define the initial region for DM-to-BH conversion. By implementing follow-up mass accretion onto the BH, we systematically validate the robustness of this estimator and ensure numerical convergence.

Upon DM-to-BH conversion, we replace the selected DM particles with a single BH particle placed at the halo center with zero initial velocity. Because our initial condition is spherically symmetric and the isolated halo carries zero net angular momentum, the BH particle remains at rest at the center throughout the simulation to within numerical precision. We re-center the BH particle to the halo center and set its velocity to zero only at simulation stop/restart boundaries---namely, at the initial DM-to-BH conversion and, in the multi-round scheme, after each discrete core-collapse round before the simulation is resumed. In the continuous-accretion scheme, the BH particle is fixed in place after the initial seeding; subsequent on-the-fly DM particle absorption (mass set to zero and transferred to the BH) does not trigger additional re-centering or velocity resets. 

To model the post-core-collapse evolution after $\rho_{\rm cen50}$ reaches the thresholds $\tilde{\rho}_{cc}$, along with the subsequent DM accretion onto the newly formed SMBH, we deploy two distinct simulation methods. We refer to these as  `multi-round core-collapse' and `black hole continuous accretion', which are detailed below.  

Before detailing the two schemes, we stress that the $Kn$-defined boundary is a fluid-scale proxy for the feeding region, not a physical accretion mechanism. A DM particle is only truly captured once its orbit is comparable to the innermost stable circular orbit (ISCO, a few Schwarzschild radii), a scale far below both our softening length and the $Kn$ boundary, and even below the Bondi accretion radius ($\sim10^{-4}r_s$; \cite{wxfeng25}). We therefore do not resolve the final capture; instead, the two schemes below bracket the range of plausible macroscopic feeding rates at the resolvable fluid scale, with the microphysical capture left for future work.

\textbf{Lower Bound: Multi-round Core-collapse.} The `multi-round core-collapse' approach iteratively repeats the core-collapse simulation: in each round, the DM halo is evolved with a central BH particle until the central density again reaches the $\tilde{\rho}_{cc}$ threshold. At that point, the DM particles newly condensed within the $Kn$-defined boundary are absorbed into the BH, discretely updating its mass at the end of that cycle. This methodology generates a sequence of discrete simulations and mirrors the primary DM-to-BH conversion: once the initially collapsed region is converted into a seed BH, the remaining halo is an SIDM system whose center is dominated by the BH potential; the still-dense DM immediately outside the converted region is then expected to re-collapse on a short timescale---further accelerated by the deepened central potential---and reproduce the SMFP conditions associated with the initial BH growth.

Previous analytical and fluid studies provide additional motivation for this closure: they indicate that the first post-seeding episode can rapidly connect a microscopic relativistic seed to the conventional $\sim1\%M_{\rm halo}$ SMFP mass scale \cite{wxfeng21,wxfeng25,meng26, gu26b}. Each subsequent multi-round collapse regenerates the same essential prerequisite---a connected, dense SMFP region---now surrounding an already massive central BH with a deeper gravitational potential. We therefore treat the eventual consumption of this rebuilt region using the same $Kn$ criterion as in the primary conversion, as an effective description of the BH accreting its surrounding SIDM. The cycles are repeated for a total post-primary evolution time of 2 Gyr, or until the depleted halo can no longer form a central region satisfying the assigned $Kn_{\rm thres}$ conversion criterion.

Because this method restricts mass updates to the end of each discrete collapse cycle and prohibits continuous, on-the-fly DM accretion, the resulting mass trajectory serves as an effective lower bound on the BH's growth history. This bound is not mathematically rigorous, as two approximations act in opposite directions. The instantaneous DM-to-BH conversion of each rebuilt SMFP region tends to overestimate the accretion rate, whereas withholding all accretion until the $\tilde{\rho}_{cc}$ threshold is reached tends to underestimate it. As estimated in Sec.~\ref{sec:BH-dark-growth}, the characteristic core-consumption time is $\mathcal{O}(1)$ Myr, far shorter than the $\sim100$ Myr gravothermal rebuilding time between successive rounds. This separation makes the instantaneous timing approximation subdominant on the halo-evolution timescale. Together with previous evidence for efficient consumption of the initial SMFP core, it motivates our expectation that the net effect is a slight underestimate. Nevertheless, because the horizon-scale capture efficiency and phase-space evolution of the rebuilt cores remain unresolved, we regard the multi-round trajectory as a physically motivated, but non-rigorous, effective lower bound.

\textbf{Upper Bound: Black Hole Continuous Accretion.} In contrast, the `black hole continuous accretion' approach treats the DM accretion onto the existing central BH (formed during the primary core-collapse simulation) as a continuous, on-the-fly process. It represents the opposite extreme of the multi-round scheme: instead of pacing accretion by discrete collapse episodes, it effectively assumes the central feeding region is replenished instantaneously, so the accretion rate is limited only by how rapidly individual particles cross the $Kn$ boundary, which is the maximum conceivable feeding rate. To ensure consistency across methods, we employ the same $Kn$ thresholds to evaluate which particles are accreted by the central BH at every timestep. For each SIDM particle, its local $Kn$ value is calculated using the 1D velocity dispersion, $\sigma_v$, and the SIDM density $\rho$, in its immediate proximity. If a particle's $Kn$ drops below the assigned threshold, its mass is transferred to the BH by setting the particle's mass to zero and adding one unit of DM particle mass to the central BH. Consequently, the BH mass is dynamically updated at every timestep. To ensure a fair comparison, these continuous-accretion simulations are also evolved for a total post-primary duration of 2 Gyr.

Because it equates the local SMFP criterion ($Kn<Kn_{\rm thres}$) with actual accretion onto the BH, this method inherently overestimates the DM accretion rate. This single approximation admits two complementary descriptions of the same source of error. Macroscopically, $Kn$ is evaluated on a per-particle basis, ignoring whether the entire central halo has re-entered a runaway core-collapse stage, equivalently bypassing the gravothermal rebuilding that genuinely funnels material onto the BH (quantified in Sec.~\ref{sec:BH-dark-growth}). Microscopically, the same gap appears as a length-scale mismatch: the resolution-limited region over which $Kn$ is evaluated ($\gtrsim0.05$ kpc $\sim10^{-2}r_s$) lies far outside the physical Bondi accretion radius ($\sim10^{-4}r_s$; see \cite{wxfeng25}), so a particle satisfying the SMFP criterion is still far from being captured. Therefore, we posit that this continuous accretion scenario establishes a loose upper bound on the BH mass growth trajectory.

\section{Primary core collapse simulation}\label{sec:primary}

\begin{figure}
    \centering
    \begin{subfigure}[t]{0.45\textwidth}
        \centering
        \includegraphics[width=\textwidth, clip,trim=0.2cm 0cm 0.2cm 0cm]{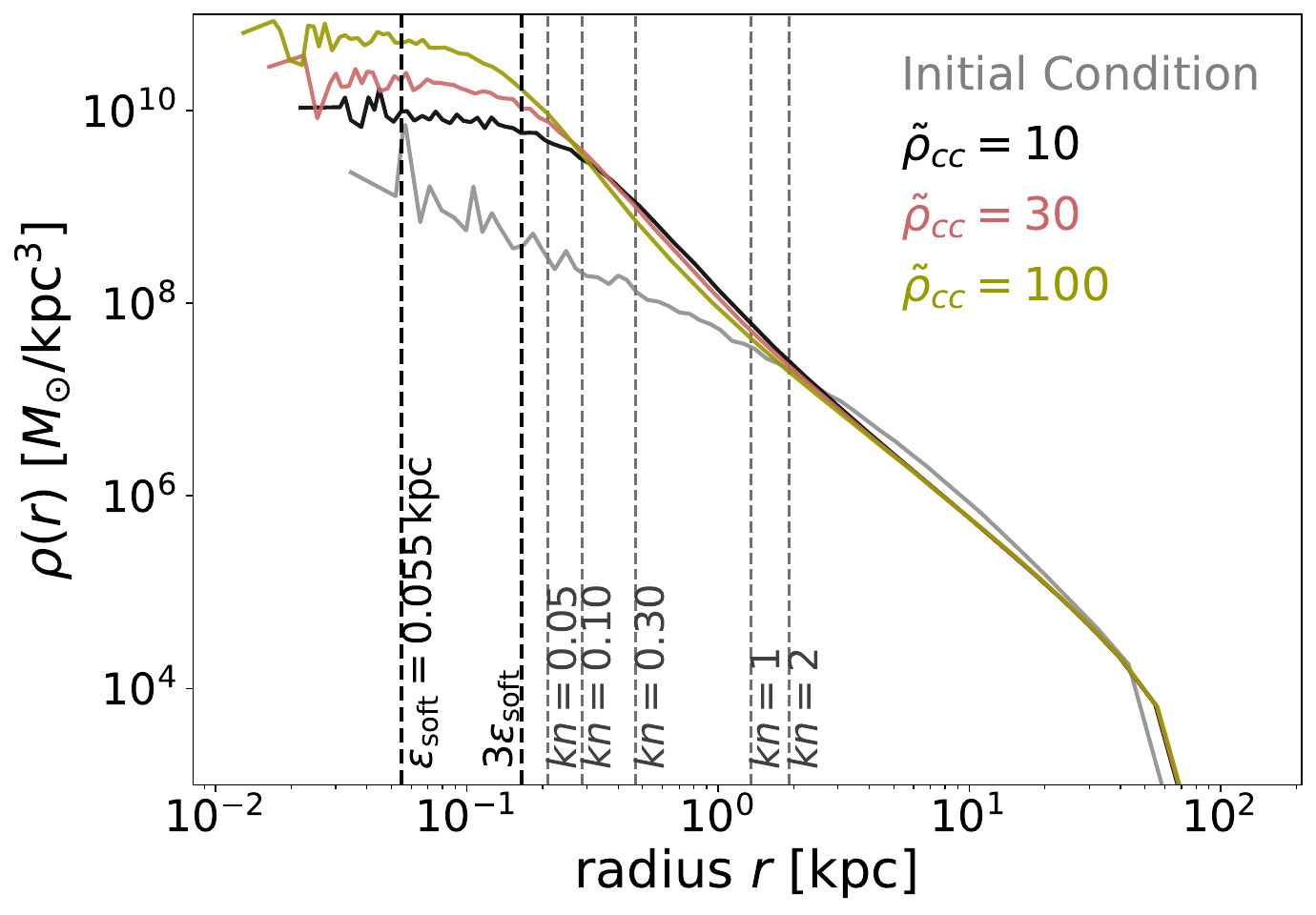}
        \caption{}
        \label{fig:primarycc-rho-r}
    \end{subfigure}
   
    \vspace{-.3em}
    \begin{subfigure}[t]{0.45\textwidth}
        \centering
        \includegraphics[width=\textwidth, clip,trim=0.2cm 0cm 0.2cm 0cm]{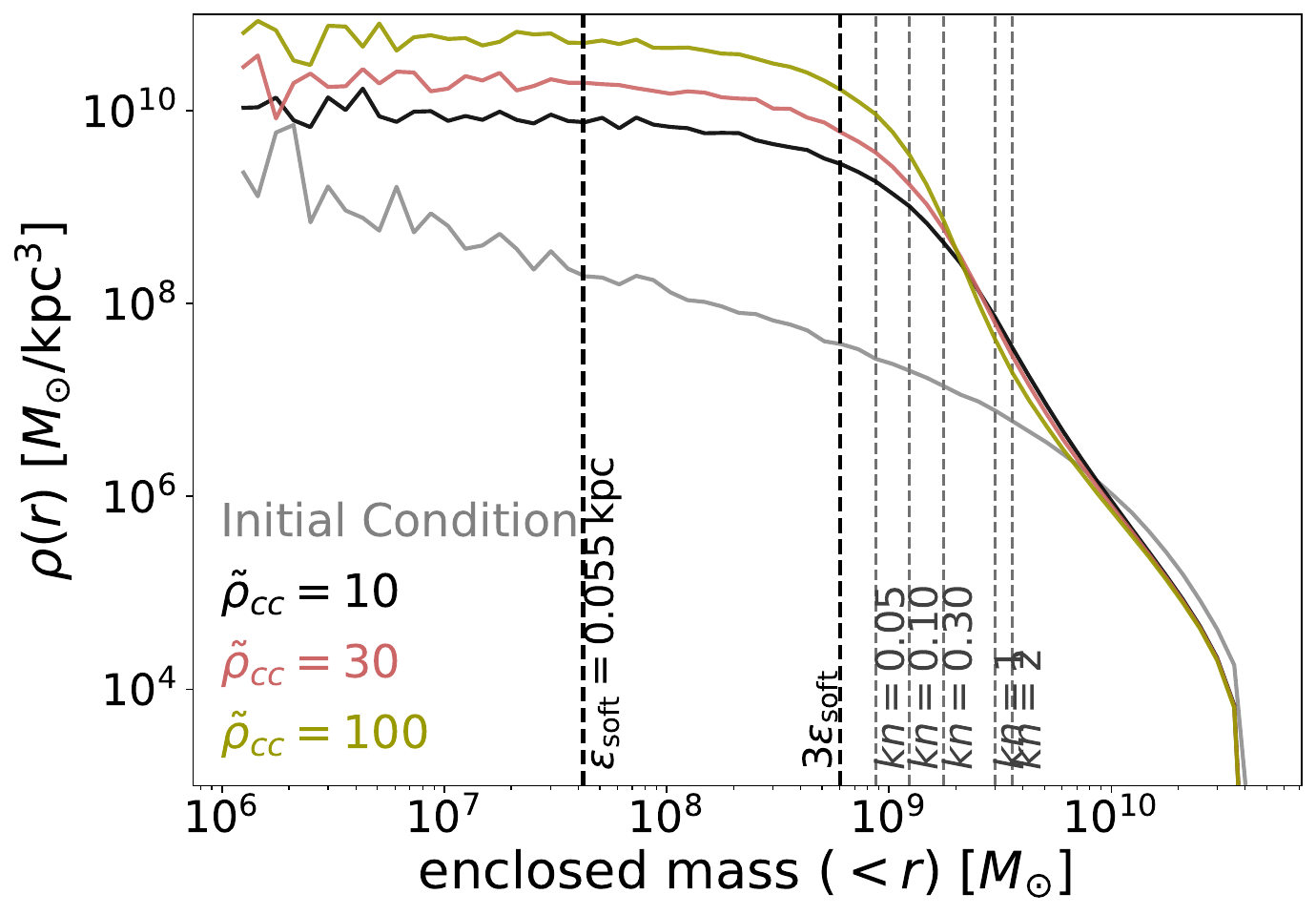}
        \caption{}
        \label{fig:primarycc-rho-m}
    \end{subfigure}
    
    \vspace{-.3em}
    \begin{subfigure}[t]{0.45\textwidth}
        \centering
        \includegraphics[width=\textwidth, clip,trim=0.2cm 0cm 0.2cm 0cm]{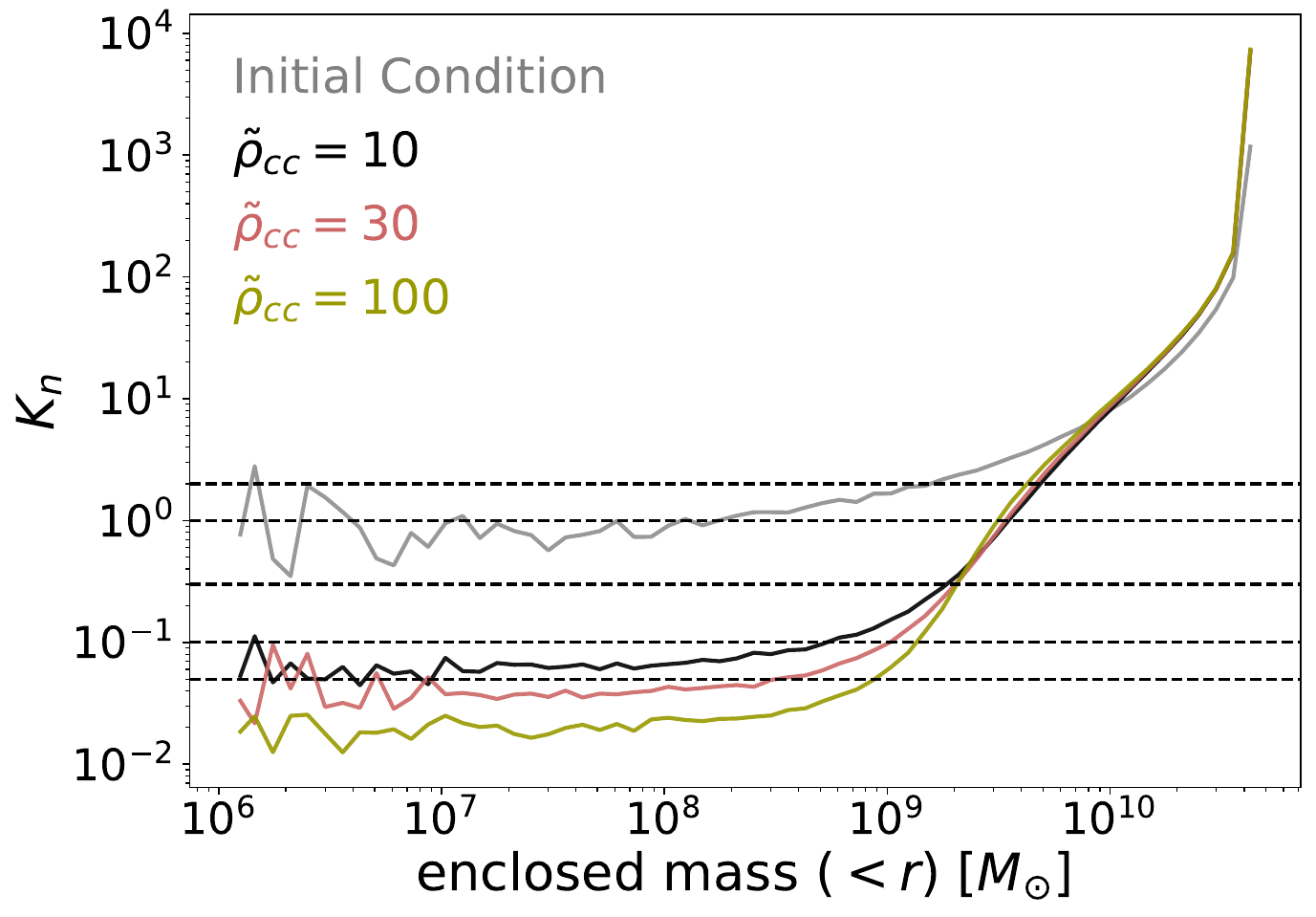}
        \caption{}
        \label{fig:primarycc-kn-m}    
    \end{subfigure}
    \caption{Structural profiles of the isolated SIDM halo during the primary core-collapse phase (evolving from an initial NFW configuration to deep core-collapse). (a) Density $\rho$ vs.\ radius $r$; (b) $\rho$ vs.\ enclosed mass $M(<r)$; (c) Knudsen number $Kn$ vs.\ $M(<r)$. The grey solid line is the initial NFW profile; black, red, and olive mark the core-collapsed states $\tilde{\rho}_{cc}=10,30,100$. The grey horizontal dashed lines in panel (c) mark the $Kn=0.05, 0.1, 0.3, 1,2$ thresholds evaluated for the central DM-to-BH mass conversion. The grey vertical dashed lines in panels (a) and (b) show the corresponding conversion boundaries, computed from the deepest core-collapse state ($\tilde{\rho}_{cc}=100$); for the shallower states the same $Kn$ thresholds would map to different enclosed masses. Black dashed lines mark the softening length $\epsilon_{\rm soft}=0.055$ kpc and $3\epsilon$, the conventional `trustworthy-region' estimate.}
    \label{fig:primarycc}
\end{figure}

In this section, we present the results from the primary core-collapse simulation, which is the evolution of the initial NFW dark halo up to the onset of gravothermal runaway. This phase determines both the core-collapse time and the structural state of the halo at the moment of BH seeding, which serves as the common initial condition for the two post-collapse accretion schemes introduced in the previous section.

Fig. \ref{fig:primarycc} illustrates the evolution of the primary core-collapse simulation, spanning from the initial NFW halo to the final collapse stages just before termination for DM-to-BH conversion, across different central density thresholds $\tilde{\rho}_{cc}$. Fig. \ref{fig:primarycc-rho-r} presents the density profiles as a function of radius, $\rho(r)$--$r$, representing the standard Eulerian perspective. In all three $\tilde{\rho}_{cc}$ scenarios, the central density rises significantly above the initial NFW profile. As expected, configurations with a larger $\tilde{\rho}_{cc}$ (corresponding to later stages of core collapse) exhibit higher central densities and correspondingly smaller core radii. In contrast, Fig. \ref{fig:primarycc-rho-m} displays the same core-collapse density profiles as a function of the enclosed mass, $M(<r)$. This Lagrangian-like representation is achieved by applying fixed mass binning at each snapshot: we sort the DM particles by their distance to the halo center and group them into pre-assigned bins containing a fixed number of particles. This ensures that each mass bin accurately captures the innermost mass fractions (e.g., the innermost $10^7M_\odot$ of DM out of the entire $3\times10^{10}M_\odot$ halo). The signature of core collapse is clearly preserved in this $\rho$--$M$ plane, where the central density substantially exceeds the initial NFW central density and scales monotonically with $\tilde{\rho}_{cc}$. Compared to the standard $\rho$--$r$ view, the $\rho$--$M$ representation offers a distinct visualization advantage: because the enclosed mass scales as $m\sim r^3$ in the inner region, the structural features of the deep core are effectively `zoomed in'. Furthermore, employing mass bins in practice mitigates the numerical artifacts associated with fixed radial binning. During the early core-expansion phase, finely spaced radial bins suffer from severe particle-discreteness noise (or even empty bins) at very small radii; conversely, coarse radial bins fail to resolve the extreme central densities during the runaway collapse stage. Fixed mass binning naturally provides an adaptive spatial resolution. We are thus motivated to primarily utilize mass bins rather than radial bins throughout the remainder of this paper, a choice that parallels the Lagrangian fluid approach employed in previous semi-analytical works (e.g., \cite{wxfeng21}).

Fig. \ref{fig:primarycc-kn-m} illustrates the evolution of the Knudsen number, $Kn$, as a function of the enclosed mass, $M(<r)$. The horizontal dashed lines correspond to our selected threshold values, $Kn_{\rm thres} \in [0.05, 0.1, 0.3, 1.0, 2.0]$, which serve to bound the central region designated for DM-to-BH conversion. Previous theoretical studies \cite{balberg02b, wxfeng21, wxfeng25} commonly adopt $Kn_{\rm thres}\sim1$ to semi-quantitatively estimate the total DM mass that eventually ends up in the BH. As shown in Fig. \ref{fig:primarycc-kn-m}, the signature of deep core collapse, which is characterized by an extremely dense, dynamically thermalized central core, becomes most prominent and stable at $Kn\lesssim0.3$. This physical behavior naturally motivates us to systematically test the convergence of the DM-to-BH conversion process across this broad range of $Kn_{\rm thres}$ values. 
We note that visible fluctuations persist at the innermost radii of the $Kn$ curves, even though our use of fixed mass bins substantially mitigates the discrete particle noise, which we argue is a numerical artifact intrinsic to the particle nature of N-body simulations. Consequently, the measured $Kn$ profiles are not strictly monotonic and may cross a given horizontal $Kn_{\rm thres}$ line multiple times in the innermost halo. This poses a practical challenge for defining a unique mass boundary. Fortunately, as shown in Fig. \ref{fig:primarycc-kn-m}, the $\rho_{cc}$ curves in the $Kn$--mass plane flatten below our chosen $Kn_{\rm thres}$ values throughout the core in most cases, so that the crossing occurs smoothly at the largest crossing point. We therefore define the boundary as the largest enclosed mass $M(<r)$ for which $Kn(M)<Kn_{\rm thres}$. This choice yields a mass envelope that effectively smooths over local fluctuations (evident in Fig. \ref{fig:primarycc-kn-m}) and thus provides a robust conversion boundary.

\section{Mass growth of SIDM seeded black hole}\label{sec:bh-mass}

\begin{figure*}
    \centering
    \begin{subfigure}[t]{0.45\textwidth}
        \centering
        \includegraphics[width=\textwidth, clip,trim=0.2cm 0cm 0.2cm 0cm]{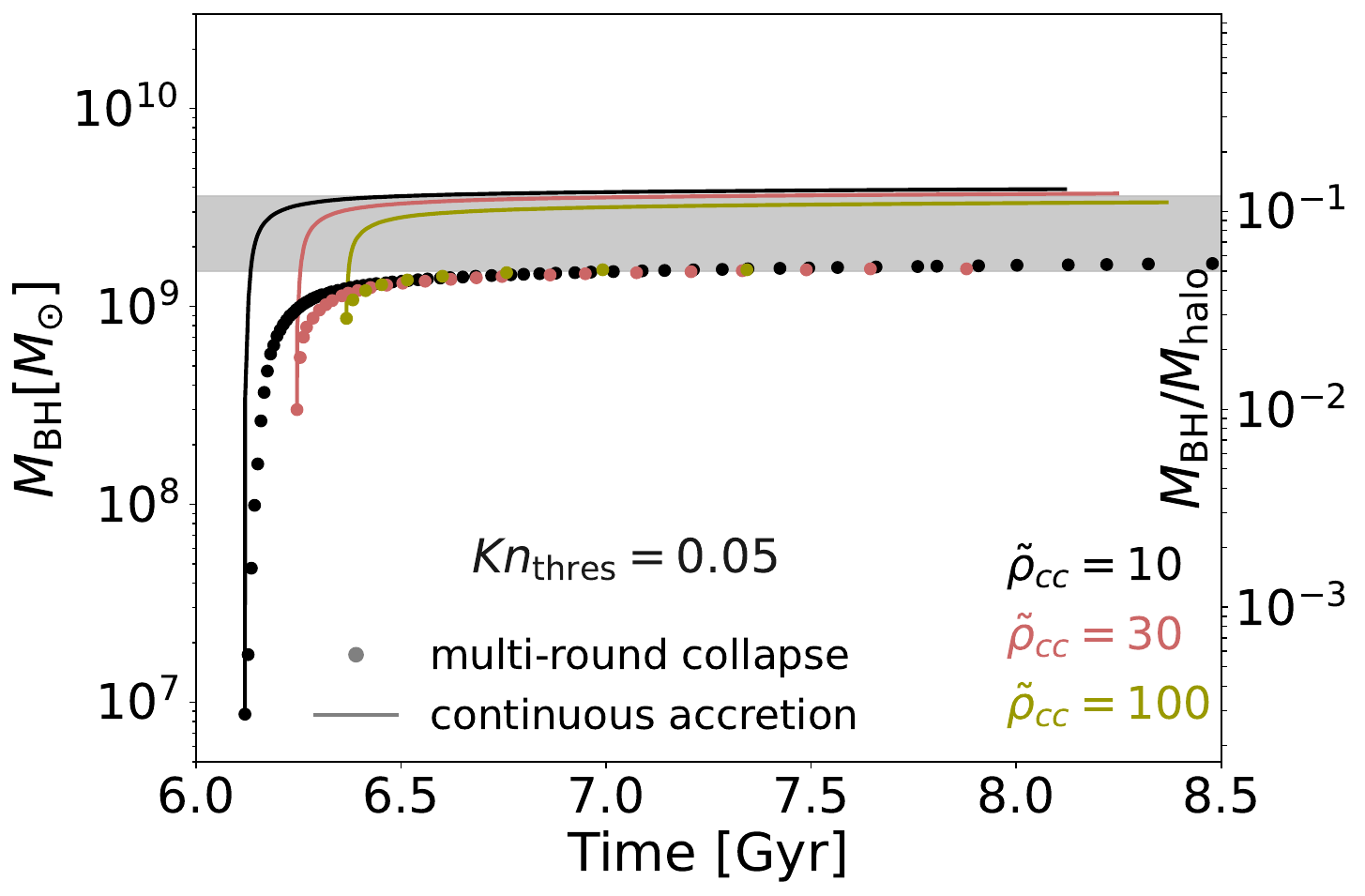}
        \caption{}
        \label{fig:multiround-kn0}
    \end{subfigure}
    ~
    \begin{subfigure}[t]{0.45\textwidth}
        \centering
        \includegraphics[width=\textwidth, clip,trim=0.2cm 0cm 0.2cm 0cm]{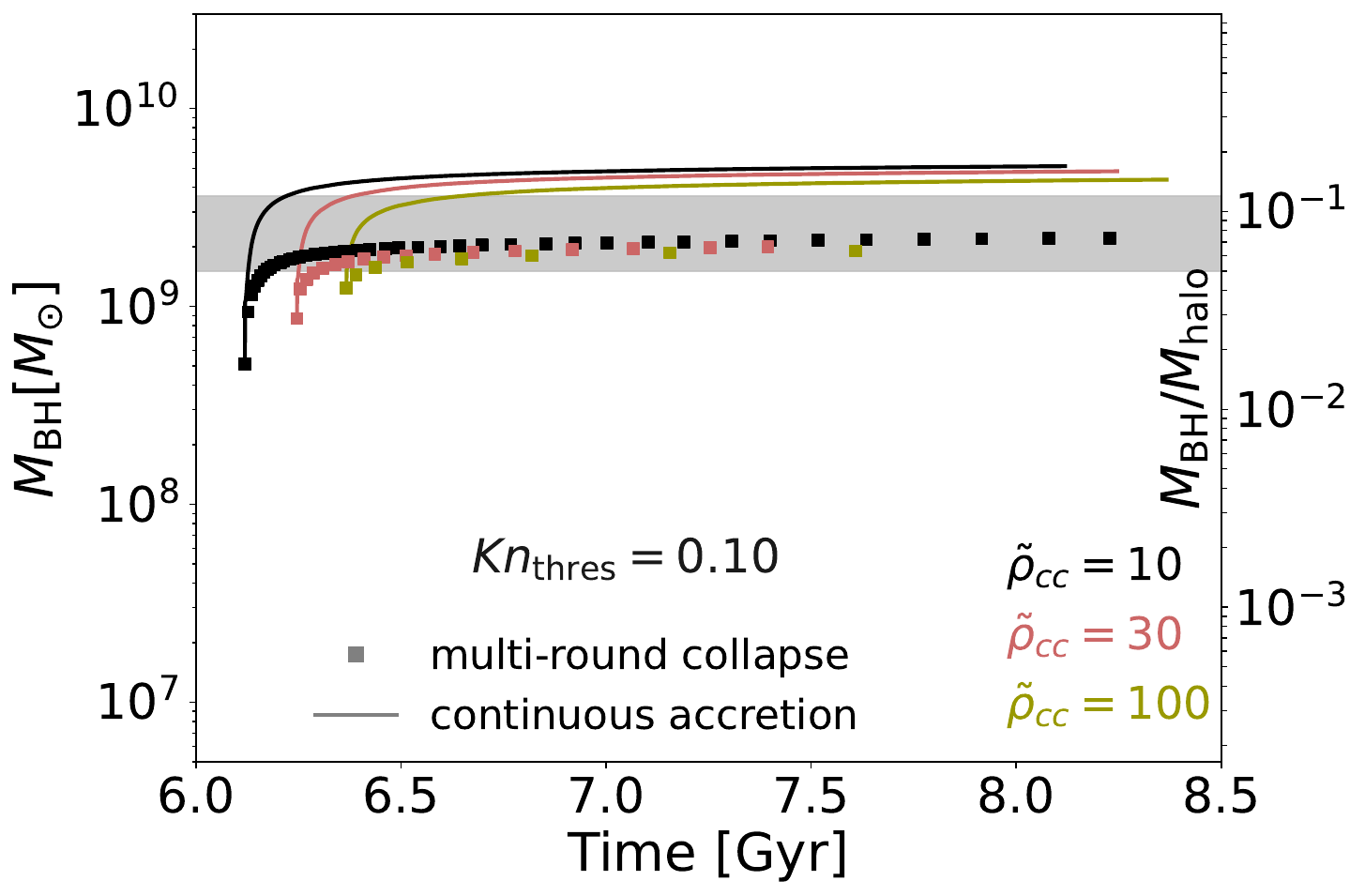}
        \caption{}
        \label{fig:multiround-kn1}
    \end{subfigure}
    ~
    \begin{subfigure}[t]{0.45\textwidth}
        \centering
        \includegraphics[width=\textwidth, clip,trim=0.2cm 0cm 0.2cm 0cm]{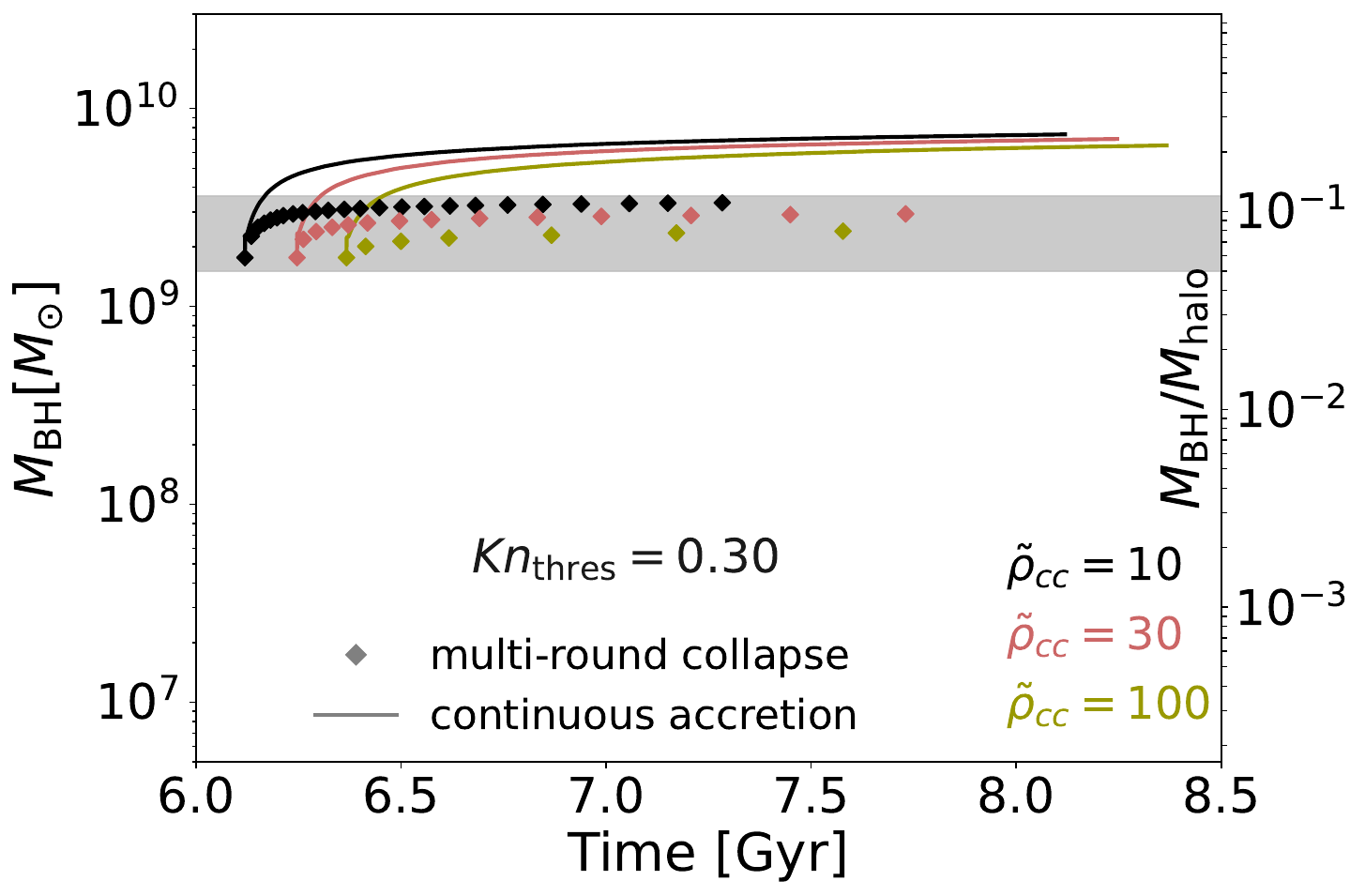}
        \caption{}
        \label{fig:multiround-kn2}    
    \end{subfigure}
    ~
    \begin{subfigure}[t]{0.45\textwidth}
        \centering
        \includegraphics[width=\textwidth, clip,trim=0.2cm 0cm 0.2cm 0cm]{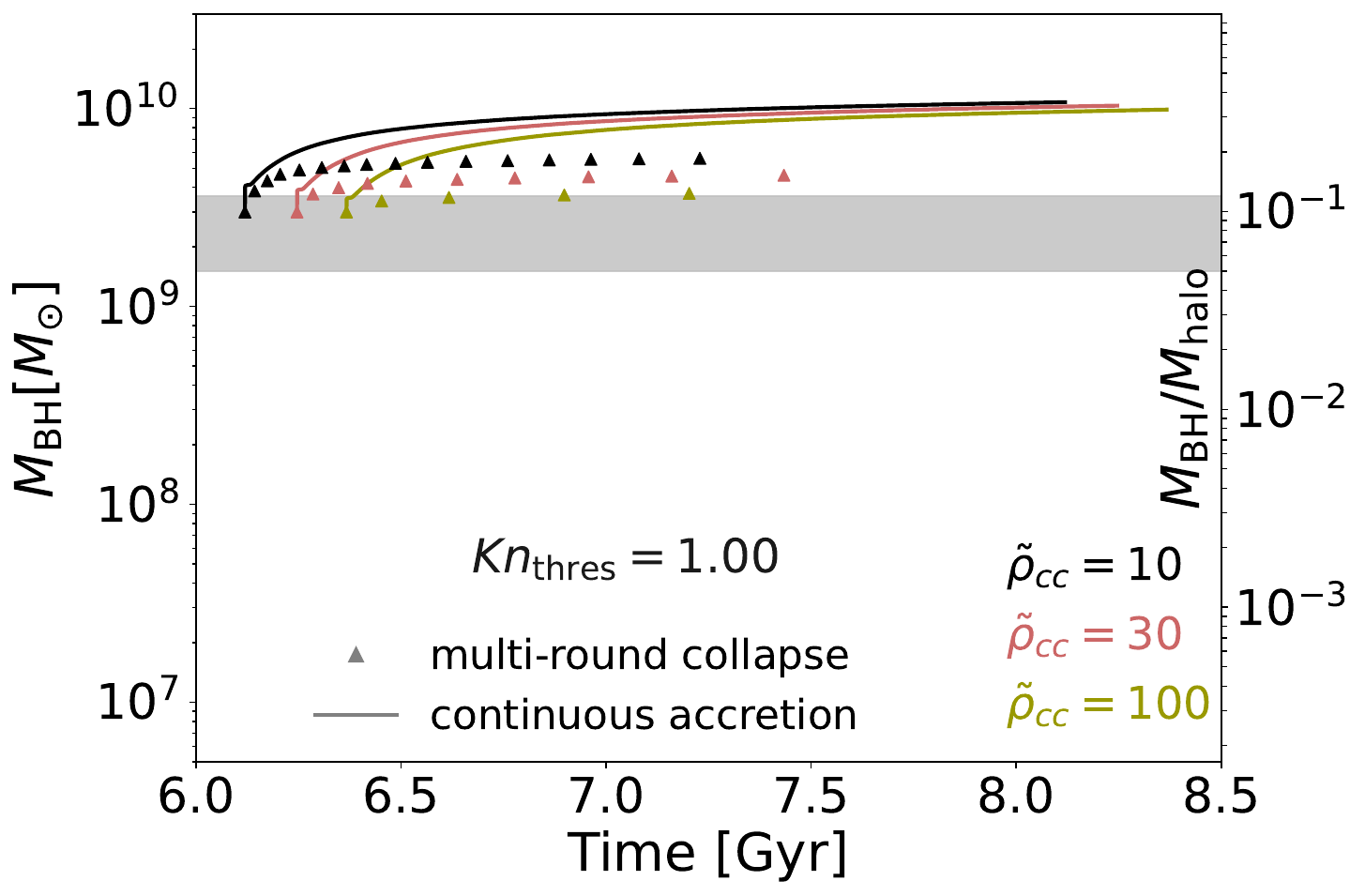}
        \caption{}
        \label{fig:multiround-kn3}    
    \end{subfigure}
    ~
    \begin{subfigure}[t]{0.45\textwidth}
        \centering
        \includegraphics[width=\textwidth, clip,trim=0.2cm 0cm 0.2cm 0cm]{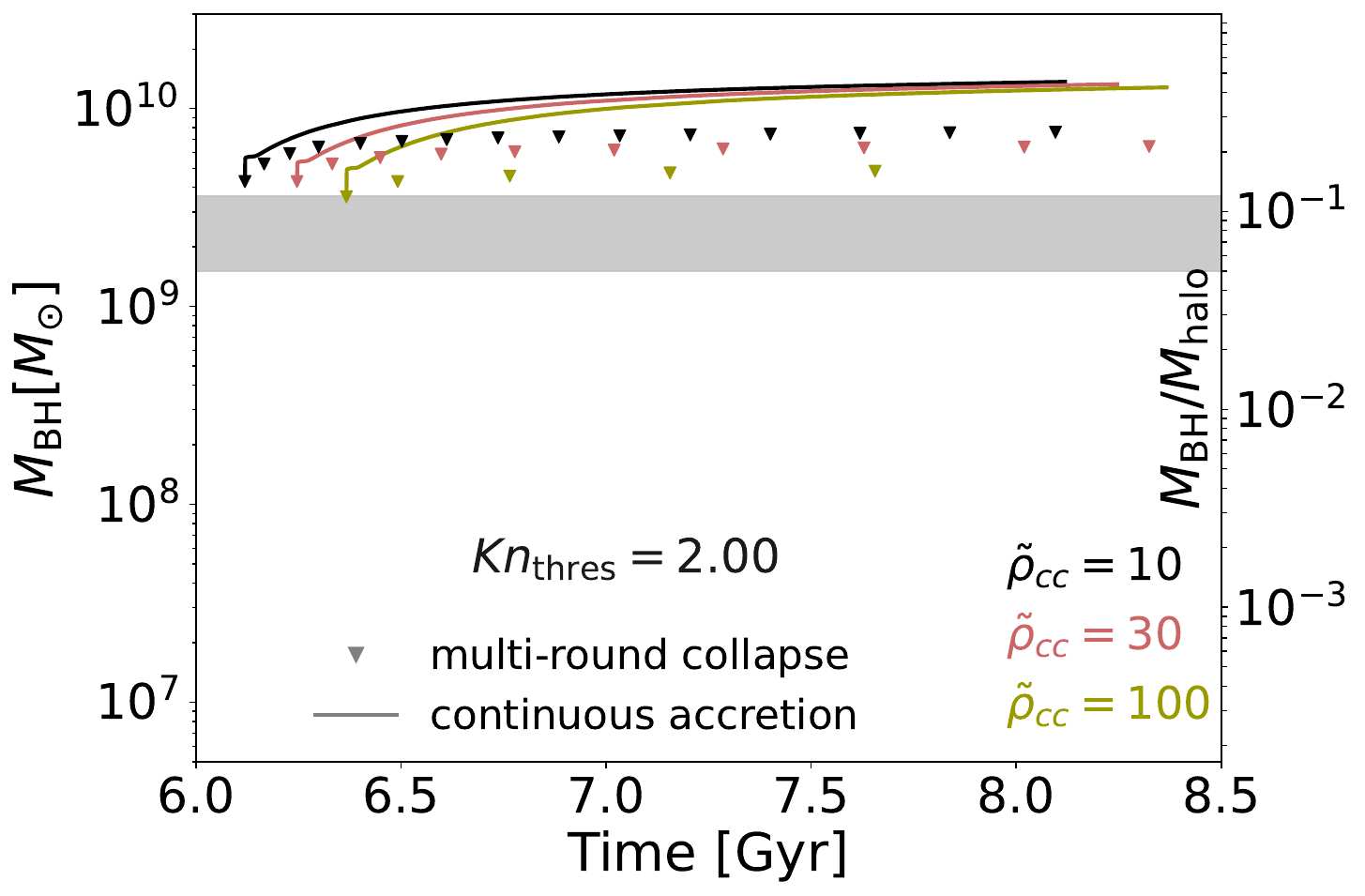}
        \caption{}
        \label{fig:multiround-kn4}    
    \end{subfigure}
    ~
    \begin{subfigure}[t]{0.45\textwidth}
        \centering
        \includegraphics[width=\textwidth, clip,trim=0.2cm 0cm 0.2cm 0cm]{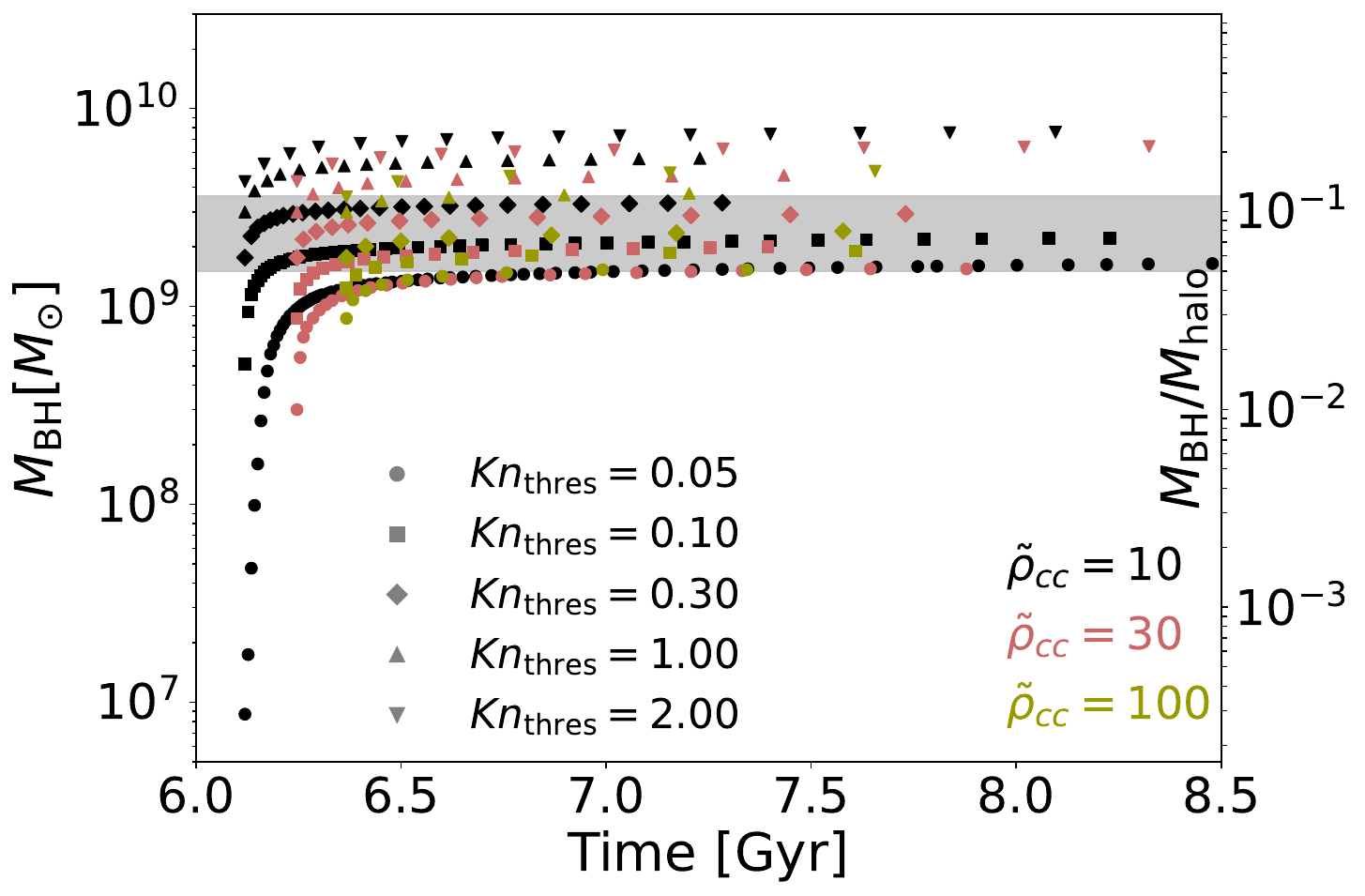}
        \caption{}
        \label{fig:multiround-all}
    \end{subfigure}
    \caption{Mass growth trajectories of the SIDM-seeded central black hole under different accretion schemes and numerical thresholds. Panels (a)--(e) illustrate the evolution across different Knudsen thresholds, $Kn_{\rm thres}=0.05, 0.1, 0.3, 1, 2$, which define the spatial boundary for DM-to-BH conversion. Scatter points track the BH mass in the multi-round core-collapse scenario; here, each marker represents a discrete simulation that corresponds to a gravothermal rebuilding cycle and the subsequent mass update. In contrast, solid curves denote the on-the-fly BH mass growth under the continuous-accretion scenario. The color coding distinguishes the manually assigned core-collapse thresholds $\tilde{\rho}_{cc}$. The shaded horizontal bands ($5\%$--$12\%$ halo mass) indicate our bounded estimate for the final BH mass achieved purely through DM accretion. Panel (f) compiles the multi-round mass trajectories across all $Kn_{\rm thres}$ values to demonstrate the asymptotic numerical convergence. }
    \label{fig:mbh-multiround}
\end{figure*}


In this section, we analyze the mass growth history of the central BH formed via SIDM core collapse, as illustrated in Fig. \ref{fig:mbh-multiround}. We employ both accretion methods detailed in the previous section and examine the numerical convergence across the parameter space of $\{\tilde{\rho}_{cc}, Kn_{\rm thres}\}$. Figures \ref{fig:multiround-kn0} through \ref{fig:multiround-kn4} compare the two accretion methods, multi-round collapse (dots) and continuous accretion (solid lines), across the three density thresholds $\tilde{\rho}_{cc}=[10, 30, 100]$. Each figure corresponds to a specific Knudsen number threshold for DM-to-BH conversion, $Kn_{\rm thres} \in[0.05, 0.1, 0.3, 1.0, 2.0]$. In these panels, the first (leftmost) dot of the multi-round scenario---or equivalently, the starting point of the continuous accretion curve---represents the formation mass and time of the initial BH seed, after which the two methods bifurcate.

\subsection{The initial conversion to BH}
 As established, a lower density threshold $\tilde{\rho}_{cc}$ is always reached earlier, guaranteeing an earlier formation time for the BH seed in all scenarios. Consequently, because the halo is intercepted at an earlier stage of collapse, one naturally expects a smaller initial BH seed mass, as fewer DM particles have condensed into the SMFP core. This expected mass hierarchy is clearly visible in Figs. \ref{fig:multiround-kn0} and \ref{fig:multiround-kn1} (where $Kn_{\rm thres}\lesssim0.1$). However, while the chronological ordering of formation times remains strictly preserved, this mass hierarchy weakens in Figs. \ref{fig:multiround-kn2} and \ref{fig:multiround-kn3}, and reverses in Fig. \ref{fig:multiround-kn4} ($Kn_{\rm thres}=2.0$). Here, the deeply collapsed $\tilde{\rho}_{cc}=100$ case paradoxically produces a smaller initial seed than the lower-threshold cases. This mass reversal can be traced back to the $Kn$--$M$ profiles in Fig. \ref{fig:primarycc-kn-m}: the curves intersect at $Kn\gtrsim0.3$. For the deepest collapse stage $\tilde{\rho}_{cc}=100$, the extremely dense thermalized core becomes spatially so compact that its outer envelope actually exhibits a lower density (and thus a higher $Kn$) at a given enclosed mass than the less-evolved halos. This crossing implies that a threshold of $Kn\gtrsim0.3$ is too large to cleanly isolate the collapsing core when the halo center is deep into the core-collapsed stage. While our goal is to demonstrate numerical convergence over the artificial choice of $\tilde{\rho}_{cc}$, this convergence must emerge from the long-term accretion history. Artificially forcing the initial seed masses to overlap by adopting an overly large $Kn_{\rm thres}$ only leads to non-convergence in the subsequent accretion phase, as demonstrated below.

\subsection{The subsequent BH mass updates}\label{sec:BH-dark-growth}

Once the DM mass within the $Kn_{\rm thres}$ boundary is converted into the initial BH seed, the divergence between the two methods, `multi-round collapse' and `continuous accretion', starts to be established.  In the multi-round collapse scenario, we iteratively repeat the core-collapse cycle: terminating the simulation upon reaching $\tilde{\rho}_{cc}$, converting the newly condensed central DM into a discrete BH mass increment, and resuming the simulation with the updated DM profile and BH mass until the collapse threshold is reached again. Consequently, in Fig. \ref{fig:multiround-all}, each dot represents the completion of one such subsequent collapse round. As shown in Figs. \ref{fig:multiround-kn0} through \ref{fig:multiround-kn4}, two prominent features emerge during the subsequent core-collapse iterations for any given $\{Kn_{\rm thres}, \tilde{\rho}_{cc} \}$ combination: the time interval between successive collapse events gradually lengthens, and the incremental BH mass gained in each round gradually decreases. This slow-down occurs because each accretion event gradually depletes the reservoir of DM available in the dense inner halo, subsequently weakening the gravothermal process and delaying the next collapse. A similar logic governs the comparison among different $\tilde{\rho}_{cc}$ cases for a fixed $Kn_{\rm thres}$: scenarios with a smaller $\tilde{\rho}_{cc}$ exhibit a higher frequency of accretion events, but each event yields a relatively smaller mass increment $\Delta M_{\rm BH}$, simply because the less stringent density threshold is more rapidly and frequently reached.  Moreover, because a larger $\tilde{\rho}_{cc}$ imposes a more stringent criterion for both triggering core collapse and executing mass updates, these larger $\tilde{\rho}_{cc}$ scenarios consistently yield a slower overall BH mass growth history compared to their lower $\tilde{\rho}_{cc}$ counterparts.

As introduced in Sec.~\ref{sec:method}, the instantaneous DM-to-BH conversion at each collapse event is a phenomenological idealization; it is nonetheless justified a posteriori by the separation of timescales between the microphysical accretion and the macroscopic refueling, which we now quantify. The DM is not absorbed truly instantaneously, but on a microphysical accretion timescale that can be estimated to order of magnitude from the Bondi rate at our macroscopic seed mass. Taking the typical seed masses and per-round increments from Fig.~\ref{fig:mbh-multiround} (several $\times10^8\,M_\odot$ and $\lesssim10^8\,M_\odot$, respectively), the core velocity dispersion $\sigma_v\sim120$--$150\,\rm km\,s^{-1}$, and the SMFP core density implied by $Kn_{\rm thres}=0.05$, the Bondi rate is $\dot M\sim10^2\,M_\odot\,{\rm yr}^{-1}$, so each accretion interval is completed within $t_{\rm acc}=\Delta M_{\rm BH}/\dot M\sim\mathcal{O}(1)\,\rm Myr$. This is roughly two orders of magnitude shorter than the $\sim100\,\rm Myr$ gravothermal rebuilding that refuels the core between rounds, so the instantaneous treatment is adequate on the timescales of interest. This estimate is independently corroborated by the spherically symmetric fluid simulations of \cite{meng26}, which resolve the microphysical accretion directly and likewise find the central DM to be consumed on Myr timescales.

In the continuous accretion scenario, the discrete `cadence' of events is naturally absent. Nevertheless, a similar overarching trend remains: at any given time, $t$, on the simulation timeline, the smaller $\tilde{\rho}_{cc}$ case always maintains a larger $M_{\rm BH}(t)$, driven by the `head start' effect of its earlier initial formation. Furthermore, the continuous accretion approach consistently produces a higher $M_{\rm BH}(t)$ trajectory than the multi-round scenario across all tested parameters. This monotonic ordering is consistent with our bracketing methodology laid out in Sec.~\ref{sec:method}: the multi-round scheme restricts accretion to discrete core-collapse events, whereas the continuous scheme bypasses the gravothermal rebuilding delay. We therefore regard the two trajectories as delineating a plausible range of BH mass growth histories, within which the true evolution is expected to lie.

\subsection{The numerical convergence}\label{sec:convergence}

As discussed in the methods section, the numerical parameters  $\{\tilde{\rho}_{cc}, Kn_{\rm thres} \}$ are not derived from first principles and thus carry the potential to introduce numerical artifacts. Therefore, a primary goal of this study is to identify an optimal parameter set that minimizes these artifacts before drawing robust conclusions regarding the BH mass growth history. While we caution the readers that numerical convergence does not always guarantee physical correctness, physical correctness strictly requires numerical convergence. Thus, establishing this convergence is crucial for extracting the most reliable results from our numerical framework. Observing Figs. \ref{fig:multiround-kn0} through \ref{fig:multiround-kn4}, the convergence of the multi-round approach across the discrete density thresholds $\tilde{\rho}_{cc}\in[10,30,100]$ noticeably improves as $Kn_{\rm thres}$ decreases, achieving excellent visual convergence at $Kn_{\rm thres}=0.05$. This indicates that at $Kn_{\rm thres}=0.05$, the long-term BH mass trajectory becomes effectively independent of the artificial termination threshold $\tilde{\rho}_{cc}$. Simultaneously, at $Kn_{\rm thres}=0.05$, the artificial `head start' effect inherent to the continuous accretion scenario is minimized because the initial BH seed mass is less severely overestimated. This configuration therefore yields the lowest upper-bound BH mass among all $Kn_{\rm thres}$ variations, while also demonstrating rapid convergence. Consequently, we identify the region bounded by these two accretion methods at $Kn_{\rm thres}=0.05$, spanning approximately 5$\%$ to 12$\%$ of the halo mass (represented by the gray horizontal bands in Figs. \ref{fig:multiround-kn0} to \ref{fig:multiround-kn4}), as the robust prediction for the final BH mass driven purely by DM accretion. The relevant timescale for the initial BH seed to grow to such macroscopic masses is approximately 1--1.5 Gyr.  

Having established convergence over $\tilde{\rho}_{cc}$ at a fixed $Kn_{\rm thres}$, we must also evaluate convergence over $Kn_{\rm thres}$ at a fixed $\tilde{\rho}_{cc}$. The goal is to determine an optimal $\tilde{\rho}_{cc}$ value to recommend for future studies. In other words, we investigate whether a specific $\tilde{\rho}_{cc}$ can enforce convergence across different $Kn_{\rm thres}$ choices, or at least exhibit asymptotic signs of converging towards the smaller $Kn_{\rm thres}$ results. To this end, Fig. \ref{fig:multiround-all} compiles all multi-round collapse trajectories spanning the entire  $\{ \tilde{\rho}_{cc}, Kn_{\rm thres}\}$ parameter space. As $\tilde{\rho}_{cc}$ increases from 10 (black points) to 100 (olive points), the overall spread of the BH mass growth curves narrows significantly. However, absolute convergence across all $Kn_{\rm thres}$ values is not yet fully realized even at the highest threshold tested $\rho_{cc}=100$. For instance, the final BH masses for the $\{\tilde{\rho}_{cc}=100, Kn_{\rm thres}=0.05\}$ and $\{\tilde{\rho}_{cc}=100, Kn_{\rm thres}=0.1\}$ cases are $1.53\times10^9M_\odot$ and $1.91\times10^9M_\odot$ respectively, leaving a residual discrepancy of $\lesssim20\%$. Pushing to thresholds higher than $\tilde{\rho}_{cc}=100$ is computationally prohibitive and physically questionable: the extreme central densities demand impractically small timesteps to resolve SIDM scattering probabilities, and the severe fluid-like nature of the deep SMFP core invalidates the discrete N-body approach.

Given this convergence analysis, we conclude that  $Kn_{\rm thres}=0.05$ constitutes the most robust numerical choice for defining the DM-to-BH conversion boundary in our framework. Although convergence over $\tilde{\rho}_{cc}$ remains asymptotic rather than absolute, larger $\tilde{\rho}_{cc}$ thresholds capture the halo in a more deeply evolved core-collapse stage. Therefore, we consider them physically more reliable and recommend adopting $\tilde{\rho}_{cc}=100$ for future isolated SIDM core-collapse simulations.

\subsection{Extrapolating to smaller $Kn_{\rm thres}$}\label{sec:kn-extrapolation}

\begin{figure}
    \centering
    \begin{subfigure}[t]{0.45\textwidth}
        \centering
        \includegraphics[width=\textwidth, clip,trim=0.2cm 0cm 0.2cm 0cm]{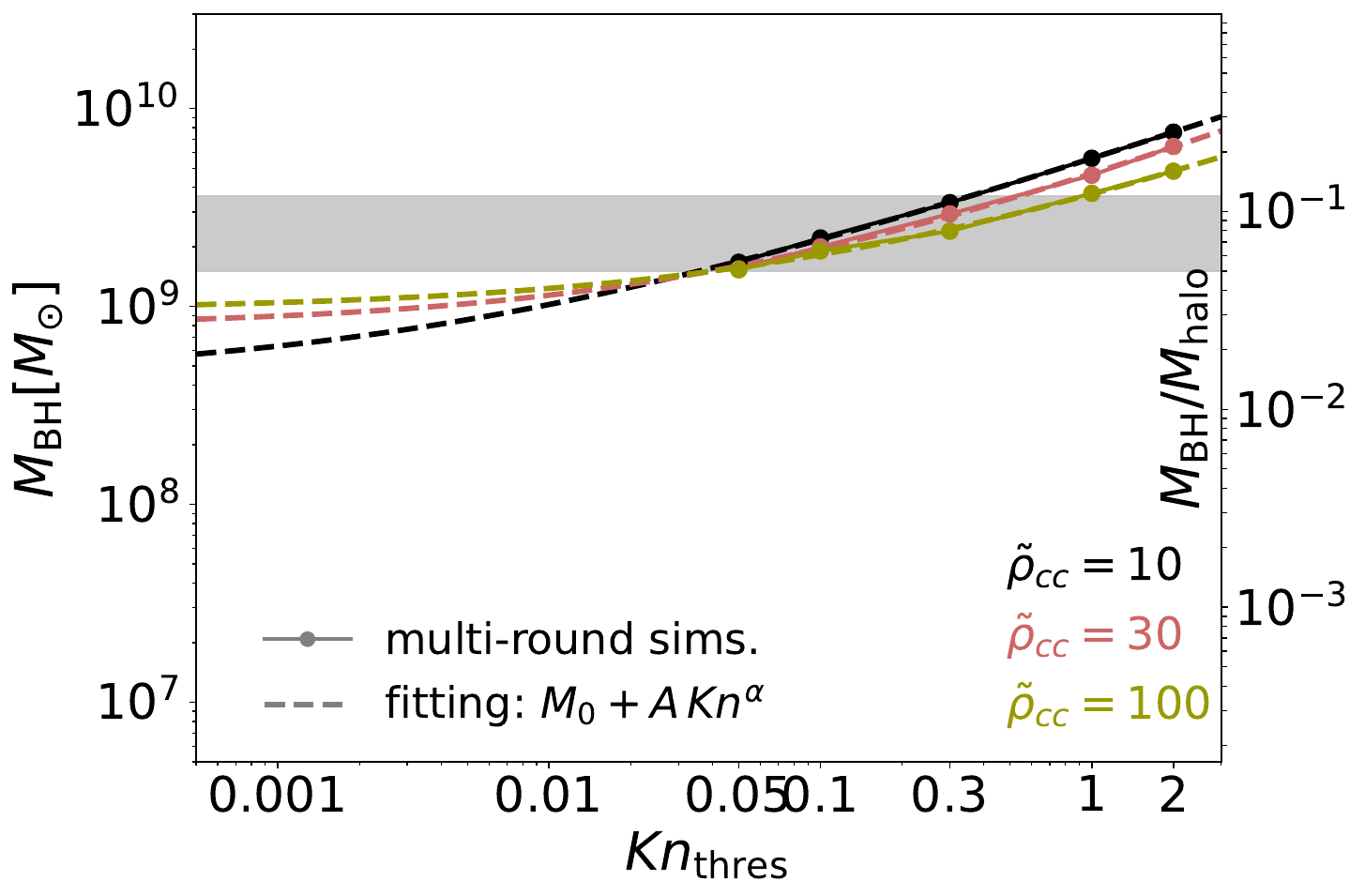}
        \caption{}
        \label{fig:mbh-kn-multiround}
    \end{subfigure}
    ~
    \begin{subfigure}[t]{0.45\textwidth}
        \centering
        \includegraphics[width=\textwidth, clip,trim=0.2cm 0cm 0.2cm 0cm]{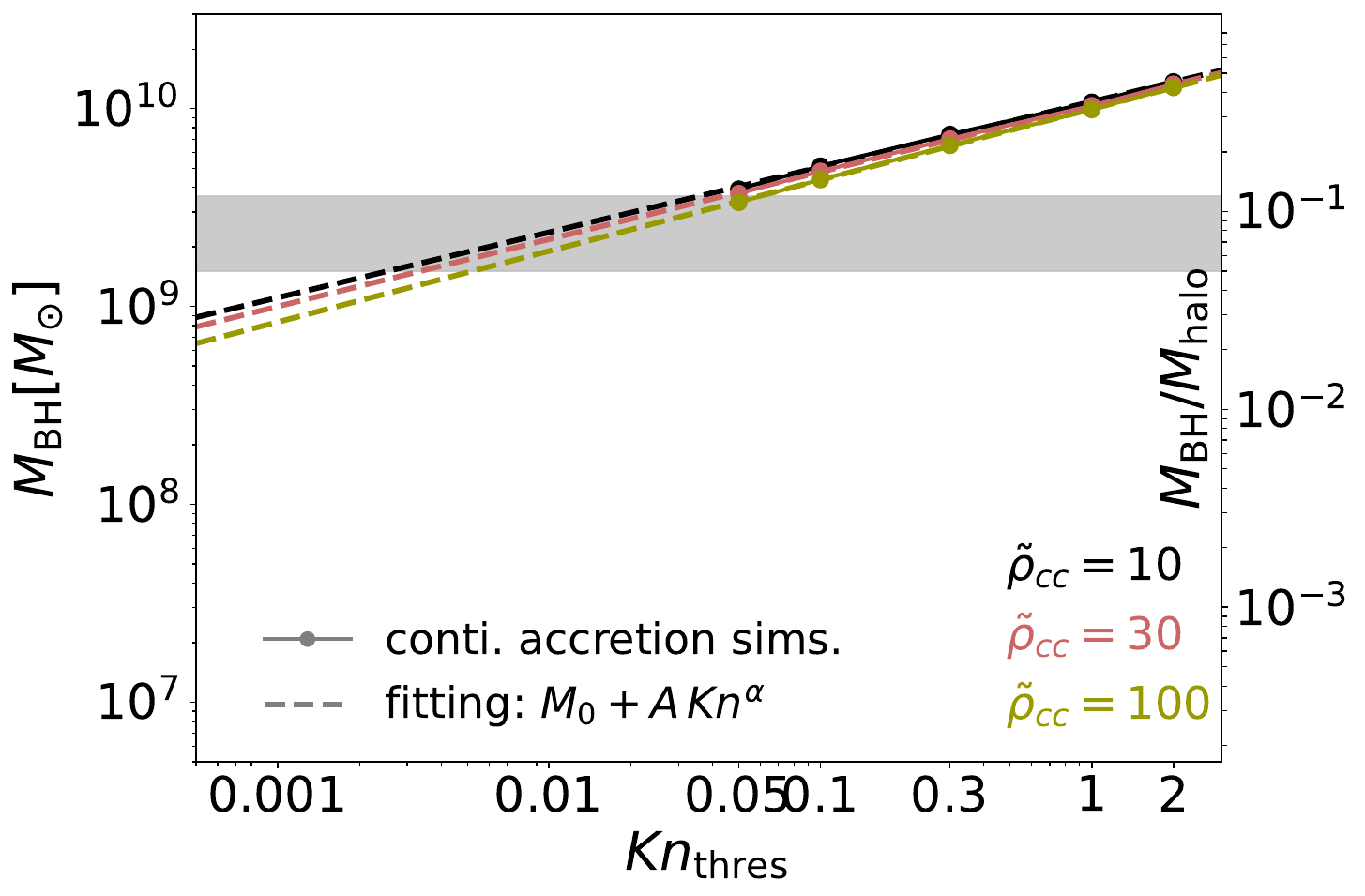}
        \caption{}
        \label{fig:mbh-kn-conti}
    \end{subfigure}
    \caption{Final black hole mass as a function of the adopted threshold $Kn_{\rm thres}$ for the multi-round (upper) and continuous-accretion (lower) scenarios, for tested core-collapse density thresholds $\rho_{cc}$. Simulation results are shown as dots, while fits of the form $M_{\rm BH}=M_0+A\,Kn^{\alpha}$ are shown as dashed lines, used to extrapolate to smaller $Kn_{\rm thres}$.}
    \label{fig:mbh-kn}
\end{figure}

Having established convergence over $\tilde{\rho}_{cc}$ at fixed $Kn_{\rm thres}$ and identified $Kn_{\rm thres}=0.05$ as the most robust choice, a natural follow-up is whether the results remain stable if the threshold could be pushed even lower. However, our current approach is severely limited at resolving smaller scales, both because of the prohibitive computational cost and the unavoidable discrete noise when approaching more fluid-like regime. These prevent us from directly resolving $Kn_{\rm thres}<0.05$. To probe this regime, we instead fit the measured $M_{\rm BH}$--$Kn_{\rm thres}$ relations and extrapolate, as shown in Fig.~\ref{fig:mbh-kn}.

In Fig.~\ref{fig:mbh-kn}, the raw data reproduce the convergence behavior established above: the multi-round final masses across the three $\tilde{\rho}_{cc}$ thresholds converge as $Kn_{\rm thres}$ decreases toward 0.05, while the continuous-accretion masses show no sign of plateauing. To push beyond the directly resolvable range, we fit the $M_{\rm BH}$--$Kn_{\rm thres}$ relation with $M_{\rm BH}=M_0+A\,Kn^{\alpha}$ in
both panels. Given only five data points per fit and three free parameters, these fits should be read as qualitative rather than quantitative; nevertheless, they still yield useful insight. This fit reproduces the convergence near $Kn_{\rm thres}=0.05$, where the three fitted curves nearly intersect; when extrapolated to smaller $Kn_{\rm thres}$, however, they separate again, with $M_0$ spanning $\sim1.5\%$--$3\%$ of the halo mass---only modestly below the established $\sim5\%$ lower bound. Given the low statistics of these fits, this re-separation should not be over-interpreted.
In contrast, the continuous-accretion final masses decrease monotonically with $Kn_{\rm thres}$ ($M_0\to0$), consistent with our reading of those results as a loose, unconverged upper bound. This contrast reinforces our conclusion that the multi-round estimate is the more robust one.

\subsection{SIDM vs. CDM in continuous accretion}

\begin{figure}
    \centering
    \begin{subfigure}[t]{0.45\textwidth}
        \centering
        \includegraphics[width=\textwidth, clip,trim=0.2cm 0cm 0.2cm 0cm]{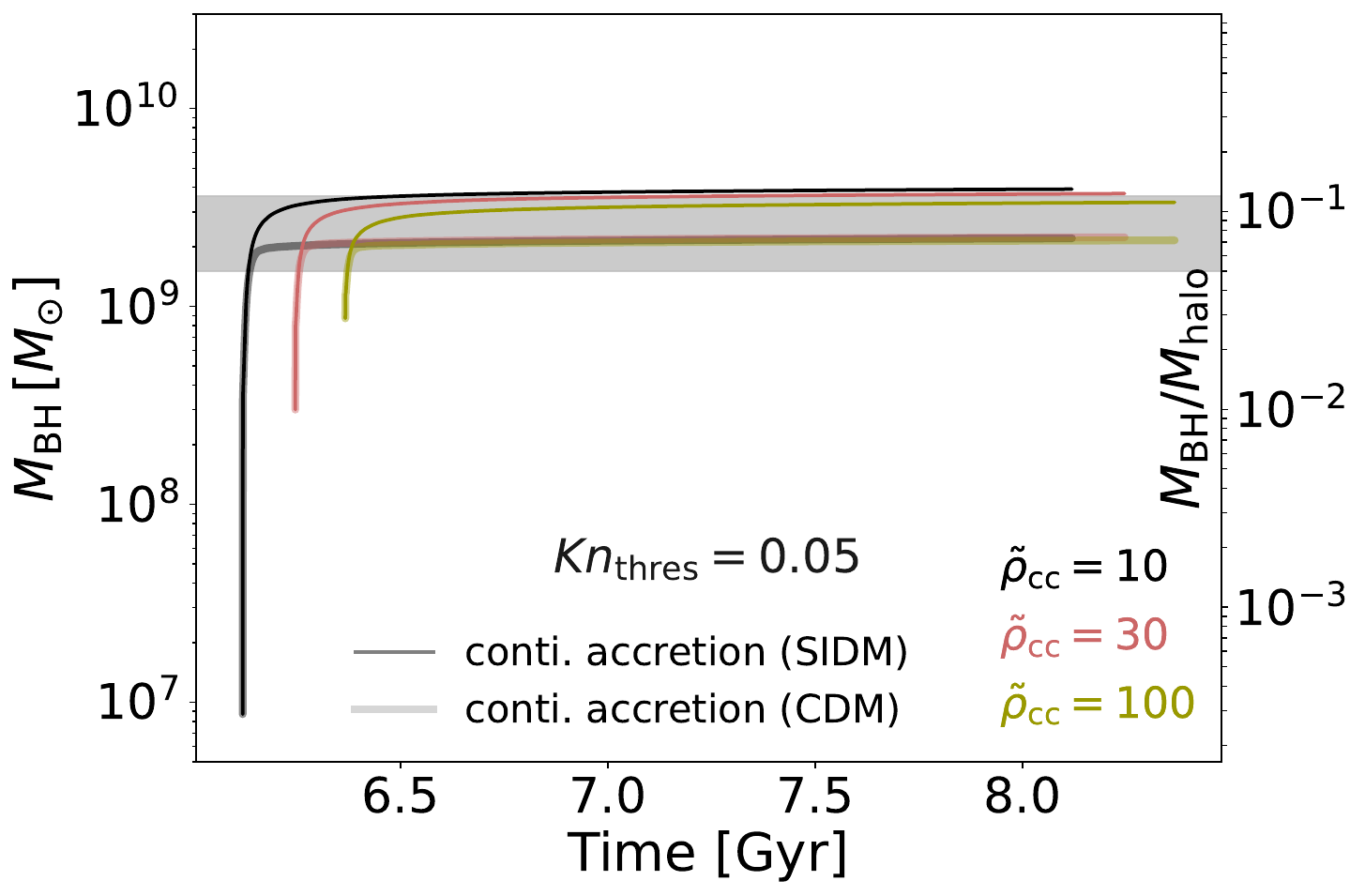}
        \caption{}
        \label{fig:sidm-cdm-1}
    \end{subfigure}
    ~
    \begin{subfigure}[t]{0.45\textwidth}
        \centering
        \includegraphics[width=\textwidth, clip,trim=0.2cm 0cm 0.2cm 0cm]{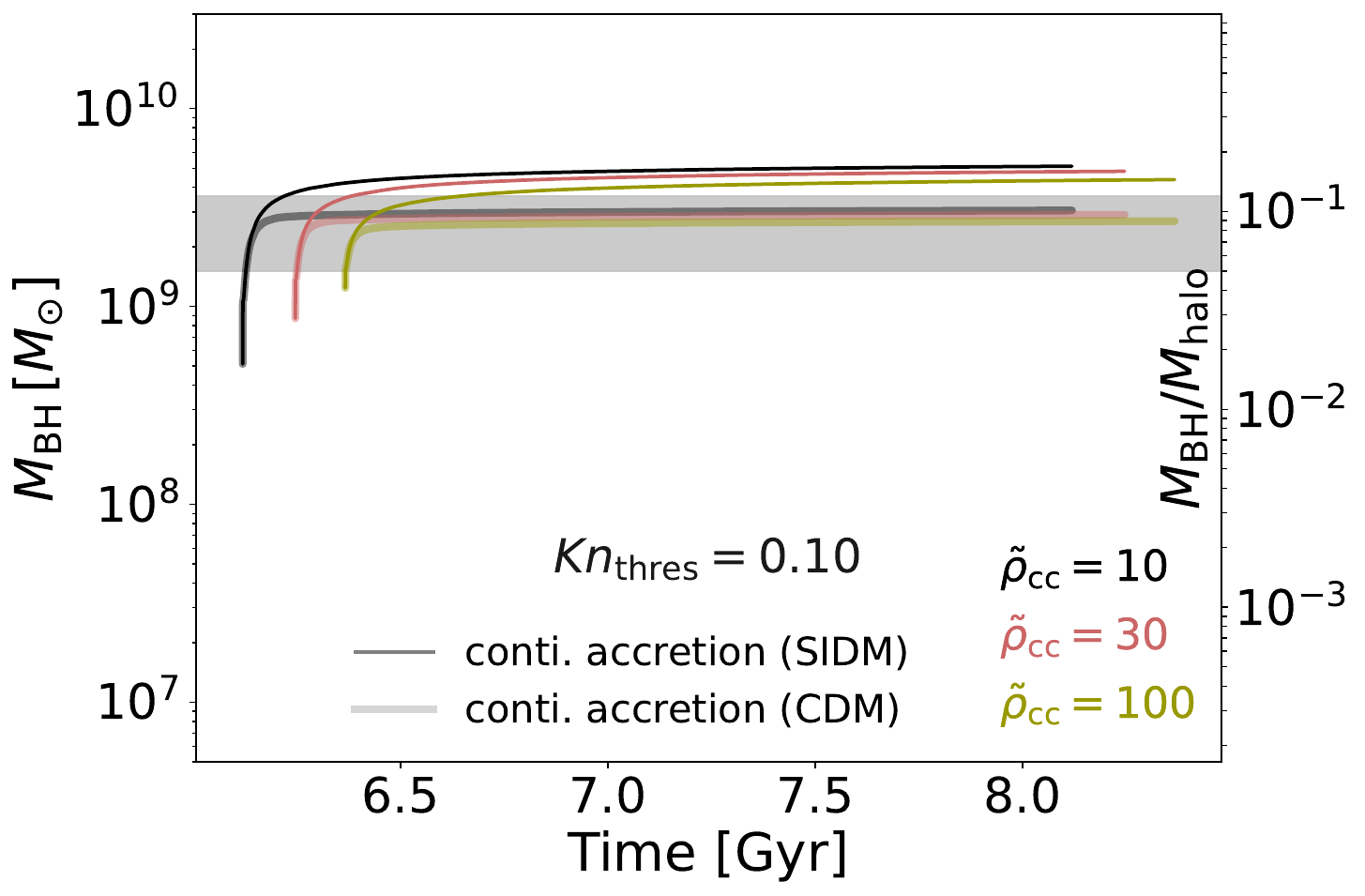}
        \caption{}
        \label{fig:sidm-cdm-2}
    \end{subfigure}
    \caption{Comparison of the BH mass growth trajectories between the SIDM (thin solid lines) and CDM (thick solid lines) control models in the continuous-accretion scenario. This comparison isolates the enhancement in DM accretion driven by SIDM gravothermal dynamics, as opposed to the purely gravitational feeding of CDM. }
    \label{fig:sidm-cdm}
\end{figure}

In the preceding sections, we established the dynamics of subsequent DM accretion onto the initial BH seeded by the primary core collapse. For consistency, both the multi-round core-collapse and continuous accretion scenarios maintained the same constant SIDM cross section throughout the entire evolution. However, this raises an important physical question: is the rapid BH mass growth driven solely by the deep central gravitational potential, or is it significantly fueled and accelerated by ongoing SIDM scatterings? This distinction is particularly crucial for realistic velocity-dependent SIDM models, where the extreme velocity dispersion near the massive BH would naturally suppress the cross section to near-zero. To investigate this limit, we perform a control simulation using the continuous accretion method: after the primary DM-to-BH conversion, we resume the simulation with the central BH intact, but explicitly turn off DM self-interactions, effectively treating the remainder of the halo as collisionless CDM. To ensure a fair comparison, we enforce the exact same $Kn_{\rm thres}$ accretion criteria for this now-effectively-CDM halo. In this setup, the nominal cross section of $\sigma/m=100\ \rm cm^2/g$ is used purely as a mathematical proxy to evaluate the local $Kn$ condition, without participating in the actual dynamical evolution. This approach allows us to test the purely gravitational baseline of our continuous accretion mechanism, thereby isolating and highlighting the specific enhancement provided by self-interactions. Note that this CDM control test can only be conducted within the continuous accretion framework; the multi-round approach inherently relies on successive gravothermal core collapse, which requires self-interactions to occur.

We present the results of this comparison in Fig. \ref{fig:sidm-cdm}, focusing on our two best-converged threshold configurations: $Kn_{\rm thres}=0.05$ and $Kn_{\rm thres}=0.1$. In both cases, the collisionless CDM scenario yields significantly less BH mass growth than the fully interacting SIDM scenario, trailing by approximately a factor of two. This confirms that while the deep central potential alone drives substantial accretion, SIDM-driven gravothermal dynamics plays a crucial role in efficiently gathering outer halo materials to feed the central BH. Furthermore, comparing the CDM accretion histories between $Kn_{\rm thres}=0.05$ and $Kn_{\rm thres}=0.1$ reveals that an overestimated initial BH seed mass persistently leads to an inflated final BH mass, independent of the underlying SIDM physics. This reinforces our earlier conclusion: employing a strict, convergent $Kn_{\rm thres}$ is a necessity for accurately modeling subsequent BH mass growth in this and future works.

We stress that for the CDM cases, this $Kn$-proxy criterion does not reproduce true collisionless accretion. A CDM particle, unlike SIDM cases which are more fluid-like in the SMFP limit, can be captured only if its orbit carries it within the innermost stable circular orbit (ISCO, a few Schwarzschild radii), some six orders of magnitude inside the $\sim0.2$ kpc boundary set by $Kn_{\rm thres}=0.05$.
The CDM control therefore overestimates purely gravitational feeding and should be read as an upper bound; correspondingly, the factor-of-two SIDM enhancement in Fig.~\ref{fig:sidm-cdm} is a conservative lower limit on the true SIDM--CDM contrast.

\subsection{Including Eddington accretion of baryons}\label{sec:eddington}

\begin{figure}
    \centering
    \begin{subfigure}[t]{0.48\textwidth}
        \centering
        \includegraphics[width=\textwidth, clip,trim=0.2cm 0cm 0.2cm 0cm]{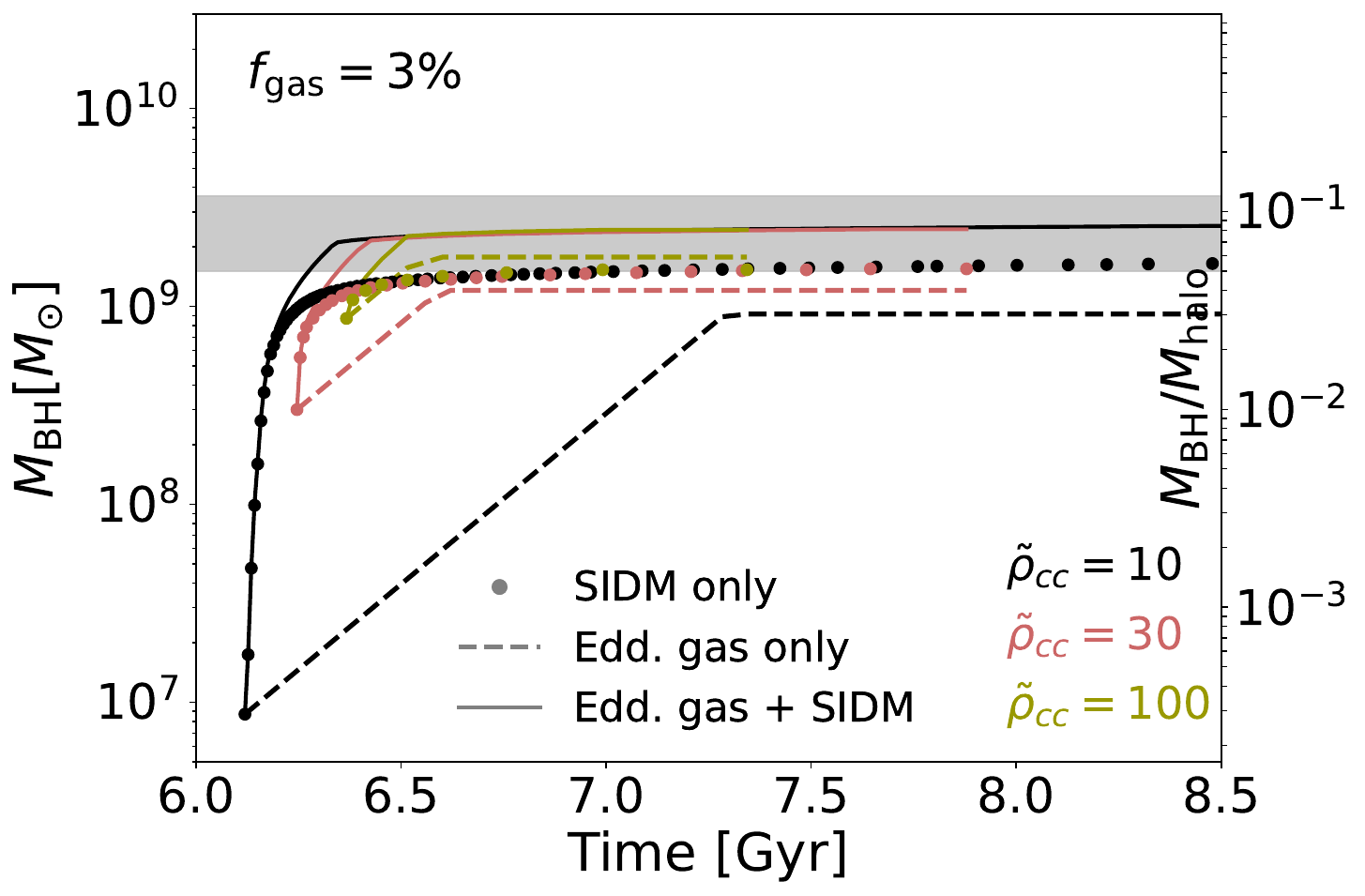}
        \caption{}
        \label{fig:edd-1}
    \end{subfigure}
    ~
    \begin{subfigure}[t]{0.48\textwidth}
        \centering
        \includegraphics[width=\textwidth, clip,trim=0.2cm 0cm 0.2cm 0cm]{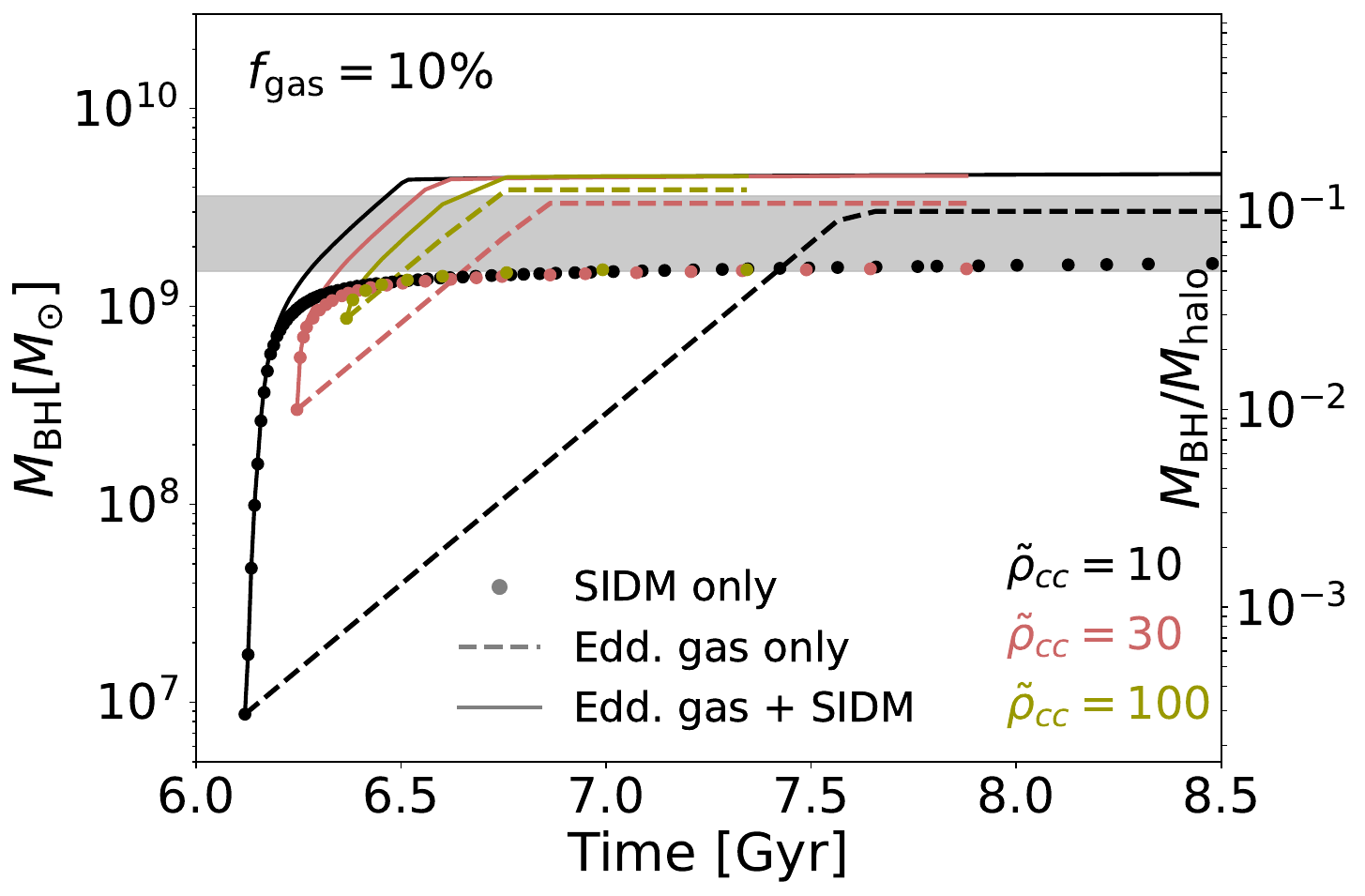}
        \caption{}
        \label{fig:edd-2}
    \end{subfigure}
    \caption{Mass growth trajectories of the central black hole, incorporating both DM accretion from the multi-round scenario and a first-order estimate of baryonic accretion. Three evolutionary tracks are compared: pure SIDM accretion (scatter points), pure baryonic accretion (dashed lines), and the combined co-evolution driven by both SIDM and baryonic accretion (solid lines). Baryonic growth is modeled by the minimum of the Eddington and Bondi accretion rates of gas, though we find the Eddington limit consistently dominates. We evaluate total available gas reservoirs of $f_{\rm gas}=3\%$ and $f_{\rm gas}=10\%$ for gas-poor and gas-rich systems, assuming a fiducial Eddington accretion efficiency of $f_{\rm Edd}=0.2$. Driven by rapid gas depletion, the final BH mass can be approximated as a simple linear superposition of the pure DM baseline and the total available gas mass. }
    \label{fig:edd}
\end{figure}

Thus far, our discussion has focused on the mass growth of the central BH formed via SIDM core collapse, considering only the accretion of DM. However, baryonic accretion constitutes another crucial ingredient. For instance, \cite{fz25} employed purely Eddington-limited baryonic accretion to model the subsequent growth of an initial SIDM-seeded BH (assumed to be $\sim1\%$ of the halo mass). Similarly, \cite{wxfeng25} utilized early-time Eddington gas accretion to grow a microscopic $\mathcal{O}(1)M_\odot$ seed, which is triggered directly by the relativistic instability \cite{sophia23}, up to $\mathcal{O}(10^4)M_\odot$. Building upon our converged results for DM-driven BH growth, we now incorporate a first-order toy model to estimate the complementary mass growth contributed by baryonic accretion.

For illustrative purposes, we introduce a reasonable simplification: modeling the baryon-fed growth as accretion from a single gas reservoir subject to both Eddington and Bondi limits. Stellar mass and other baryonic channels are not explicitly included. We model this gas reservoir as a uniform-density sphere with a radius of $0.37$ kpc (corresponding to the half-light radius of the host halo's galaxy utilized in \cite{zzc24}) and a total mass of $f_{\rm gas}\times M_{200c}$, where the gas mass fraction $f_{\rm gas}$ serves as a free parameter. Here $f_{\rm gas}$ should be interpreted as the total gas budget accumulated over the halo's cosmic life, i.e., a late-time, low-redshift value, rather than the instantaneous gas content at any single epoch. Our fiducial choices of $3\%$ and $10\%$ are thus representative of the gas fractions retained by low-redshift dwarfs, whereas a value closer to the cosmic baryon fraction would apply at the high redshifts of the LRD population \cite{fz25}. Following \cite{wxfeng25}, we evaluate two distinct limits for baryonic accretion: the radiation pressure limit (Eddington) and the hydrodynamic feeding limit (Bondi). The Eddington accretion rate is given by

\begin{equation}
    \frac{dM_{\rm Edd}}{dt} = \frac{1-\epsilon}{\epsilon}f_{\rm Edd}\frac{4\pi G M_{\rm BH}}{c(\sigma_{\rm radiation}/m)},
\end{equation}

where we adopt a standard radiative efficiency of $\epsilon=0.1$, a fiducial accretion efficiency of $f_{\rm Edd}=0.2$ following the setup of \cite{fz25} (with the maximal case $f_{\rm Edd}=1$ explored as an extreme upper limit in the appendix), and the Thomson scattering cross section $\sigma_{\rm radiation}/m=0.4\ \rm cm^2/g$. The Bondi accretion rate is given by

\begin{equation}\label{eqn:bondi}
     \frac{dM_{\rm Bon}}{dt} = 4\pi \lambda_s \frac{G^2M_{\rm BH}^2}{v_{\rm sound}^3}\rho_{\rm gas},
\end{equation}

where $\lambda_s=0.25$ is an $\mathcal{O}(1)$ accretion eigenvalue, and $v_{\rm sound}$ is the sound speed of the gas \cite{morton21}:

\begin{equation}
    v_{\rm sound}=\sqrt{\frac{\gamma k_B T_{\rm vir}}{\mu m_p}}.
\end{equation}

Here, $k_B$ is the Boltzmann constant, $m_p$ is the proton mass, $\gamma=5/3$ is the adiabatic index for a monatomic gas, and $\mu$ is the mean molecular weight. Assuming a fully ionized primordial gas, we set $\mu=0.6$. The virial temperature, $T_{\rm vir}$, assuming an isothermal sphere, is given by:

\begin{equation}
    T_{\rm vir} = \frac{1}{3} \frac{\mu m_p}{k_B} \frac{GM_{200c}}{r_{200c}}.
\end{equation}

Note that while our Eddington accretion prescription follows the standard setup, our application of the Bondi formula differs from that of \cite{wxfeng25}. Specifically, \cite{wxfeng25} employed a \textit{dark} Bondi accretion scheme to model the influx of SIDM particles onto the central BH, whereas we utilize the standard Bondi formula to model the accretion of baryonic gas. Because both the Eddington and Bondi rates represent physical upper limits on gas accretion, the effective baryonic accretion rate is governed by the minimum of the two, capped by the remaining gas mass:

\begin{equation}\label{eqn:BH-gas}
    \Delta M_{\rm BH-gas}=\min(\dot{M}_{\rm Bon}\,\Delta t,\ \dot{M}_{\rm Edd}\,\Delta t,\ m_{\rm gas}(t)),
\end{equation}

where $\Delta t$ is the discrete mass-update interval---one multi-round core-collapse cycle---and $m_{\rm gas}(t)$ is the gas mass remaining at the start of that interval. At the conclusion of each cycle, we update the BH mass and deduct the accreted increment $\Delta M_{\rm BH-gas}$ from the remaining gas reservoir (and also re-calculate the gas density in Eqn. \ref{eqn:bondi}). The explicit inclusion of the available gas mass, $m_{\rm gas}(t)$ introduces a key physical difference from the analytical Eddington modeling in \cite{wxfeng25}. In their framework, a simple unconstrained exponential growth is sufficient because the initial seed is microscopic and gas depletion is negligible in the early stages. In our scenario, however, as we will demonstrate below, the finite gas reservoir eventually becomes the fundamental limiting factor for baryonic accretion.

The corresponding mass growth of the BH via baryonic accretion is shown in Fig. \ref{fig:edd}.
In this analysis, we examine two scenarios. First, the initial central BH---seeded by the primary core-collapse simulation---grows exclusively by accreting gas, without any further accretion of SIDM. This is achieved by updating the BH mass using only the $dM_{\rm BH-gas}$ component at the end of each discrete simulation round. The results of this gas-only scenario are represented by the dashed lines in Fig. \ref{fig:edd}. Because we empirically find that the Eddington limit is consistently more stringent than the Bondi limit in Eqn. \ref{eqn:BH-gas}, Eddington radiation pressure effectively becomes the sole limiting mechanism on gas accretion. Thus, we denote this baryonic channel as `Edd. gas' in the figure legends. Second, we incorporate both DM and baryonic accretion. To achieve this within our multi-round collapse framework, we compute the differential between adjacent $M_{\rm BH}(t)$ data points from the DM-only simulations. This naturally yields the effective DM accretion increment, $dM_{\rm BH-DM}$, across each discrete interval bounded by the resumption and termination of the multi-round simulations. We then evaluate the total mass increment $dM_{\rm BH}=dM_{\rm BH-gas}+dM_{\rm BH-DM}$ at each discrete mass-update interval, feeding the updated BH mass back into Eqn. \ref{eqn:BH-gas} to simultaneously track both the DM and baryonic contributions. The resulting mass growth trajectories are shown as solid lines in Fig. \ref{fig:edd}. We note a methodological caveat: because the $dM_{\rm BH-DM}$ increments are extracted from pure DM-only simulations, our post-processing approach incorporates the boosted gas accretion due to the underlying DM accretion, but neglects the reciprocal feedback.  In reality, the rapid BH mass gain from gas accretion would deepen the central potential, potentially accelerating the gravothermal collapse and the subsequent DM accretion. Nevertheless, we argue that this missing positive feedback does not severely compromise our toy model estimation. As evident in Fig. \ref{fig:edd}, the gas accretion saturates very rapidly, limiting the time window over which this bias accumulates.

Fig. \ref{fig:edd} displays two panels corresponding to a relatively gas-poor ($f_{\rm gas}=3\%$) and a gas-rich ($f_{\rm gas}=10\%$) scenario. In both cases, the Eddington-only curves saturate at $M_{\rm BH,ini}+M_{\rm gas}$, fundamentally capped by the total available gas reservoir. In contrast, the combined Eddington+SIDM curves saturate at $M_{\rm BH,DMO}+M_{\rm gas}$, where $M_{\rm BH,DMO}$ represents the final BH mass achieved via SIDM-only accretion described in previous sections. This indicates that, to a first approximation, the final BH mass can be treated as the linear superposition of the SIDM-only theoretical limit and the total available gas mass. This rapid saturation mechanism is driven by the rapid initial growth of the BH when Eddington accretion is activated, as we can also see from the steep slopes of Eddington cases in Fig. \ref{fig:edd}. The early-time competition between gas and DM accretion is governed by an interplay between the initial BH mass ($\dot {M}_{\rm Edd}\propto M_{\rm BH}$) and the local DM density. For a shallower collapse ($\tilde{\rho}_{cc}=10$), the smaller initial BH seed yields a lower Eddington rate, while the surrounding DM envelope retains a relatively high density, allowing DM accretion to initially dominate (black curves). Conversely, for deeper collapse stages ($\tilde{\rho}_{cc}\geq30$) the larger initial BH seed exponentially boosts the Eddington rate. Simultaneously, because the core is heavily depleted into the BH, the local DM density in the immediate envelope drops, suppressing early DM accretion. Coupled with the macroscopic bottleneck of gravothermal rebuilding (the DM reservoir effect), the continuous Eddington gas accretion easily outpaces the delayed DM feeding in these deeply collapsed scenarios, rapidly consuming the available baryon reservoir.

This result qualitatively validates the purely baryonic approach explored in \cite{fz25}, whose Eddington-based gas growth prescription we also follow, albeit with an explicit cap on the available gas reservoir. However, our analysis highlights that neglecting the DM-contributed mass growth can lead to a significant underestimation of the final BH mass, typically by several tens of percent, and up to a factor of two, depending on the halo's gas fraction and SIDM cross section. More importantly, this toy model illustrates a synergistic picture: while baryonic accretion is a crucial ingredient for early rapid growth, the purely DM-driven mass limit serves as the underlying baseline. Fortunately, because the gas reservoir is rapidly consumed, the final BH mass can be conveniently approximated by a simple linear superposition of the two components. We emphasize that this additive $+f_{\rm gas}$ contribution is a \emph{supply-limited} upper bound, rather than a rate-limited one: the central BH is such an efficient accretor that gas arriving near the center is readily captured, so the realized mass is expected to lie close to this bound rather than far below it. Two neglected effects act on the trajectory without altering this endpoint. First, in reality the gas is fed into the halo gradually over cosmic time rather than being available from the outset; this slows the early growth and flattens the curves in Fig.~\ref{fig:edd}, but leaves the final BH mass unchanged, since the total budget is the same. Second, gas outflow and feedback would expel a fraction of the reservoir from the halo, lowering the endpoint itself; quantifying this is beyond the scope of this work and constitutes the primary uncertainty on the baryonic contribution. This additive property greatly simplifies future theoretical modeling, making it straightforward to adjust $f_{\rm gas}$ and compare with high redshift supermassive black holes.

\subsection{Comparison with other works}

\begin{figure*}
    \centering
        \centering
        \includegraphics[width=\textwidth, clip,trim=0.2cm 0cm 0.2cm 0cm]{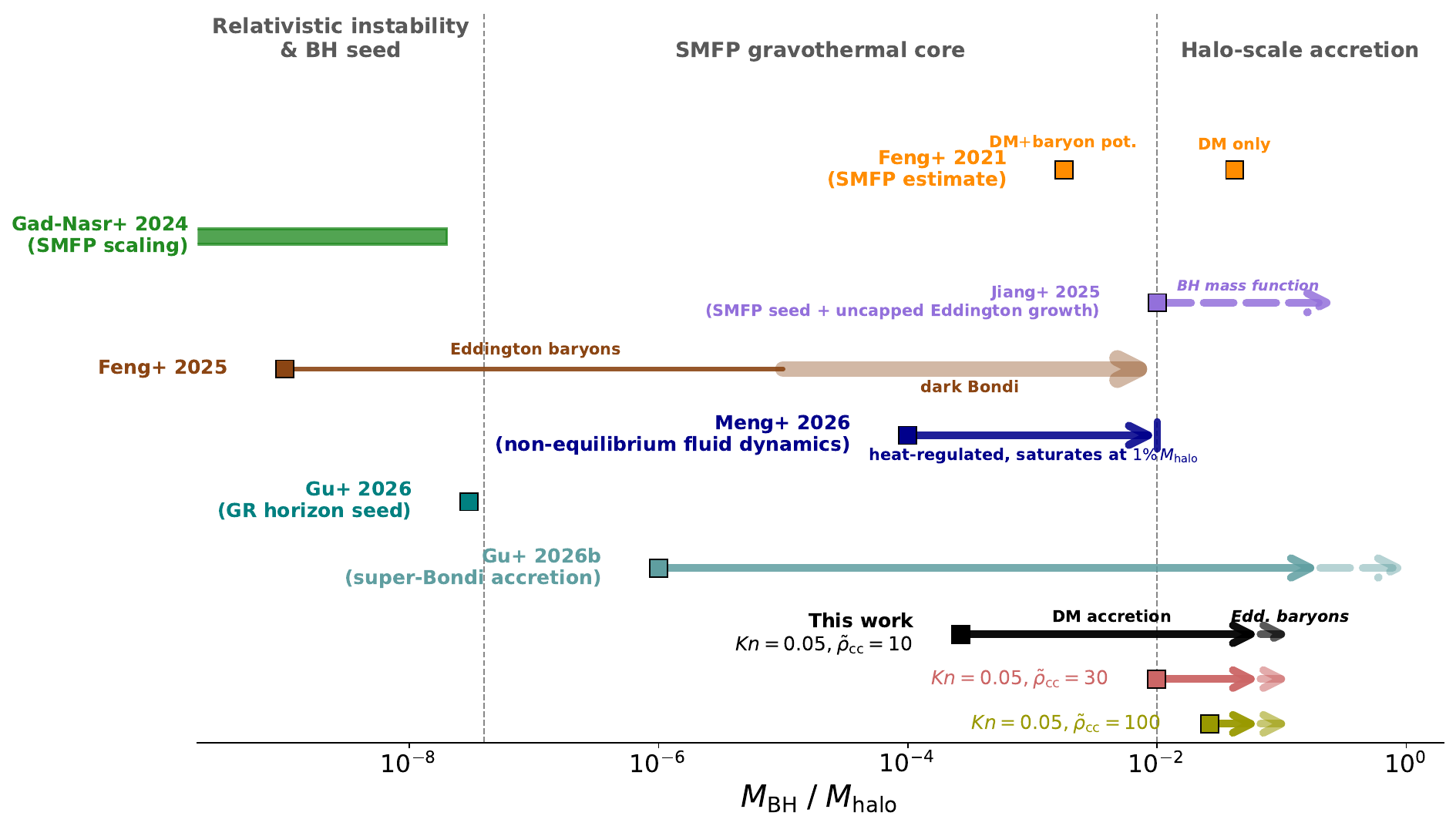}
    \caption{A schematic overview comparing contemporary theoretical works for SIDM-seeded black holes with our macroscopic co-evolution results in this work. For each referenced study, we illustrate the predicted BH mass scales along the BH growth trajectory, highlighting the physical mechanisms focused on and the key approaches employed. The references listed here are (top to bottom): Feng+ 2021 \cite{wxfeng21}, Gad-Nasr+ 2024 \cite{sophia23}, Jiang+ 2025 \cite{fz25}, Feng+ 2025 \cite{wxfeng25}, Meng+ 2026 \cite{meng26}, and Gu+ 2026a,b \cite{gu26, gu26b}.}
    \label{fig:schematic}
 \end{figure*}

In this section, we compare our results on SIDM-formed BHs with other relevant works, providing a conceptual summary of contemporary progress on this front. Fig. \ref{fig:schematic} presents a schematic overview of these works, illustrating their key mechanisms, approaches, and the resulting SIDM-BH mass scales. To keep the comparison focused, Fig.~\ref{fig:schematic} is restricted to elastic SIDM models; related but distinct channels, such as the gravothermal collapse of (ultra-)strongly self-interacting dark matter (uSIDM) \cite{pollack15, mgroberts25, mgroberts26}, or dissipative (and atomic) dark matter \cite{hyxiao21, buckley24, twshen25}, are not included.

The theoretical foundation for forming a BH via SIDM core collapse stems from the gravothermal catastrophe, where the central core achieves increasingly higher densities while shrinking in size \cite{balberg02b}. As observed in semi-analytical fluid models \cite{wxfeng21} (see their Fig.~1) and our N-body analysis (Fig.~\ref{fig:primarycc-rho-m}), the central density rises rapidly across the late-collapse stages (from $\tilde{\rho}_{cc}=10$ to $\tilde{\rho}_{cc}=100$). At the same time, the $\rho$--$M$ and $Kn$--$M$ profiles of these different stages---although distinct within the central core---cross at nearly the same enclosed mass, slightly outside the core. The physical picture of this phenomenon is a uniform-density sphere in the central halo that rapidly shrinks in size, effectively detaching from the surrounding halo. 
Consequently, this characteristic ``crossing mass'' envelope, which lies just inside the nominal $Kn<1$ boundary (in our case at $Kn\approx0.3$) and is conventionally labeled as the SMFP mass, has historically been used as a first-order estimate for the phenomenological mass of the resulting supermassive BH.
This corresponds to the orange points in Fig.~\ref{fig:schematic}. Notably, \cite{wxfeng21} found that the inclusion of a baryonic potential not only shortens the core-collapse timescale but also significantly affects this SMFP BH mass estimate. This aspect has not been extensively studied and represents a potentially major systematic uncertainty in SIDM BH research, which we will discuss in more detail alongside other variables in Sec.~\ref{sec:summary}.

While this phenomenological SMFP estimate provides a reasonable order-of-magnitude approximation, it does not capture the full physical picture. It leaves at least two fundamental questions unanswered: (a) how does the BH grow from the initial true seed---formed via relativistic instability during the gravothermal runaway process---up to this $\sim1\%$ halo mass level? and (b) is this $\sim 1\%$ halo mass scale the end of the story, leaving the BH and the remaining halo in peaceful coexistence, or will the system continue to evolve?

To answer question (a), we must first establish a rigorous understanding of the seed mass from a relativistic perspective, determining the true foundational mass for subsequent macroscopic accretion and evolution. By establishing scaling relations between different phases of late core-collapse and the deeply SMFP stage, \cite{sophia23} (green block in Fig. \ref{fig:schematic}) bridged the gap to find the relevant mass at the onset of relativistic instability (core speed $\sim c/3$). They found a microscopic seed mass in the range of $\mathcal{O}(10^{-9})\sim\mathcal{O}(10^{-8})$ of the halo mass (i.e., stellar-mass seed BHs), depending on the cross-section models. Taking another rigorous approach, \cite{gu26}  (blue in Fig. \ref{fig:schematic}) explicitly incorporated general relativistic calculations into the late-time core-collapse evolution. Building upon and going beyond the quasi-equilibrium treatment in \cite{wxfeng22}, \cite{gu26} introduced an SIDM heat-flux term into the energy-momentum tensor using the Misner-Sharp formalism to compute the non-equilibrium evolution. With this framework, they demonstrated that intense outward heat flux drives significant mass outflow during the final plunge, arriving at an apparent horizon BH seed mass of merely $\mathcal{O}(10^{-8})$ of the halo mass. The next critical question is how this microscopic $\mathcal{O}(10^{-8})M_{\rm halo}$ seed grows into a macroscopic $\mathcal{O}(10^{-2})M_{\rm halo}$ supermassive BH. 
\cite{wxfeng25} (brown arrow in Fig. \ref{fig:schematic}) demonstrates this through a two-phase accretion relay. Following the formation of the relativistic seed, baryonic Eddington accretion dominates the first phase, lifting the BH mass to $\sim10^{-5}M_{\rm halo}$. Once the BH reaches this threshold, dark Bondi accretion takes over. 
Because the Bondi rate scales as $dM_{\rm Bon}/dt \propto M_{\rm BH}^2$, it drives explosive growth in the second phase, rapidly increasing the BH mass to the $\sim 1\%$ halo mass scale. Complementing this analytical picture, recent fluid calculations by \cite{meng26} explicitly resolved the non-equilibrium fluid dynamics of this feeding process. They demonstrated that SIDM heat conduction effectively regulates the extreme accretion compared to the adiabatic dark Bondi limit, causing the BH mass growth to saturate at this $\sim 10^{-2}M_\mathrm{halo}$ scale over Myr timescales. Phenomenologically, this plateau (e.g., Fig. 2 of \cite{meng26}) marks the rapid consumption and exhaustion of the initial SMFP core.

To answer question (b), \cite{fz25} takes the $1\%M_{\rm halo}$ SMFP mass as the starting point for subsequent Eddington gas accretion. By feeding this growth recipe into semi-analytical merger trees, they established the predicted BH mass function at high redshifts $z\gtrsim5$. Their results successfully match the observed abundance of the JWST Little Red Dots. However, the Eddington accretion modeling in \cite{fz25} does not explicitly impose a strict upper bound on the available gas budget, hence the open-ended purple arrow in Fig. \ref{fig:schematic}. While their study focuses on the high-redshift Universe, where the limited cosmic time window might seemingly alleviate the issue of over-accretion, the exponential nature of Eddington growth from an already massive seed (e.g., $10^{-2}M_{\rm halo}$) can still rapidly exhaust the baryon budget.

In this work, we address question (b) with a global N-body framework that tracks the post-core-collapse accretion of individual DM particles onto the BH, rather than relying on analytical estimates of the SMFP mass. The key distinction from previous work lies in how the DM fuel supply is treated.
\cite{wxfeng25} assumed a constant background DM density at the SMFP core level, so their analytical dark Bondi accretion proceeds at a fixed density and runs away explosively. \cite{meng26}, in contrast, simulates only the local core region and reaches a saturation wall once their finite initial core is consumed. Both describe a single accretion episode with no replenishment of the fuel. Our N-body scheme, by contrast, continuously rebuilds the core through multi-round gravothermal collapse: the accretion rapidly drains the inner DM
density on $\sim$Myr timescales, but each rebuilding cycle restores the fuel supply on $\sim100$ Myr timescales. Adopting a stringent threshold $Kn=0.05$, we obtain a converged long-term DM-accreted mass of $\sim 5\%$--$12\%\,M_{\rm halo}$ that is insensitive to the precise core-collapse criterion and systematically exceeds the $\sim 1\%$ single-episode limit.  Including a capped gas reservoir then yields a final composite BH mass of $5\%$--$12\%+f_{\rm gas}$ of the host halo mass, providing a self-consistent basis for predicting SIDM-seeded BH mass functions.

The most closely related work to our approach is the very recent \cite{gu26b}, which extends the Misner--Sharp framework of \cite{gu26} past horizon formation to follow the seed's subsequent dark-matter accretion over a full halo. Their treatment is complementary to ours in one important respect: by evolving a true apparent horizon as an absorbing inner boundary, they self-consistently resolve the innermost capture that our $Kn$-based continuous-accretion (upper bound) scheme can only approximate from a larger, $Kn$-based capture boundary. In the regime where their fluid treatment remains valid, they find a super-Bondi inflow that drains the envelope and grows the seed to of order ten percent of the halo mass, qualitatively consistent with our $\sim5\%$--$12\%\,M_{\rm halo}$ estimate and similarly leaving a depleted inner halo of the kind we quantify in Sec.~\ref{sec:halo}; extrapolated beyond this regime, their solution would consume nearly the entire halo. However, as \cite{gu26b} pointed out, this rests on the gravothermal fluid closure and its local-thermal-equilibrium assumption---an inherent limitation of the gravothermal-fluid approach in general---which is not necessarily valid once the inner halo is nearly depleted. Notably, unlike the LMFP collapse phase of standard gravothermal approach, which is calibrated against N-body simulations \cite{koda11, mace25}, the super-Bondi phase has no such calibration yet. Our N-body treatment, which explicitly resolves the discrete gravothermal rebuilding, should therefore yield the more conservative long-term BH mass by accretion of DM.

\section{Evolution of the Dark Matter Halo Before and After Black Hole Formation}\label{sec:halo}

In the previous section, we focused on a detailed analysis of the BH mass growth trajectory. We now turn our attention to the dark matter halo component, examining its structural evolution both leading up to the primary core-collapse and following the DM-to-BH conversion. To deliver a comprehensive physical picture, we focus on two primary aspects: in Sec. \ref{halo:rho}, we analyze the density evolution, utilizing both the central density $\rho_{\rm cen50}$ as a tracer of core-collapse and the halo-wide density profiles; and in Sec. \ref{halo:vel}, we investigate the radial velocity and velocity anisotropy. The overarching goal is to provide a detailed dynamical investigation into how the halo behaves as it approaches primary core-collapse, during the iterative DM-to-BH conversion, and eventually after $\sim10\%$ of its total mass has been accreted by the central BH.

\subsection{Density Evolution}\label{halo:rho}

\subsubsection{The central density $\rho_{\rm cen50}$} \label{sec:density1}

\begin{figure*}
    \centering
    \begin{subfigure}[t]{0.32\textwidth}
        \centering
        \includegraphics[width=\textwidth, clip,trim=0.2cm 0cm 0.2cm 0cm]{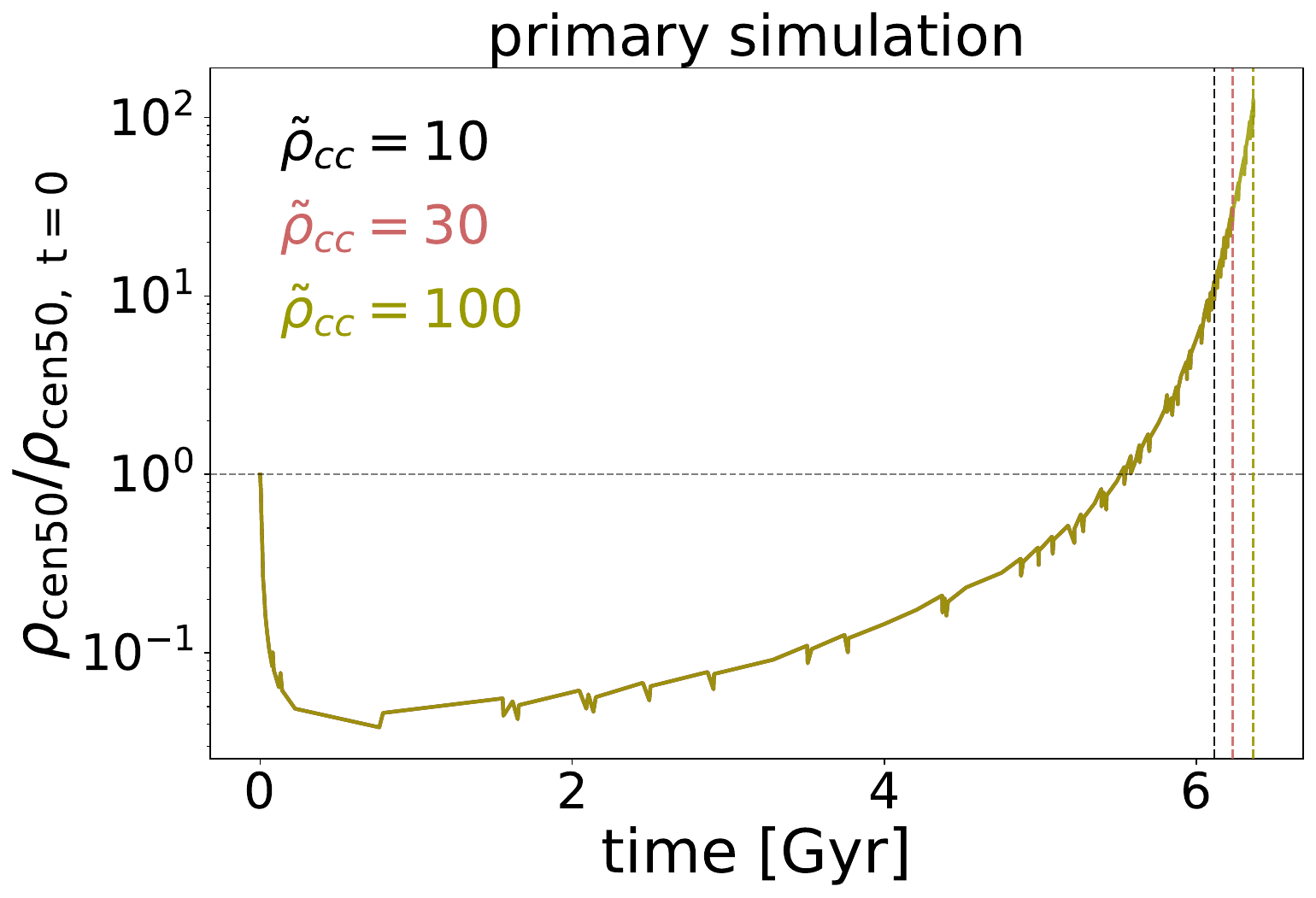}
        \caption{}
        \label{fig:rhocen-primary}
    \end{subfigure}
    ~
    \begin{subfigure}[t]{0.32\textwidth}
        \centering
        \includegraphics[width=\textwidth, clip,trim=0.2cm 0cm 0.2cm 0cm]{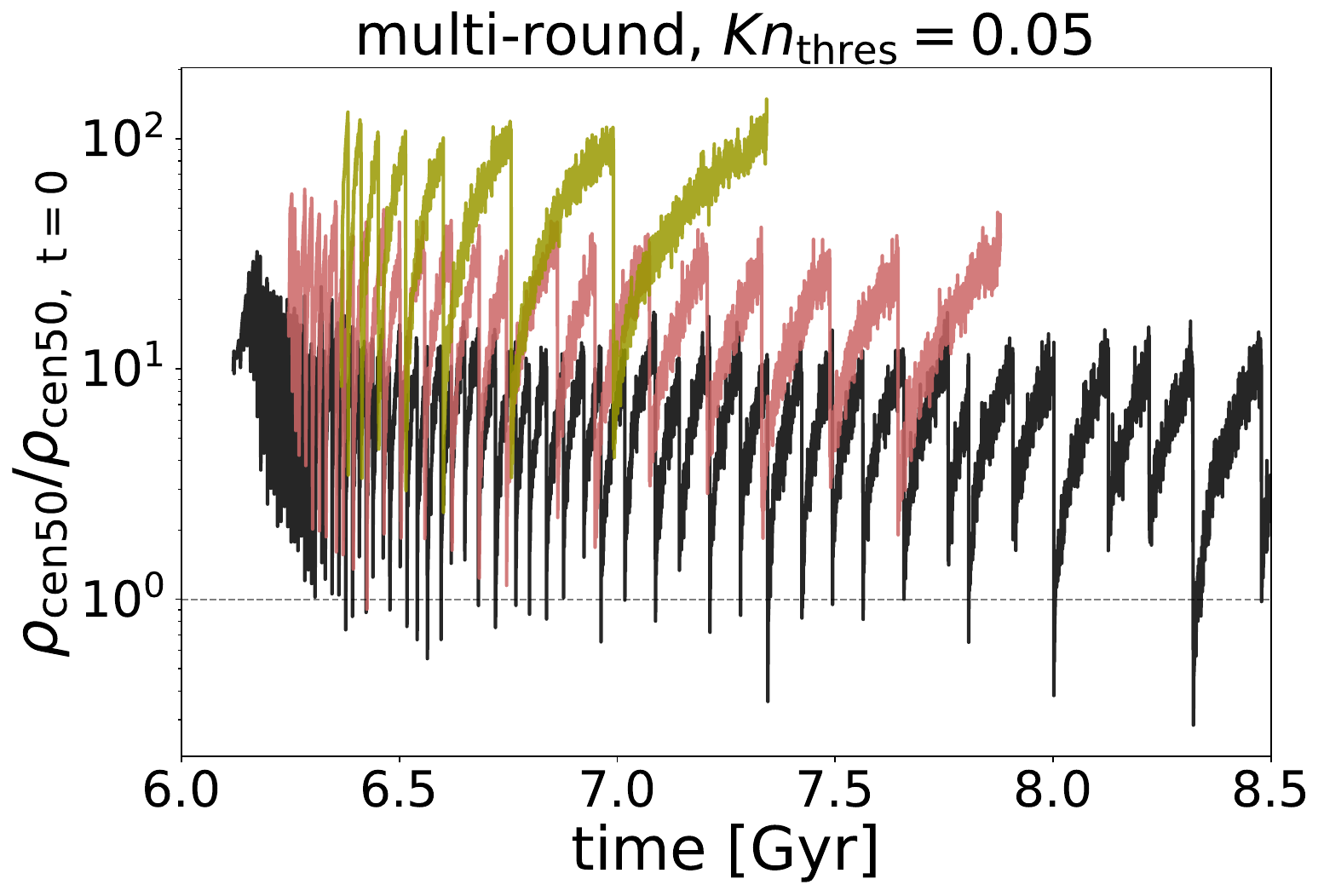}
        \caption{}
        \label{fig:rhocen-multi-round}
    \end{subfigure}
    ~
    \begin{subfigure}[t]{0.32\textwidth}
        \centering
        \includegraphics[width=\textwidth, clip,trim=0.2cm 0cm 0.2cm 0cm]{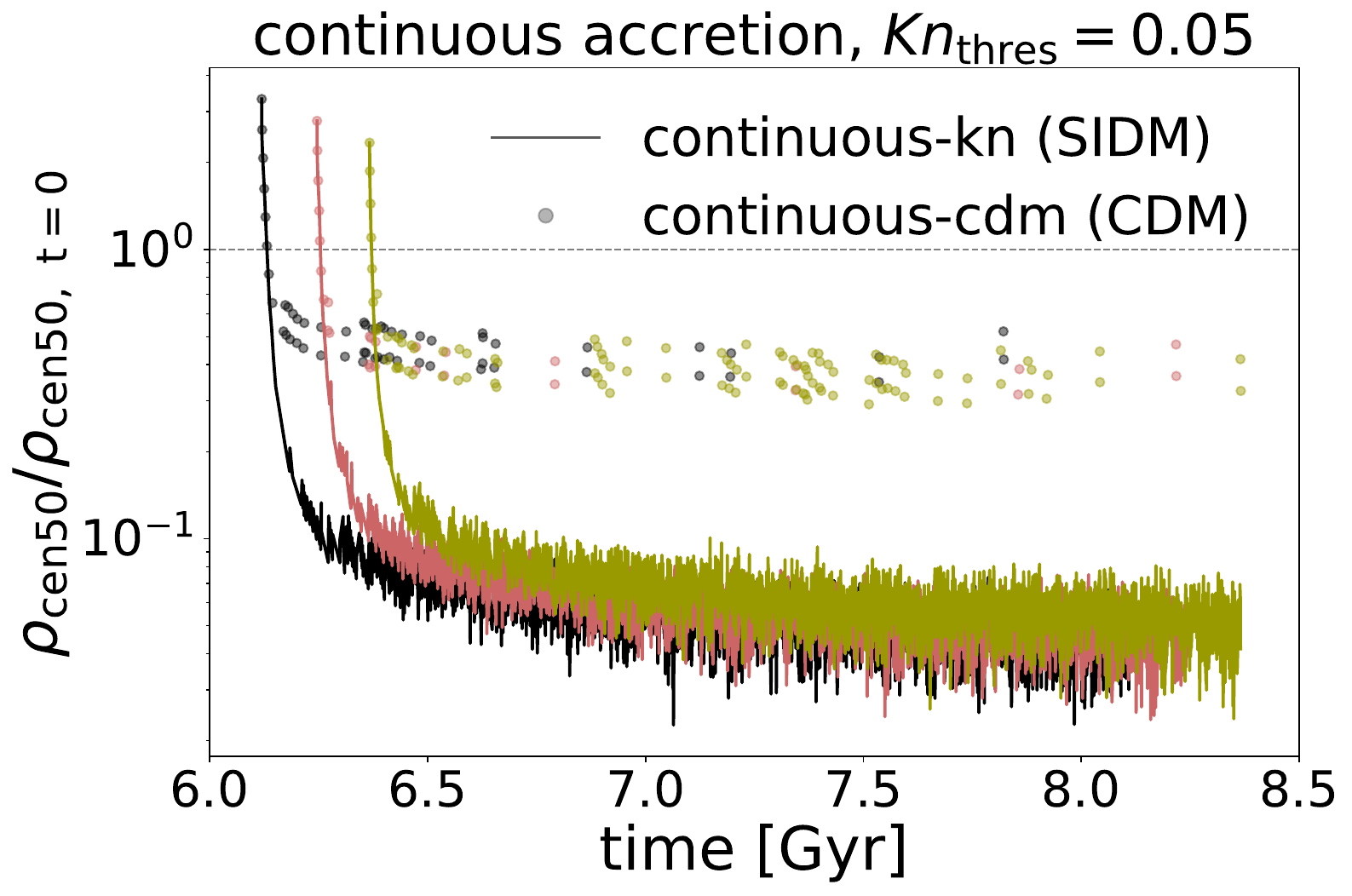}
        \caption{}
        \label{fig:rhocen-multi-conti}
    \end{subfigure}
    \caption{Time evolution of the innermost DM density $\rho_{\rm cen50}$ across two evolutionary phases: (a) the primary simulation up to the initial core-collapse; post-primary evolution under (b) the multi-round scenario and (c) the continuous-accretion scenario. The color coding distinguishes the core-collapse density thresholds $\tilde{\rho}_{cc}$. Vertical dashed lines denote the corresponding core-collapse times of the primary simulation. In panel (b), $\rho_{\rm cen50}$ repeatedly hits the $\tilde{\rho}_{cc}$ ceiling followed by sharp drops, visually capturing the iterative gravothermal rebuilding cycles caused by discrete BH mass conversions. Panel (c) contrasts this with the continuous-accretion scenario and a collisionless CDM control, demonstrating the overestimated BH mass growth in this scenario. }
    \label{fig:rhocen}
\end{figure*}

We first present the evolution of the central density $\rho_{\rm cen50}$ in Fig. \ref{fig:rhocen}, which serves as the primary tracer for the SIDM core-collapse process and the behavior of DM particles in the innermost halo before and after BH formation. Figure \ref{fig:rhocen-primary} displays $\rho_{\rm cen50}$ during the primary simulation. Starting from the initial NFW distribution, the halo rapidly forms a density core within the first Gyr and subsequently enters the rapid core-collapse stage at $\sim6\  \rm Gyr$. This exhibits the well-known self-similar gravothermal evolution of an isolated SIDM halo, which follows a nearly universal trajectory\footnote{We note that the inclusion of a baryonic potential can break this self-similarity; for instance, \cite{zzclg} demonstrated that Milky Way-mass systems can bypass the core-expansion phase and enter early core-collapse directly, with the central DM density growing monotonically more cuspy.}.  As shown in Fig. \ref{fig:rhocen-primary}, the evolutionary tracks of $\rho_{\rm cen50}$ for different $\tilde{\rho}_{cc}$ cases are identical up to their respective termination points. This is expected, as they represent the same halo simulated with an identical SIDM cross section of $100 \ \rm cm^2/g$, differing only in the chosen core-collapse termination criterion, marked by the vertical dashed lines. 

In Fig. \ref{fig:rhocen-multi-round}, we plot $\rho_{\rm cen50}$ as a function of time for the multi-round simulations, where the DM enclosed within the best converged $Kn=0.05$ boundary is iteratively converted into the central BH. As depicted, the central DM density experiences a sharp reset every time it hits the corresponding $\tilde{\rho}_{cc}$ threshold, triggering the simulation's termination and subsequent DM-to-BH conversion. Notably, the time intervals between these resets---which essentially represent the manual removal of the local SMFP core (as defined by the $Kn$ threshold)---gradually lengthen over time. This physically reflects the dark matter reservoir depletion discussed in previous sections: as increasing amounts of DM are accreted by the BH, the remnant DM density in the inner halo steadily drops (as will be detailed soon in the discussion of Fig. \ref{fig:density-profiles}). Consequently, the halo requires increasingly longer macroscopic timescales to undergo gravothermal rebuilding and trigger the next round of core-collapse. 

For comparison, we show the post-primary $\rho_{\rm cen50}$ evolution for the continuous-accretion scenario in Fig. \ref{fig:rhocen-multi-conti}. Here, the central DM density declines monotonically without recovery. This clearly demonstrates that by instantaneously removing DM particles based solely on their local $Kn$ values, the continuous-accretion scheme bypasses the macroscopic gravothermal rebuilding bottleneck. Thus, it strictly overestimates the DM accretion rate, reaffirming that this continuous method should only serve as an upper-bound estimate for BH mass growth.

\subsubsection{Density profiles} \label{sec:density2}

\begin{figure*}
    \centering
    \begin{subfigure}[t]{0.45\textwidth}
        \centering
        \includegraphics[width=\textwidth, clip,trim=0.2cm 0cm 0.2cm 0cm]{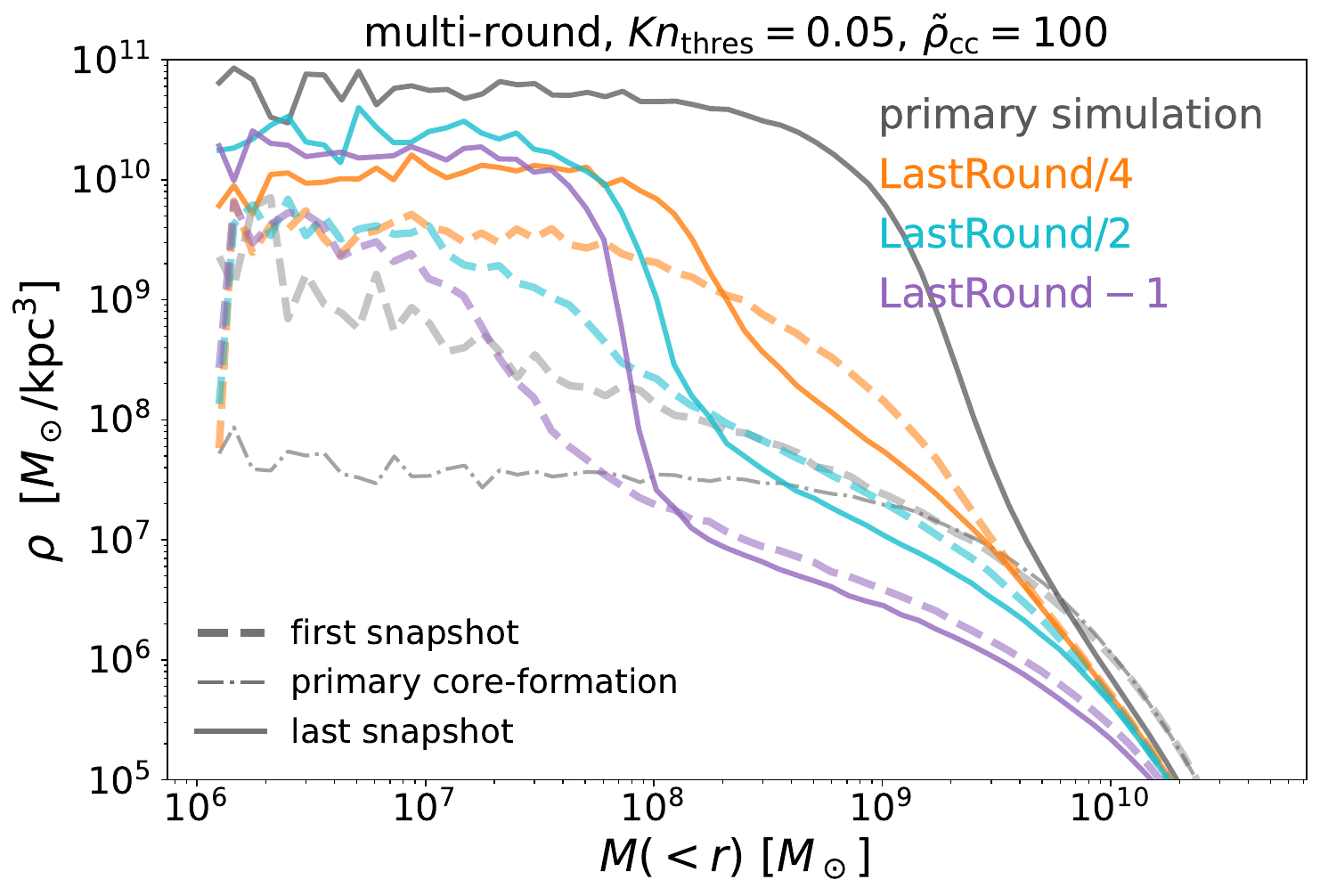}
        \caption{}
        \label{fig:density-rho-m}
    \end{subfigure}
    ~
    \begin{subfigure}[t]{0.45\textwidth}
        \centering
        \includegraphics[width=\textwidth, clip,trim=0.2cm 0cm 0.2cm 0cm]{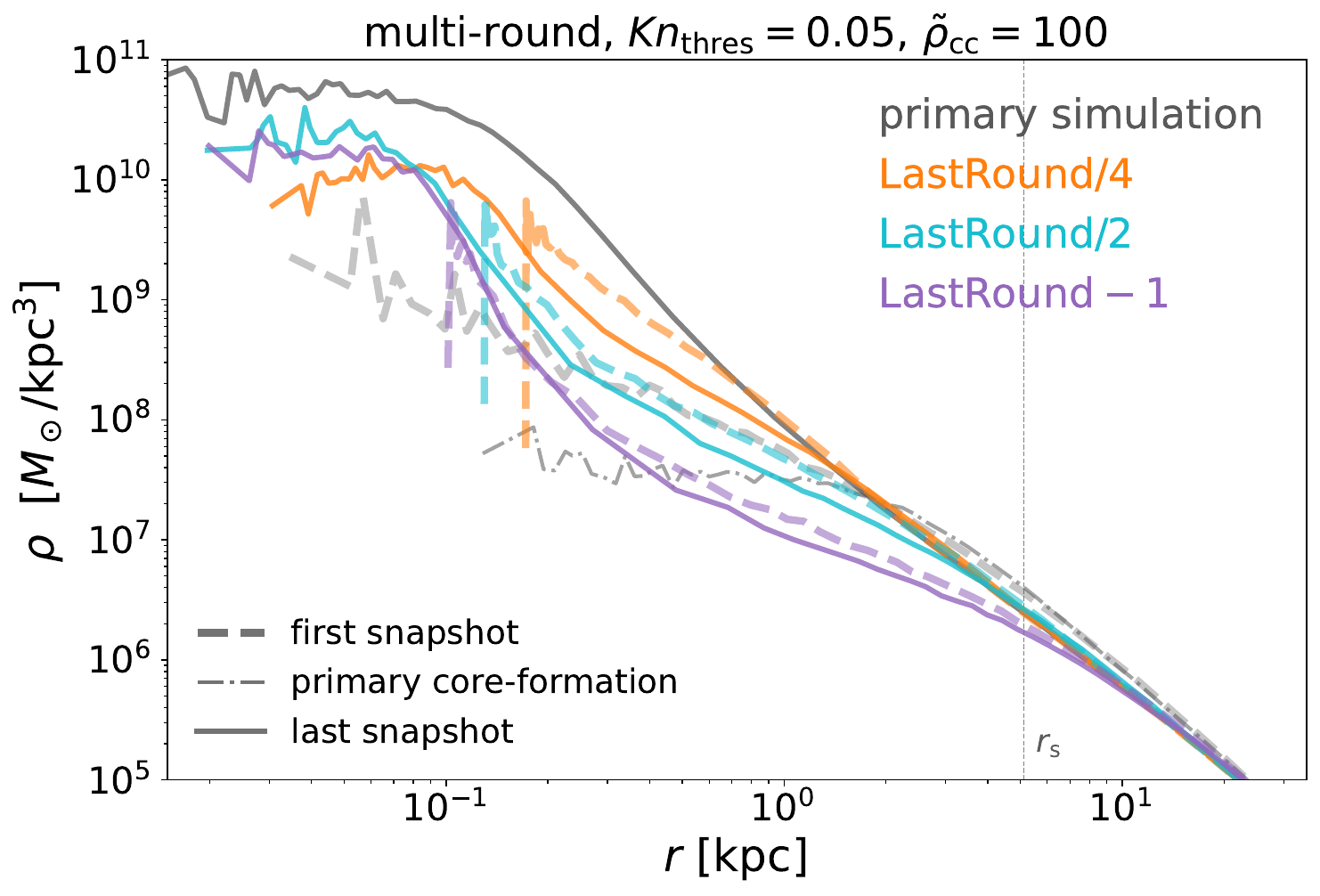}
        \caption{}
        \label{fig:density-rho-r}
    \end{subfigure}
    ~
    \begin{subfigure}[t]{0.45\textwidth}
        \centering
        \includegraphics[width=\textwidth, clip,trim=0.2cm 0cm 0.2cm 0cm]{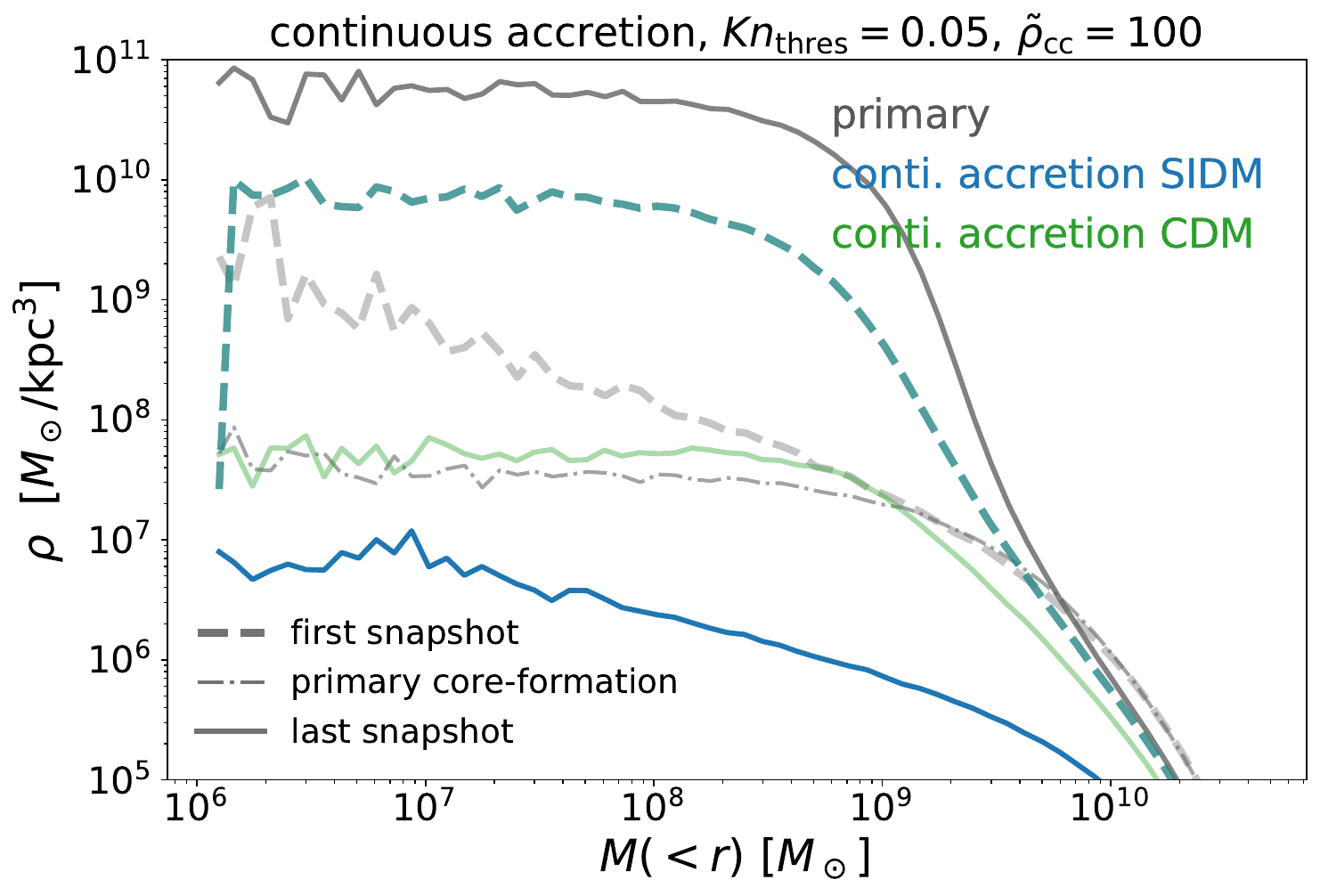}
        \caption{}
        \label{fig:density-rho-m-conti}    
    \end{subfigure}
    ~
    \begin{subfigure}[t]{0.45\textwidth}
        \centering
        \includegraphics[width=\textwidth, clip,trim=0.2cm 0cm 0.2cm 0cm]{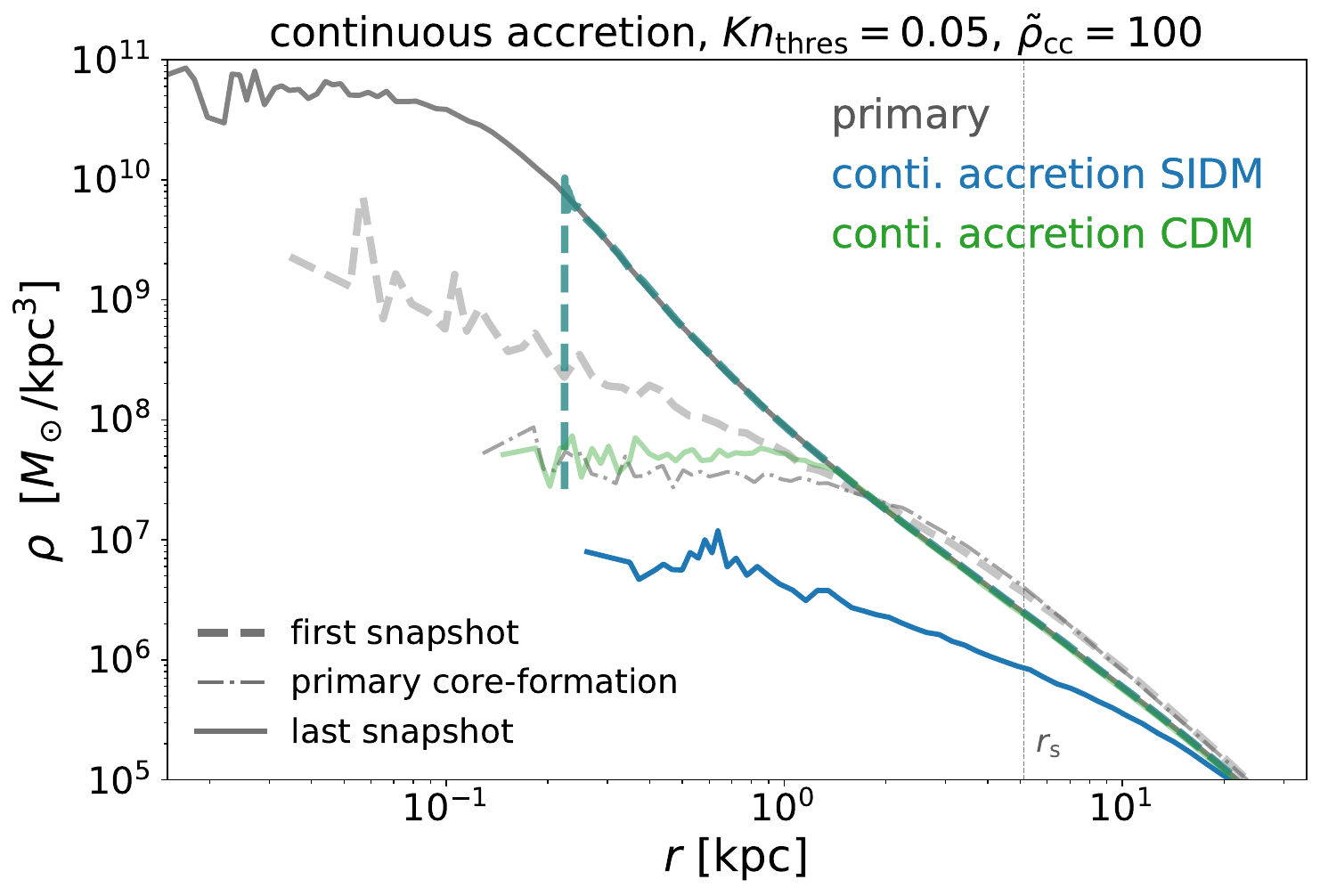}
        \caption{}
        \label{fig:density-rho-r-conti}    
    \end{subfigure}
    \caption{Density profiles of the DM halo in our best-converged configuration ($\{Kn_{\rm thres}=0.05, \tilde{\rho}_{cc}=100 \}$). Panels (a) and (b) present the multi-round scenario viewed in the Lagrangian-like $\rho-M(<r)$ and Eulerian $\rho-r$ planes respectively; panels (c) and (d) display the continuous-accretion scenario in the same two parameter spaces. In (a) and (b), we show three representative discrete rounds of simulations following the primary simulation (grey), indexed by $\rm LastRound/4$ (orange), $\rm LastRound/2$ (cyan), and $\rm LastRound-1$ (purple). Within each selected round, the first snapshot (dashed lines) and the final snapshot (solid lines) are plotted to demonstrate the gravothermal rebuilding cycle. In (c) and (d), we show the continuous-accretion scenario for both the CDM (blue) and SIDM (green) models, comparing the immediate post-seeding state (dashed lines) against the final state (solid lines) after 2 Gyr of on-the-fly DM accretion onto the existing BH (solid lines). This set of plots visually encapsulates the macroscopic supply chain of post-primary evolution: the outer halo feeds the inner halo, and the inner halo feeds the BH. This set also highlights the fundamental difference between the multi-round and continuous-accretion scenarios: the former properly accounts for gravothermal rebuilding as the key process, while the latter bypasses it, artificially overestimating both the BH mass growth and the resulting central halo density depletion. }
    \label{fig:density-profiles}
\end{figure*}

In Fig. \ref{fig:density-profiles}, we turn our attention to the detailed temporal evolution of the DM density distributions during the post-primary core-collapse era. In Figs. \ref{fig:density-rho-m} and \ref{fig:density-rho-r}, we present the density profiles in the $\rho-M(<r)$ and $\rho-r$ planes, respectively, utilizing our best-converged $\{ \tilde{\rho}_{cc}=100, Kn_{\rm thres}=0.05 \}$ configuration as an illustrative example. In each panel, alongside the primary simulation, we show three subsequent multi-round iterations: indexed as $\texttt{LastRound}/4$ (orange), $\texttt{LastRound}/2$ (cyan), and $\texttt{LastRound}-1$ (purple), where $\texttt{LastRound}$ denotes the final index and thus the total number of subsequent simulations run in this multi-round approach (see Sec. \ref{sec:method} for the termination criterion of the multi-round runs).  For each selected round, we display both the initial (dashed lines) and final (solid lines) snapshots. 
These three rounds are chosen to demonstrate two complementary points. First, within each round the dashed-to-solid transition shows the gravothermal rebuilding cycle: the void left by the previous conversion is re-filled and re-collapses. Second, across successive rounds (smaller to larger indexed) the cumulative depletion of the inner halo becomes evident, as more DM material is irreversibly lost to the BH.

We identify several prominent features in Figs. \ref{fig:density-rho-m} and \ref{fig:density-rho-r}: (a) At the onset of each post-primary round (dashed lines), the innermost density bins exhibit a sharp drop compared to the surrounding DM distribution. This results from the instantaneous removal of the densest SMFP DM material at the end of the previous round, effectively creating a central void; (b) Starting from these initial configurations, the halo evolves via SIDM scatterings until core-collapse is triggered again (solid lines). During this rebuilding phase, the density $\rho$ at larger radii (or larger enclosed masses $M(<r)$) decreases as material flows inward to replenish the previously depleted central void. (c) As the multi-round sequence progresses, despite the innermost region repeatedly reaching high densities via gravothermal collapse, the overall macroscopic DM density within $\sim 2r_s$ (enclosing $\sim 30\% M_{200c}$) suffers a severe reduction. This density drops by as much as $\gtrsim1$ dex compared to the CDM counterpart (grey dashed lines), even falling far below the level of the initial SIDM core-formation phase. We can summarize this macroscopic structural evolution as a continuous supply chain: \textit{the extended DM halo continuously feeds the central core, which in turn continuously feeds the central BH. Consequently, if an SIDM-seeded BH is to grow to $\sim10\%$ of the halo mass, the inevitable structural cost is a severely depleted inner halo ($\lesssim2r_s$) with densities much lower than CDM predictions.} 

\begin{figure*}
    \centering
        \centering
        \includegraphics[width=0.93\textwidth, clip,trim=0.2cm 0cm 0.2cm 0cm]{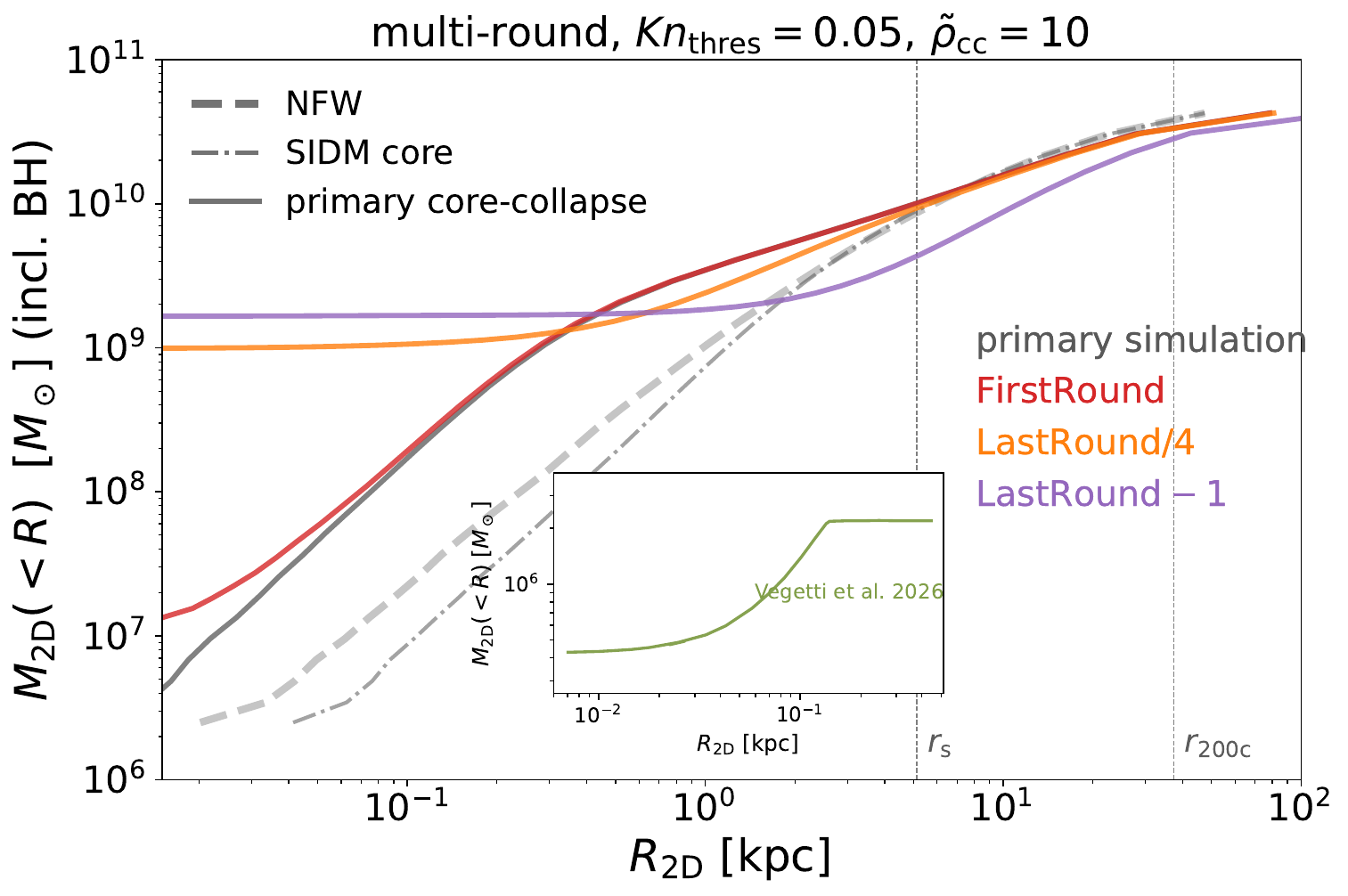}
    \caption{Projected 2D enclosed mass profiles $M_{2D}(<R_{2D})$ of the isolated halo across distinct evolutionary phases: NFW, SIDM core, SIDM primary core-collapse, and our post-primary multi-round evolution. These are presented for a qualitative morphological comparison with the anomalous strong lensing substructure, detection V of JVAS B1938+666, recently discovered by Vegetti et al. 2026 \cite{vegetti26}. Their best-fit lens model, comprising an unresolved point mass and a uniform 2D disk (UD+PM), appears to be best mirrored by our multi-round results in late rounds of simulations. }
    \label{fig:vegetti}
 \end{figure*}

Notably, a recent strong lensing study \cite{powell25,vegetti26} reported an unprecedented halo density distribution in detection V\footnote{Throughout this work, ``detection V'' refers to the newly detected substructure V of the JVAS B1938+666 lens system (mass $\sim10^6\,M_\odot$), not the earlier-discovered and more widely studied substructure A \cite{vegetti12} of the same lens system (mass $\sim10^9\,M_\odot$).} of the lens system JVAS B1938+666 that remarkably mirrors this picture: an unresolved central point-mass surrounded by an extended, constant-surface-density region that is significantly less concentrated than CDM expectations. By explicitly tracking the post-collapse BH feeding and the resulting gravothermal depletion, our multi-round evolutionary framework provides a natural, self-consistent physical explanation for their findings. 
To facilitate a direct comparison with the uncommon lens perturber (detection V) discovered in \cite{vegetti26}, Fig. \ref{fig:vegetti} presents our halo's 2D enclosed cylindrical mass $M_{2D}(<R)$ as a function of projected radius $R_{2D}$, plotted alongside their best-fit Uniform Disk plus Point Mass (UD+PM) model. Although an exact one-to-one mass scaling is precluded by the vast difference in system masses (a $1.8\times10^6M_\odot$ subhalo in \cite{vegetti26} vs. the $3\times10^{10} M_\odot$ isolated halo in this work), the morphological similarity between the two enclosed mass profiles is striking. We note that a recent study \cite{xyzhang26} utilizing 1D gravothermal fluid simulations showed that the primary core-collapse phase of an SIDM halo can transiently produce a dense secondary core that approximates the central compactness of this perturber. However, as such an ultra-dense fluid core is expected to eventually trigger relativistic instability and form a seed BH, exploring the long-term, post-collapse macro-structure becomes essential. When compared with the standard NFW profile, the SIDM core, and the highly concentrated SIDM primary core-collapse phase, our post-primary framework, which explicitly accounts for the central BH's accretion of surrounding SMFP dark matter, is the most natural scenario that durably reproduces the extended, flattened density plateau favored by the lensing data. Note that we adopt the $\tilde{\rho}_{cc}=10$ case for this illustration: at $Kn_{\rm thres}=0.05$ the multi-round trajectory is well converged across
$\tilde{\rho}_{cc}$, so the late-round morphology is insensitive to this
choice, while the smaller initial seed by $\tilde{\rho}_{cc}=10$ cleanly separates the early rounds from the late rounds, most clearly displaying the full evolutionary sequence.

For comparison, the corresponding post-collapse density evolution under the continuous-accretion scenario is presented in Figs. \ref{fig:density-rho-m-conti} and \ref{fig:density-rho-r-conti}. We observe that the DM density depletion remains most prominent within the inner $\sim 2r_s$ region, but with a significantly larger magnitude compared to the multi-round scenario. Because the continuous-accretion scheme intrinsically overestimates the DM accretion rate onto the BH, this exacerbated drop should be viewed as an upper bound on the severity of the central density depletion (equivalently, a lower bound on the residual central density).

\subsection{Velocity and anisotropy}\label{halo:vel}

\begin{figure*}
    \centering
    \begin{subfigure}[t]{0.45\textwidth}
        \centering
        \includegraphics[width=\textwidth, clip,trim=0.2cm 0cm 0.2cm 0cm]{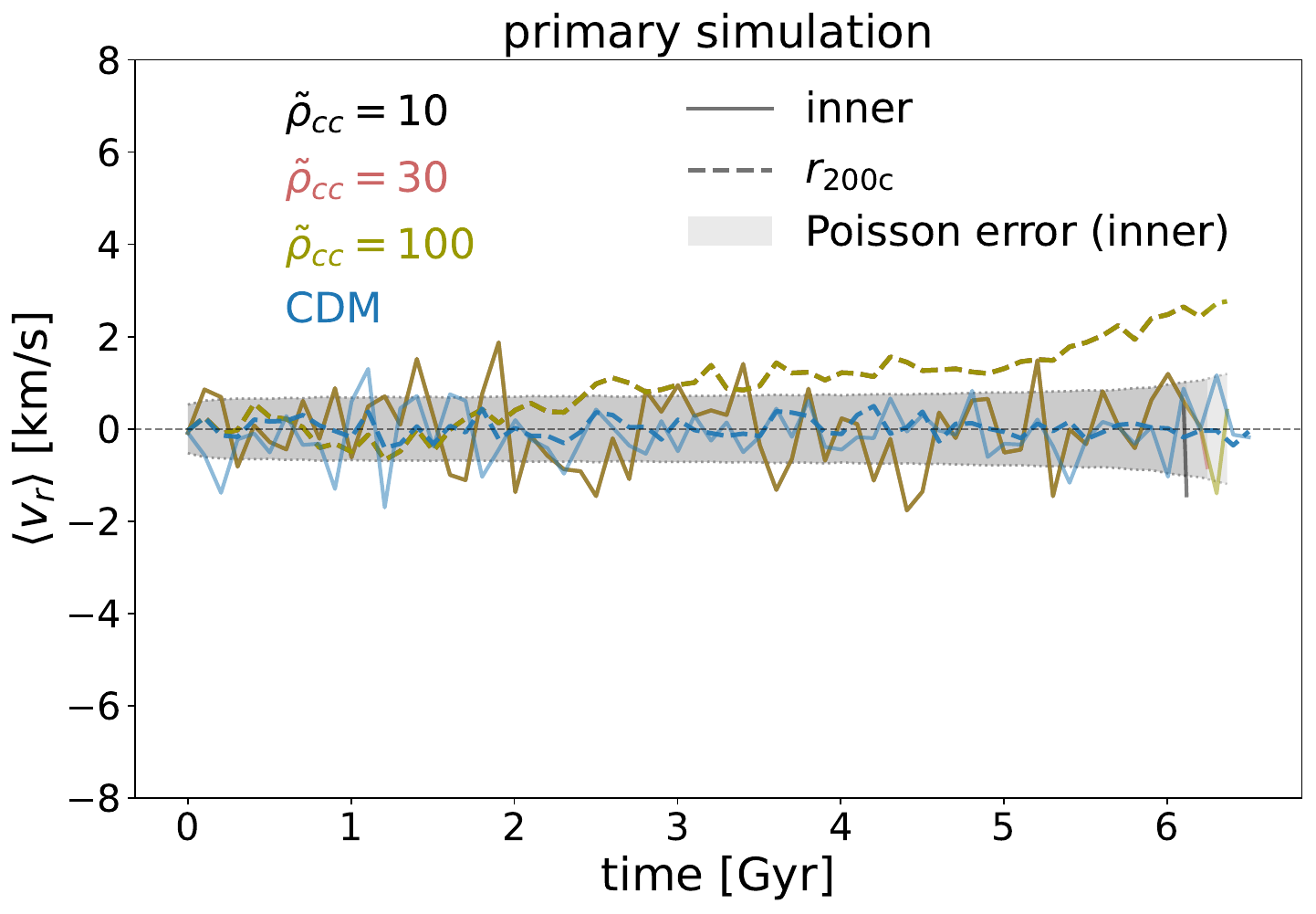}
        \caption{}
        \label{fig:primary-meanvr}
    \end{subfigure}
    ~
    \begin{subfigure}[t]{0.45\textwidth}
        \centering
        \includegraphics[width=\textwidth, clip,trim=0.2cm 0cm 0.2cm 0cm]{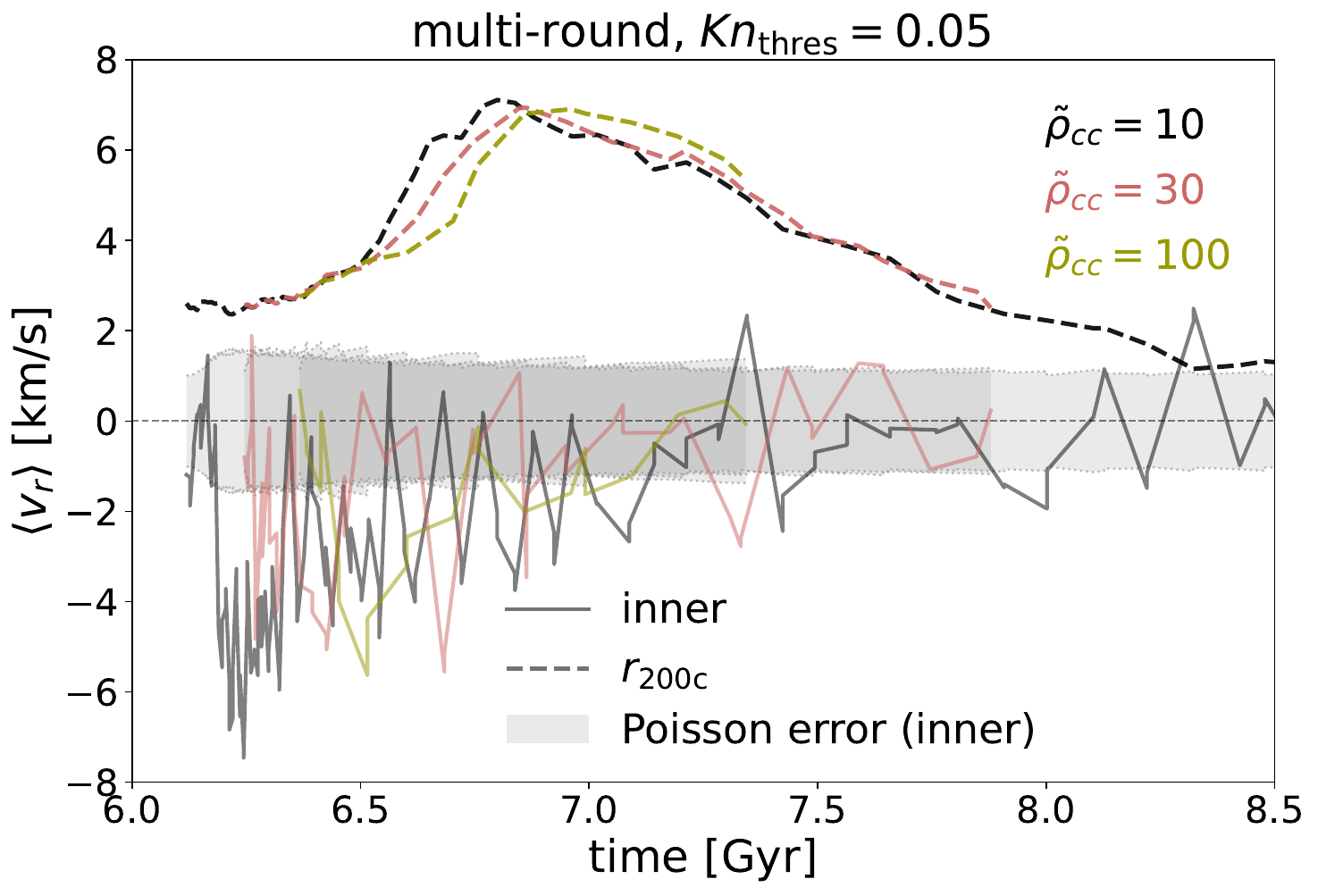}
        \caption{}
        \label{fig:multi-round-meanvr}
    \end{subfigure}
    ~
    \begin{subfigure}[t]{0.45\textwidth}
        \centering
        \includegraphics[width=\textwidth, clip,trim=0.2cm 0cm 0.2cm 0cm]{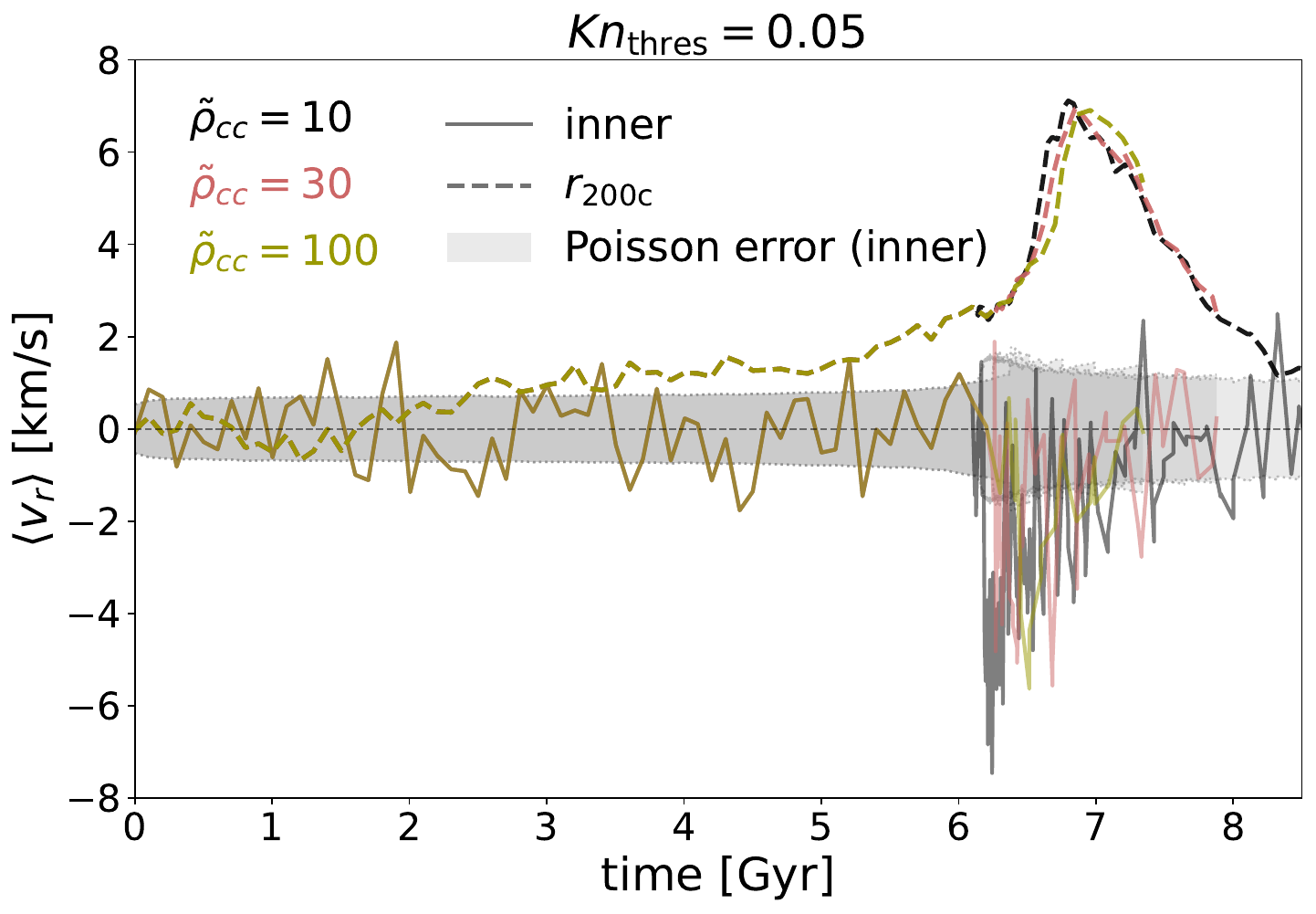}
        \caption{}
        \label{fig:full-meanvr}    
    \end{subfigure}
    ~
    \begin{subfigure}[t]{0.45\textwidth}
        \centering
        \includegraphics[width=\textwidth, clip,trim=0.2cm 0cm 0.2cm 0cm]{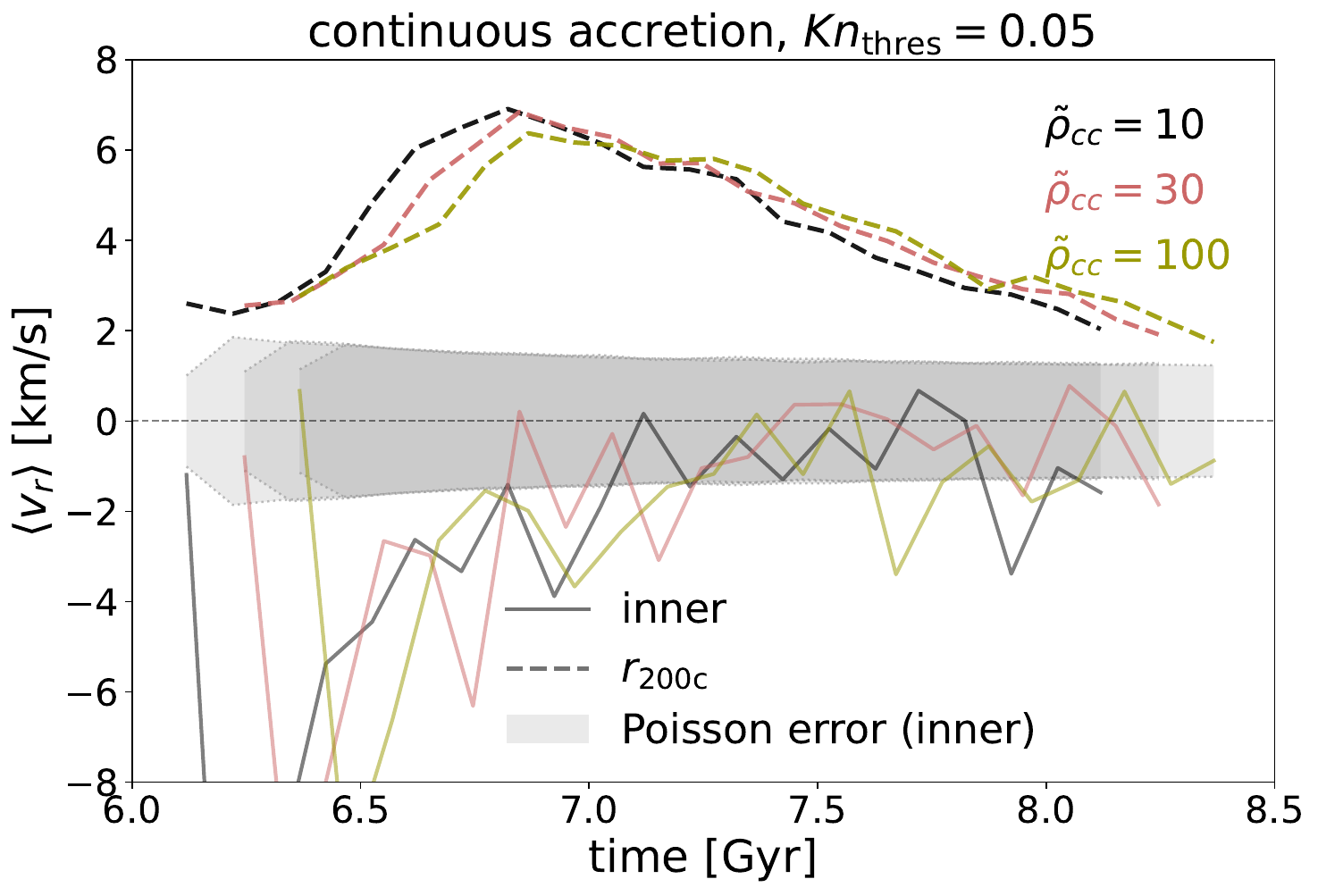}
        \caption{}
        \label{fig:conti-meanvr}    
    \end{subfigure}
    \caption{Time evolution of the mean radial bulk velocity $\langle v_r \rangle$ for the isolated SIDM halo, evaluated at both the deep core (solid lines; tracked via the innermost $N_{\rm in}=4000$ particles) and the halo periphery (dashed lines; tracked via the $N_{200c}=30000$ particles closest to the virial radius $r_{200c}$). Panels (a)–(d) demonstrate the kinematics across distinct evolutionary phases or approaches: (a) the primary simulation up to initial core-collapse; (b) the post-primary evolution under the multi-round scenario; (c) the combined, long-term trajectory of both the primary and multi-round phases; and (d) the post-primary evolution under the continuous-accretion scenario. This kinematic diagnosis reveals the underlying macroscopic mass flows: an outward flux at the boundary driven by central gravothermal heating, and a post-primary-core-collapse, inward flow in the core to replenish the severe pressure deficit induced by central BH accretion. The shaded gray band marks the $\pm 1\sigma$ Poisson uncertainty of the innermost-particle subset, centered on zero.}
    \label{fig:halov-meanvr}
\end{figure*}

\begin{figure*}
    \centering
    \begin{subfigure}[t]{0.45\textwidth}
        \centering
        \includegraphics[width=\textwidth, clip,trim=0.2cm 0cm 0.2cm 0cm]{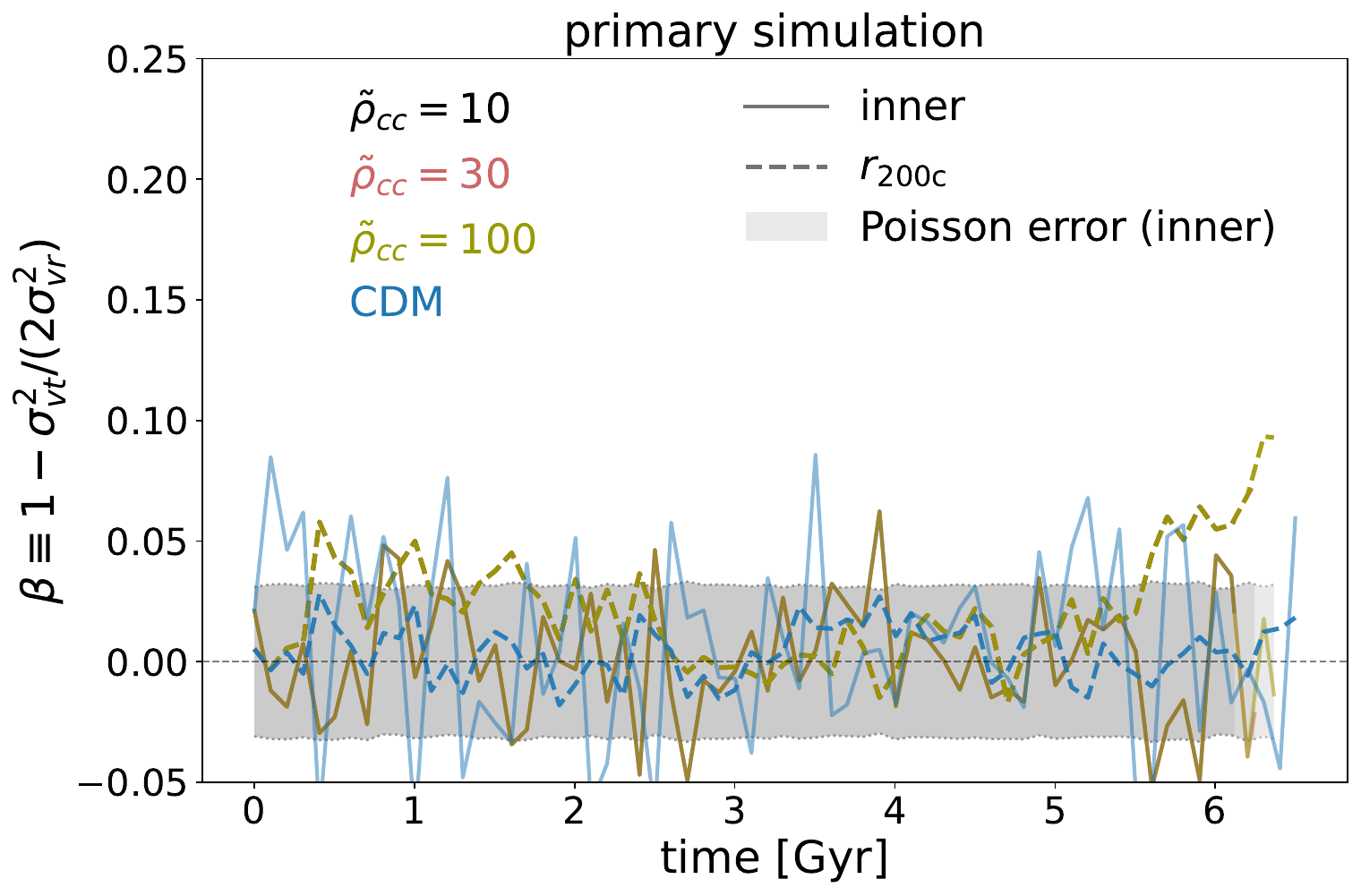}
        \caption{}
        \label{fig:primary-beta}    
    \end{subfigure}
    ~
    \begin{subfigure}[t]{0.45\textwidth}
        \centering
        \includegraphics[width=\textwidth, clip,trim=0.2cm 0cm 0.2cm 0cm]{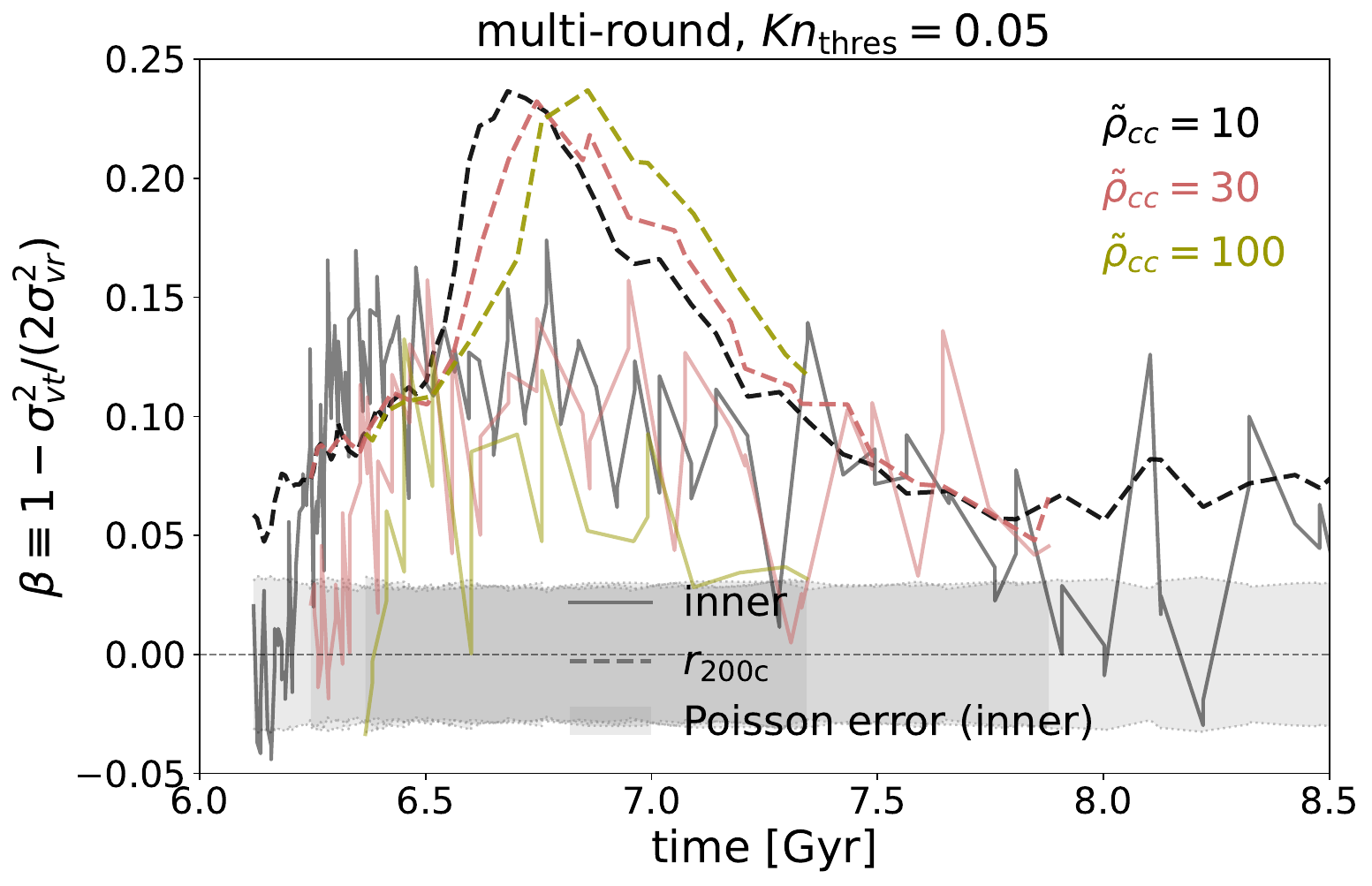}
        \caption{}
        \label{fig:multi-round-beta}    
    \end{subfigure}
    ~
    \begin{subfigure}[t]{0.45\textwidth}
        \centering
        \includegraphics[width=\textwidth, clip,trim=0.2cm 0cm 0.2cm 0cm]{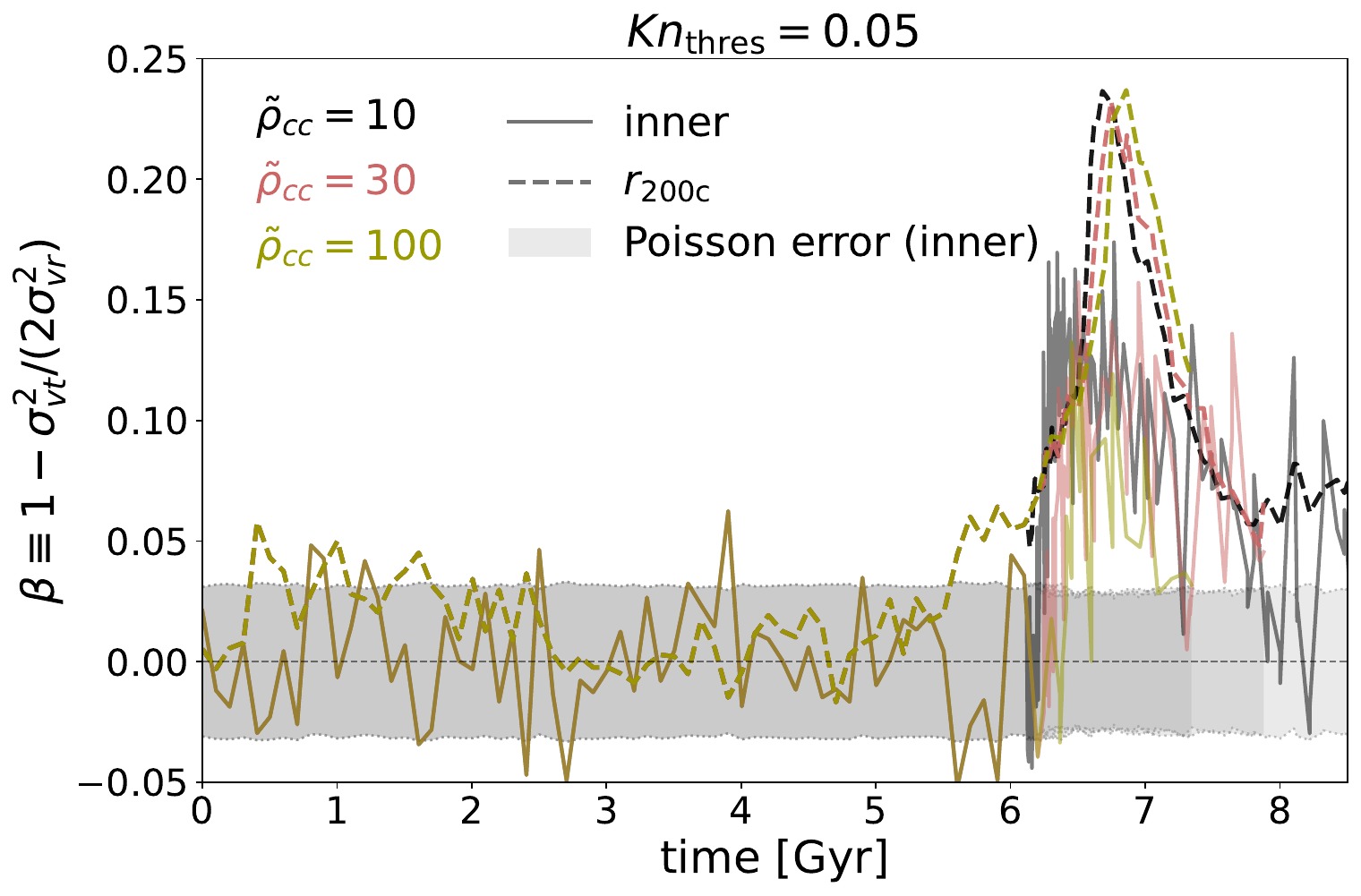}
        \caption{}
        \label{fig:full-beta}
    \end{subfigure}
    ~
    \begin{subfigure}[t]{0.45\textwidth}
        \centering
        \includegraphics[width=\textwidth, clip,trim=0.2cm 0cm 0.2cm 0cm]{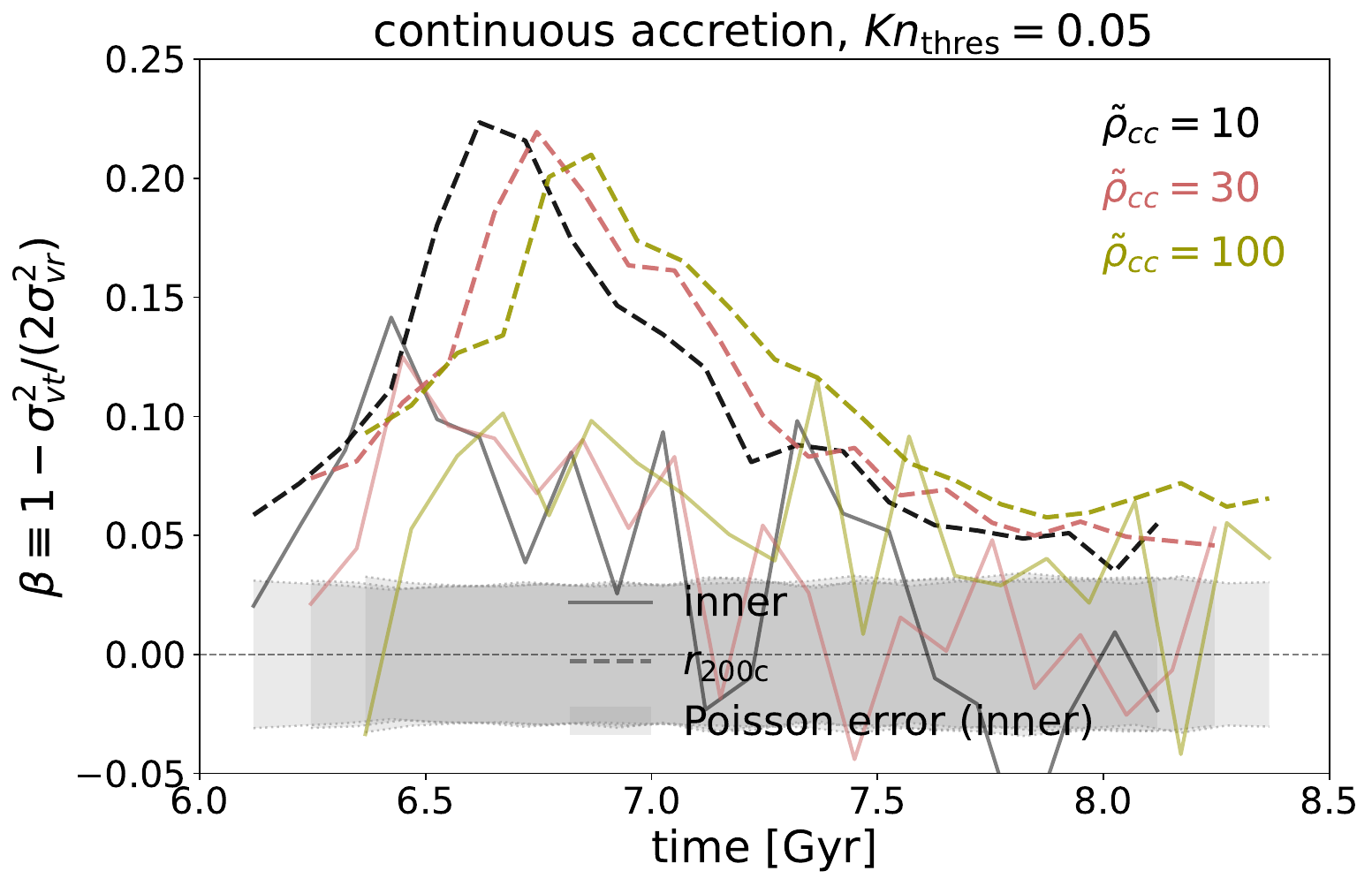}
        \caption{}
        \label{fig:conti-beta}
    \end{subfigure}
    \caption{Time evolution of the velocity anisotropy $\beta$, evaluated across the same phases and radial shells as in Fig. \ref{fig:halov-meanvr}. Despite an initially isotropic state $\beta=0$, a positive radial anisotropy $\beta>0$ gradually emerges at the periphery during and after primary core-collapse. In the post-primary phases, this radial preference also emerges in the deep core. In both regions, this dynamically induced anisotropy acts as a kinematic footprint of incomplete local thermalization. The shaded gray band marks the $\pm 1\sigma$ Poisson uncertainty of the innermost-particle subset, centered on zero.}
    \label{fig:halov-beta}
\end{figure*}

We now turn our attention to the evolution of the DM kinematics, examining the periods both before and after the DM-to-BH conversion. To robustly characterize the overall velocity distribution, we focus on two complementary kinematic indicators: the mean radial velocity $\langle v_r \rangle$ (Fig. \ref{fig:halov-meanvr}), which traces the instantaneous macroscopic bulk flow of mass; and the velocity anisotropy parameter $\beta\equiv 1-\sigma^2_{v_t}/(2\sigma^2_{v_r})$ (Fig. \ref{fig:halov-beta}). Our kinematic analysis tracks two representative subsets of particles: the innermost $N_{\rm in}$ particles (solid lines) and the $N_{200c}$ particles located closest to the halo virial radius $r_{200c}$ (dashed lines). To strike a balance between mitigating Poisson noise and ensuring that these subsets genuinely reflect the local physics of their respective regions (e.g., $N_{\rm in}$ strictly tracing the physical density core shown in Fig. \ref{fig:density-rho-m}),  we select $N_{\rm in}=4000, N_{200c} = 30000$. 

During the primary simulation, one would theoretically expect the innermost halo to exhibit a positive $\langle v_r \rangle$ during the core-expansion phase and a negative $\langle v_r \rangle$ during the runaway core-collapse phase. However, as seen in Fig. \ref{fig:primary-meanvr}, these macroscopic radial flows are not clearly resolved in our data. This limitation is primarily driven by the particle discreteness noise in the deep core, limited by $N_{\rm in}$. For instance, given a core-expansion timescale of $\sim 0.5$--$1\ \rm Gyr$ (Fig. \ref{fig:rhocen-primary}) and a physical core radius of $\sim1\ \rm kpc$ (Fig. \ref{fig:density-rho-r}), the expected macroscopic radial bulk velocity is only on the order of $\sim 1$--$2\ \rm km/s$. This weak signal is entirely submerged beneath the local fluctuations of the $N_{\rm in}=4000$ particles. Capturing these sub-kpc kinematics would require significantly higher mass resolution to suppress the noise.  Conversely, at the halo boundary (dashed lines) where the much larger particle sample of $N_{200c} = 30000$ effectively beats down the noise, we observe a clear outward radial bulk motion of DM particles. The amplitude of such outward motion steadily increases over time, driven by the continuous outward heat and mass transfer induced by central SIDM scattering.

During the post-primary evolution, both the multi-round (Fig. \ref{fig:multi-round-meanvr}) and continuous-accretion (Fig. \ref{fig:conti-meanvr}) approaches exhibit consistent kinematic trends: (a) At the halo's outer boundary $r_{200c}$, the outward bulk velocity $\langle v_r \rangle$ continues to grow at an accelerated rate compared to the primary core-collapse phase (see Fig. \ref{fig:full-meanvr}). This acceleration persists until $\langle v_r \rangle$ hits a peak when the DM pool in the halo center begins to gradually deplete. As the central density drops, the outward heat and mass flux driven by SIDM scatterings naturally weakens. Consequently, $\langle v_r \rangle$ decreases but maintains a positive value, indicating that a macroscopic radially outward bulk motion is sustained. (b) Conversely, in the halo center, the instantaneous conversion of DM into incremental BH mass effectively creates a void in the halo center and shuts down the pressure term of SIDM. Driven by this gradient, the surrounding inner DM acquires a net radially inward bulk motion to replenish the void. In the multi-round scenario, this inward flow is iteratively triggered and reset, becoming less prominent over time as the macroscopic rebuilding cadence slows down.

Turning to the velocity anisotropy parameter $\beta$ in the primary simulation (Fig. \ref{fig:primary-beta}), it is important to first note that our halos are initialized with an isotropic velocity distribution (initial $\beta=0$). For the collisionless CDM companion, $\beta$ trivially remains zero at all radii because the halo remains in dynamical equilibrium. For the SIDM halo, however, the inner particles (solid lines) maintain a nearly constant $\beta \approx 0$ throughout the primary simulation for a fundamentally different reason. This is because frequent, isotropic SIDM collisions within the dense core rapidly thermalize the orbits, actively erasing any kinematically induced anisotropy. In contrast, for the outer particles near $r_{200c}$, the anisotropy $\beta$ builds up as the halo approaches core-collapse ($\gtrsim5\ \rm Gyr$). This rise is a consequence of the gravothermally induced outward mass flux, in correspondence with the outward radial bulk motion in Fig. \ref{fig:primary-meanvr}, which preferentially populates radial orbits. Because the collision rate in the dilute outer halo is too low to isotropize these orbits, the radial anisotropy continuously accumulates.

Turning to the post-primary evolution, the anisotropy $\beta$ at $r_{200c}$ exhibits an accelerated increase. Consistent with the amplified outward bulk motion $\langle v_r \rangle$ (Fig. \ref{fig:full-meanvr}), this rise is driven by the enhanced, continuous mass outflux in this post-primary stage. More intriguingly, in the halo center, $\beta$ increases to become positive, indicating a preference for radial orbits. Mirroring the inward bulk flow seen in the  $\langle v_r \rangle$ panel (Fig. \ref{fig:full-meanvr}), this central anisotropy is another kinematic response to the pressure deficit created by the void. The fact that $\beta$ is no longer trivially 0 in the halo center may imply that local thermalization is incomplete during the time window between successive mass conversions. This may hint that the multi-round core-collapse is strongly driven by the dynamical infall of DM replenishing the void, rather than being a purely gravothermal evolution. Whether this subtle, dynamically driven deviation from perfect thermalization could alter the BH mass accretion increments at the microphysical scale remains an open question for future studies.

In summary, our kinematic analysis at both the inner and outer halo radii reveals the macroscopic bulk radial flow and the dynamical establishment of velocity anisotropy from an initially strictly isotropic distribution.
For the outer halo ($r_{200c}$) the net radially outward bulk motion and the gradual buildup of positive anisotropy $\beta>0$ during core-collapse align with recent findings by \cite{gurian25}. Utilizing a Direct Simulation Monte Carlo (DSMC) framework, they demonstrated that the velocity distribution in the intermediate mean free path (IMFP) regime naturally develops a positive radial preference due to outward heat flux. Conversely, during the post-primary core-collapse phase, we observe a net inward radial flow and a non-trivial positive $\beta$ in the halo center as matter falls in to replenish the central void. These dynamically induced anisotropies highlight a critical limitation of traditional gravothermal fluid models \cite{balberg02, essig19, nishikawa20, sophia23}, which inherently assume velocity isotropy. Applied to our post-primary evolution, this suggests that the inward-falling, incompletely thermalized DM would systematically delay subsequent core-collapse events. Therefore, the isotropic gravothermal fluid framework may overestimate the rebuilding efficiency of the core and the subsequent core-collapse, which highlights the necessity of N-body (or other new SIDM approaches) for accurately modeling post-collapse SIDM-BH co-evolution.

\section{Summary and discussion}\label{sec:summary}

In this work, we investigate the post-core-collapse evolution of an isolated SIDM system, focusing on the macroscopic co-evolution of the SIDM-seeded BH and its host dark matter halo. We first evolve an initially NFW dark halo to the core-collapse stage using N-body simulations (denoted as the primary simulation), and then convert the collapsed central DM particles into a massive black hole particle. To model subsequent accretion, we employ two complementary DM-to-BH conversion schemes: (a) the multi-round core-collapse scenario, in which accretion is restricted to discrete events triggered only when the runaway core-collapse criterion is met, thereby accounting for the gravothermal-rebuilding delay and providing a lower bound on BH growth; and (b) the continuous-accretion scenario, in which the BH accretes DM on the fly and bypasses the macroscopic rebuilding bottleneck entirely, providing an upper bound. A central focus of our study is to establish numerical convergence over two key artificial parameters in these approaches: the core-collapse threshold $\tilde{\rho}_{cc}$ (the ratio between the halo's instantaneous central density and its initial NFW state value); and the spatial boundary defining the `central DM region' available for conversion, parameterized by the standard Knudsen number threshold $Kn_{\rm thres}$. Through systematic convergence tests, we demonstrate that at $Kn_{\rm thres}=0.05$, the manually assigned core-collapse thresholds ($\tilde{\rho}_{cc}=10, 30, 100$) asymptotically converge toward a common, threshold-independent BH mass growth trajectory. We thus recommend this setup for future studies. As a result, we provide a dynamically bounded estimate for the final SIDM-formed central BH, converging to a mass range of $5\%-12\%$ of the halo mass $M_{200c}$ when considering only DM accretion. We further provide a first-order estimate of the baryon contribution via combined Eddington and Bondi accretion. Because exponential baryonic accretion rapidly depletes the finite gas reservoir, the ultimate BH mass can be treated as a simple linear superposition of the initial gas mass fraction $f_{\rm gas}$ onto the $5\%$--$12\%M_{200c}$ baseline.

Our approach complements, rather than competes with, existing studies. Compared to other related works \cite{wxfeng21, wxfeng22, wxfeng25, fz25, sophia23, gu26} on SIDM-seeded supermassive black holes, our study does not delve into the microscopic physics or the exact onset of relativistic instability that triggers initial BH formation. Instead, taking the successful formation of such a BH seed as a given baseline, our work serves as a first self-consistent and quantitative description of how the BH and its host halo dynamically co-evolve on macroscopic timescales.

Beyond the BH mass trajectory, we investigate the structural evolution of the surrounding dark matter halo during this extended accretion phase. We demonstrate that gravothermal dynamics establish a continuous supply chain: the inner halo $\lesssim2r_s$ continuously channels material to rebuild the collapsing SMFP core, which is subsequently consumed by the central BH. Consequently, the macroscopic DM density within this $\sim2r_s$ region suffers a severe depletion, dropping by $\gtrsim1$ dex compared to its CDM counterpart. This severe post-collapse depletion provides a plausible theoretical explanation for the anomalous subhalo density distribution of detection V in the strong lensing system JVAS B1938+666 \cite{vegetti26}. Their best-fit lens model, an unresolved central point mass surrounded by an extended, constant-surface-density plateau that is significantly less concentrated than CDM expectations, qualitatively matches our physical picture of an SIDM-seeded BH and its exhausted dark matter reservoir (see Fig. \ref{fig:vegetti}). 

Furthermore, we analyze the kinematic evolution of the halo, evaluating the macroscopic radial flow and velocity anisotropy at both the deep core and the halo periphery. Throughout both the primary core-collapse and post-primary phases, the periphery ($r_{200c}$) establishes a radially outward bulk flow, driven by the continuous outward heat and mass transfer induced by central SIDM scatterings. Conversely, during the post-primary phase, the deep core develops a dynamically triggered inward mass flux. This inflow is driven by the thermal pressure deficit in the halo center since the central BH continuously accretes, forcing outer DM to fall inward to replenish the void. Tracing the velocity anisotropy $\beta$ from an initially isotropic state $\beta=0$, we find diverging kinematic features. During the primary collapse, the inner halo maintains $\beta \approx 0$ due to rapid collisional thermalization, despite the drastic structural evolution; whereas the dilute outer halo establishes a positive radial anisotropy $\beta>0$ because the local collision rate is insufficient to fully isotropize the outward flux. Crucially, during the post-primary phase, the deep core also establishes a non-trivial positive $\beta>0$.  This central anisotropy suggests incomplete thermalization even in the deep core, such that the rapid dynamical infall of DM to feed the central void outpaces local collisional relaxation.

Below, we outline the current limitations of our framework and highlight possible pathways for future exploration:

\begin{itemize}
    \item \textit{Angular momentum and the innermost capture:} Our $Kn$-based conversion is a fluid-scale feeding criterion and should not be read as resolving the final, sub-resolution capture onto the BH; whether an individual particle is accreted still depends, microscopically, on its angular momentum. However, the classical loss-cone argument for collisionless particles---which constructs orbits from angular-momentum conservation on individual particles---does not directly apply, because frequent self-scatterings in the collapsing core continuously exchange angular momentum between particles. This is made clear in Fig.~\ref{fig:particle-track}: individual particles cross into the SMFP core only after a final scattering event that biases them onto near-radial, low-angular-momentum orbits. Self-scattering as an angular-momentum transport mechanism has been discussed in \cite{wxfeng21}, which showed that SIDM viscosity rapidly dissipates the angular-momentum remnant of a collapsing core; we note, however, that their analysis concerns the dissipation of a \emph{bulk} spin, whereas our halos are initialized with zero net angular momentum, so only the microscopic scattering channel is directly shared. Whether self-scattering is sufficient to bring individual particles below the ISCO capture condition at the unresolved innermost scale is not covered in this work and remains to be carefully studied in the future.

	\item \textit{Expanding to a broader parameter space:} In this first, demonstrative study, we restricted our BH-halo co-evolution analysis to a single isolated halo. This choice was driven by the severe computational demands of the N-body framework, which requires executing and systematically evaluating a vast number of discrete simulations to establish numerical convergence. Having asymptotically achieved this convergence, we are now positioned to expand our framework across a broader parameter space. Our future works will systematically vary the initial halo mass, concentration, and SIDM cross section models to assess the universality of the $5\%$--$12\%M_{200c}$ mass bounds derived here. While traditional gravothermal fluid models predict a strict self-similarity for SIDM halo evolution \cite{essig19, o22, sq22, sq23, sophia23}, our findings of the dynamically induced anisotropy indicate that such strictly isotropic, equilibrium-based self-similarity may partially break down in the post-collapse regime. Therefore, dedicated N-body simulations or other advanced kinetic treatments would be indispensable for validating the self-similarity in post-collapse regimes.

	\item \textit{The gravitational influence of the baryonic potential:} To account for baryonic mass growth, we incorporated a first-order estimate by dynamically updating the Eddington and Bondi accretion rates at the end of each discrete DM accretion cycle. However, the true interplay between baryons and SIDM is more complex. Setting aside the intricacies of baryonic feedback processes is known to drastically accelerate the SIDM gravothermal core-collapse \cite{essig19, wxfeng21, ymzhong23, zzclg}. Furthermore, semi-analytical fluid models \cite{wxfeng21} have demonstrated that the inclusion of a compact baryon potential can significantly suppress the theoretical SMFP core mass, reducing it from $\sim4\%$ to $\sim 0.2\%$ of the total halo mass. It remains an open and critical question whether our N-body co-evolution framework will exhibit a similarly proportional suppression in the final BH mass bound when the central baryonic potential is explicitly included.

	\item \textit{Subhalos, satellite galaxies, and dwarf AGNs:} While this proof-of-concept study focuses on an isolated halo, much of the contemporary interest in the SIDM paradigm is driven by observed anomalies in low-mass systems. These include the diversity of dwarf satellite galaxies \cite{oman15, tulin17, bullock17, santos20, hayashi20, li20, sameie20, sales22, nadler23}, substructure strong lensing \cite{gilman19, meneghetti20, dg21, dnyang21, dg22, meneghetti23, birendra23, zzc23, dutra24, enzi24, dmkong24, dmkong25, lei25, sbli25, tajalli25, vegetti26, hou26, mace26}, and dark substructures' perturbations to stellar streams \cite{bonaca18, xyzhang24, nibauer25, menker26}. Previous theoretical studies on the evolution of SIDM subhalos and satellite galaxies \cite{zzc22, zzc23, zzc24, dnyang23, dnyang24b, sashimi-sidm, nadler23symphony, nadler23b} have demonstrated that the interplay between a subhalo's internal gravothermal evolution and environmental interactions with the host halo introduces significant diversity into its evolutionary trajectory. While this diversity provides a plausible solution to the aforementioned small-scale anomalies, it severely complicates universal theoretical scaling relations. Therefore, extending our macroscopic co-evolution framework to subhalo environments is essential. It remains an open question whether the central density depletion driven by BH feeding---an inevitable consequence of the core-collapse seeding scenario demonstrated here---will compromise or uniquely preserve previous SIDM solutions to these small-scale problems. On the other hand, recent observations of AGNs in dwarf systems \cite{reines13, reines19, greene20, mezcua24} offer new observational counterparts to this BH formation scenario, complementing the traditional high-redshift, massive halo hosts invoked for Little Red Dots \cite{lrd, wxfeng21, fz25, twshen25, mgroberts26}.

	\item \textit{Empirical modeling of the core-collapsed state:} Recent efforts \citep[Whisnant et al. in prep.]{robertson16, sq22, fischer23b, tran24, sbli26, hou25} have sought to empirically model and fit the density profiles of core-collapsed SIDM halos, which lays the foundation for matching with observations such as dwarf kinematics and substructure lensing. However, these current fitting templates all assume a dense, steeply cuspy final state in the halo center. Our results demonstrate that if gravothermal collapse successfully seeds a central massive BH, the ensuing macroscopic supply chain will almost inevitably cause a severe density depletion in the inner halo, rendering the widely assumed ultra-dense and cuspy core-collapsed state a mere transient phase. Consequently, we suggest that future phenomenological fitting frameworks incorporate this BH-DM co-evolution. 

	\item \textit{Can core-collapse halt itself before BH formation?} For simplicity of modeling, we chose a constant cross section of $100\ \rm cm^2/g$ in this demonstrative study. However, cross sections $\gtrsim\mathcal{O}(1)\ {\rm cm^2/g}$ are strongly disfavored by observations on cluster scales, where characteristic velocities reach $\gtrsim 1000$ km/s \cite{rocha13, Peter13, kim17, tulin17, elbert18, robertson19}. Conversely, resolving the small-scale anomalies via core-collapse in dwarf-sized systems within a Hubble time generally requires much larger cross sections of $\mathcal{O}(100)\ {\rm cm^2/g}$ \cite{essig19, zzc23, zzc24}. A natural solution to reconcile these mass-scale discrepancies is to invoke velocity-dependent cross sections that monotonically decrease at higher velocities. From a particle physics perspective, such velocity dependence is highly motivated by light-mediator models; for instance, DM self-scattering through a dark Yukawa potential naturally transitions into a Rutherford-like regime at high velocities, yielding a steep $\sigma/m \propto v^{-4}$ scaling \cite{jfeng10, kaplinghat16, tulin17}. This strong velocity dependence, however, poses a potential challenge to the very foundation of the BH seeding scenario. As gravothermal collapse proceeds, the central velocity dispersion increases dramatically, eventually approaching the relativistic threshold of $\sim c/3$ required to trigger the relativistic instability required for seeding BH \cite{wxfeng21, wxfeng22, sophia23}. In a velocity-dependent model, the local scattering cross section would plummet as the particles heat up. This raises a critical question: in certain regions of the SIDM parameter space, could the plunging cross section bottleneck the heat transfer, effectively halting or ``locking'' the core-collapse process? If BH formation is indeed prevented, the ultimate fate of such a stalled halo remains unclear, with perhaps a highly dense, hot quasi-equilibrium state. We argue that investigating the late-stage thermodynamics of velocity-dependent SIDM represents a crucial frontier for future theoretical efforts (Zeng et al. in prep.). Likewise, our use of $Kn$ as a tracer of the core-collapse depth relies on the inverse scaling $Kn\propto\sigma^{-1}$, which guarantees a monotonic relation only for a constant cross section. In certain velocity-dependent or resonant \cite{dg22, tran25} models, where $\sigma(v)$ varies steeply or exhibits peaks in the $\sigma$--$v$ plane, this monotonicity is lost and $Kn$ may no longer be a reliable metric for how deep the core-collapse state is. A more suitable metric would need to be devised for such models in future work.

    \item \textit{Positive vs. negative heat capacity across scales.}
    The BH growth problem is intrinsically multiscale.
    In the BH-dominated micro-regime, the system effectively possesses a positive heat capacity: recent high-resolution fluid simulations \cite{meng26} show that SIDM heat conduction puffs up the inner layers and causes the local accretion flow to self-regulate, saturating at $\sim 1\%M_\mathrm{halo}$ over Myr timescales.
    This supports our phenomenological DM-to-BH conversion---not as a model of the unresolved feeding flow, but as an effective representation of its net outcome after a single core-collapse episode.
    The surrounding SIDM halo, by contrast, is a self-gravitating, negative-heat-capacity system whose gravothermal evolution persistently transports mass and energy inward over Gyr timescales.
    The local conduction therefore governs how efficiently the BH consumes each delivery, while the halo-scale gravothermal evolution determines how rapidly that reservoir is replenished---a supply-chain effect beyond fixed-boundary fluid simulations that eventually drives the BH mass to $\sim 5\%$--$12\%\,M_\mathrm{halo}$.
    A complete theory will require hybrid schemes that couple the local, conduction-regulated feeding flow with this global gravothermal supply chain, bridging these positive and negative heat capacity regimes.

     \item \textit{Knudsen number for SIDM vs. baryons.}
     Our conversion criterion is built on the SIDM Knudsen number $Kn\equiv\lambda/H=t_{\rm sca}/t_{\rm dyn}$, the ratio of the scattering timescale to the dynamical timescale. This two-timescale structure is not unique to SIDM: it is also the collisionality diagnostic of baryonic accretion, where one can define $Kn\equiv\lambda/L$ from the Coulomb mean free path relative to the system size and treat the plasma as collisionless when $Kn\gtrsim1$ \cite{gammie25}. The baryonic analogue of our threshold is thus the ratio of the Coulomb collision time to the accretion (infall) time. An order-of-magnitude estimate gives $t_{pp}/t_a\sim \mathcal{O}(10^{4})$ for protons and $t_{ee}/t_a\sim\mathcal{O}(1)$ for electrons \cite{mahadevan97}, a four-dex gap reflecting the proton-electron mass ratio and the two-temperature nature of the plasma \cite{gammie25}. These numbers are not directly comparable to our $Kn_{\rm thres}=0.05$: the SIDM and electromagnetic cross sections have unrelated physical origins, and the baryons have radiative cooling while SIDM does not. The meaningful analogy therefore lies only in the shared two-timescale structure. The crucial distinction is functional: in SIDM, self-scattering is the sole energy-transport channel, so $Kn$ directly tracks the core-collapse depth and hence the DM-to-BH conversion rate; in baryonic accretion, the inflow is instead regulated by angular-momentum transport, so its Knudsen number controls only thermalization and radiative efficiency. Our criterion is therefore a DM-specific proxy, and a baryonic analogue that directly uses $Kn$ would be inappropriate.

    \item \textit{The unresolved dark matter spike:} Our N-body framework cannot resolve the immediate vicinity of a central massive BH, where a density spike is expected to form even in standard CDM \cite{gondolo99, gnedin03, sadeghian13, ijohn23}. In our simulations this region is represented by a single unresolved massive particle, embedded in a surrounding DM halo whose density is eventually substantially reduced. This does not necessarily conflict with the spike prediction: rather, the spike forms on scales below our resolution and is indistinguishable from the unresolved point mass in our results. The physics of such spikes and cusps has been studied in weakly collisional (cored, $Kn\gg1$) SIDM, where conduction-fluid models predict a power-law cusp around the central BH \cite{shapiro14, shapiro18, sabarish25}. A more complete multi-scale picture would then be `inner halo $\to$ spike $\to$ BH', of which we resolve the outer, gravothermal link, while \cite{shapiro14, shapiro18, sabarish25} model the innermost, weakly-collisional link.
\end{itemize}

In summary, this work serves as a first quantitative, self-consistent framework to model the post-core-collapse co-evolution of the SIDM halo and its central black hole. By establishing bracketed mass growth trajectories for the SIDM-seeded BH, we illustrate how gravothermal dynamics drive a continuous macroscopic supply chain that substantially alters the inner host halo. The resulting central density depletion offers a natural physical interpretation for the anomalous substructure (detection V) recently detected in the strong lensing system JVAS B1938+666 \cite{vegetti26}.  Ultimately, our findings suggest that integrating this BH-halo co-evolution is a vital component for making physically consistent predictions of SIDM core-collapse states and accurately interpreting observational data.

\begin{acknowledgments}
We thank Kimberly Boddy, Hai-Bo Yu, Fangzhou Jiang, Frank van den Bosch and Shashank Dattathri for useful discussion.

The simulations in this work were mostly conducted on the Grace Cluster of High Performance Research Computing at the Texas A\&M University, while some early stage simulations were conducted at Ohio Supercomputer Center \cite{osc}.

\end{acknowledgments}

\appendix

\section{Supplementary figures}

\begin{figure*}
    \centering
    \begin{subfigure}[t]{0.45\textwidth}
        \centering
        \includegraphics[width=\textwidth, clip,trim=0.2cm 0cm 0.2cm 0cm]{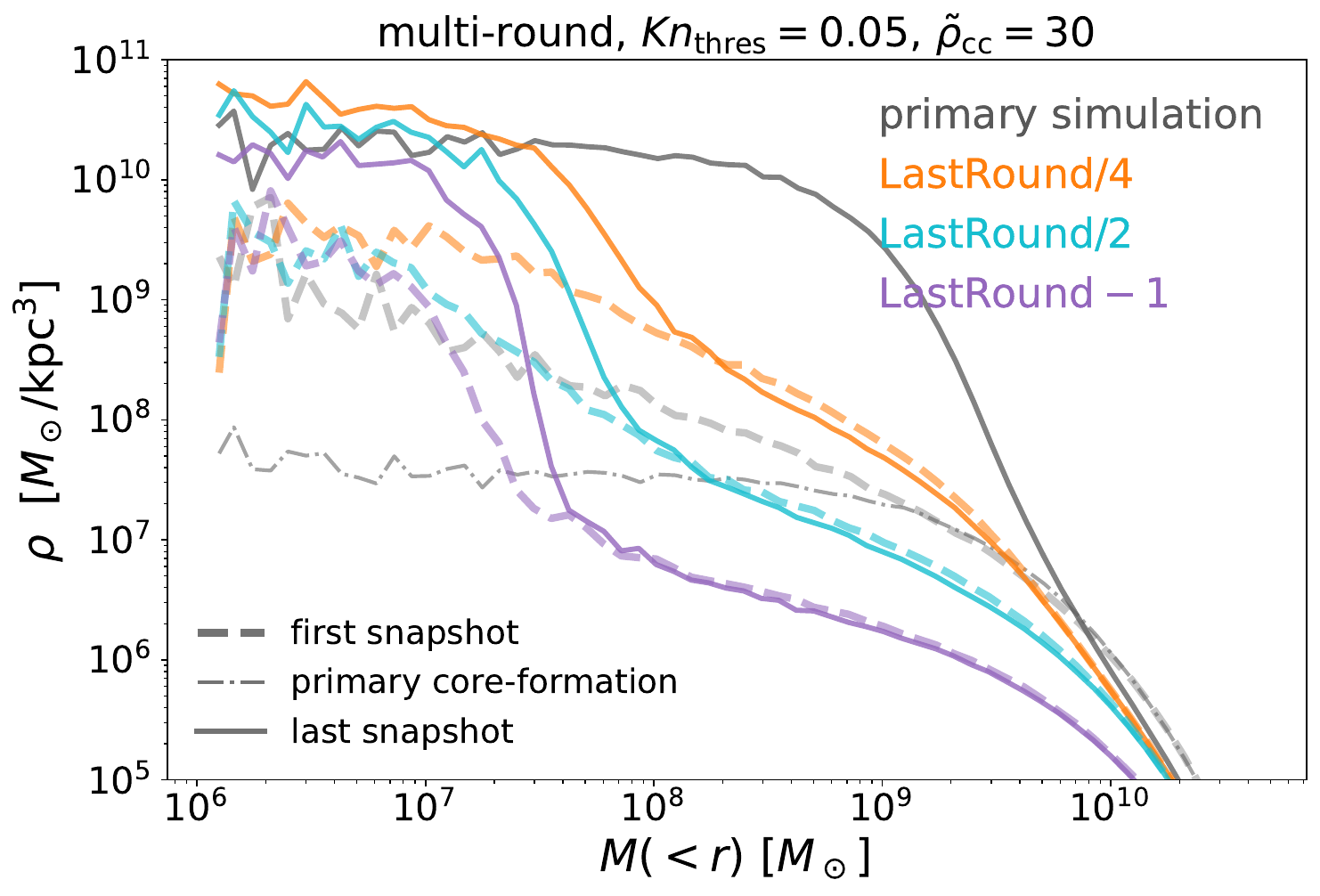}
        \caption{}
        \label{fig:density-rho-m-rhocc30}
    \end{subfigure}
    ~
    \begin{subfigure}[t]{0.45\textwidth}
        \centering
        \includegraphics[width=\textwidth, clip,trim=0.2cm 0cm 0.2cm 0cm]{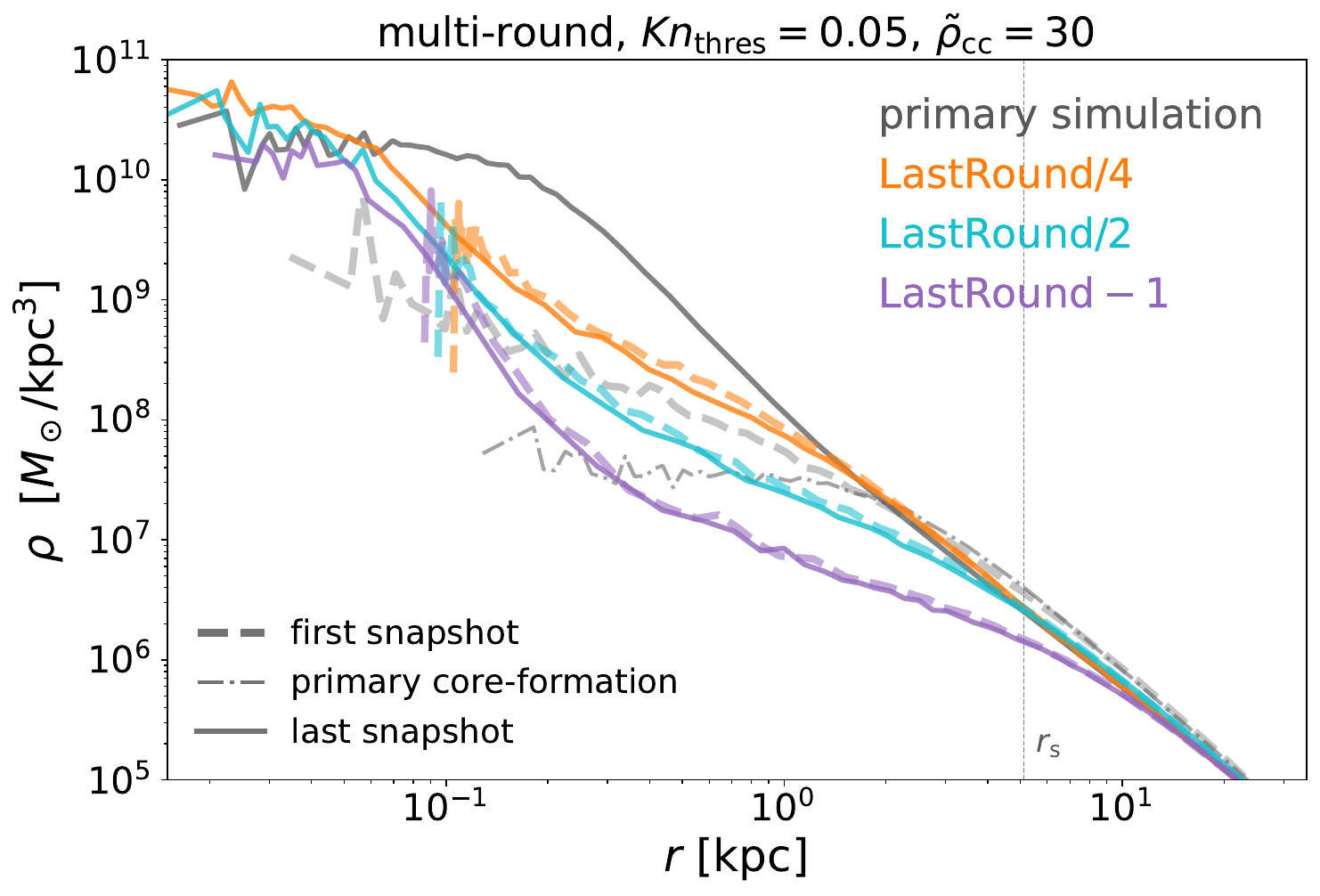}
        \caption{}
        \label{fig:density-rho-r-rhocc30}
    \end{subfigure}
    ~
    \begin{subfigure}[t]{0.45\textwidth}
        \centering
        \includegraphics[width=\textwidth, clip,trim=0.2cm 0cm 0.2cm 0cm]{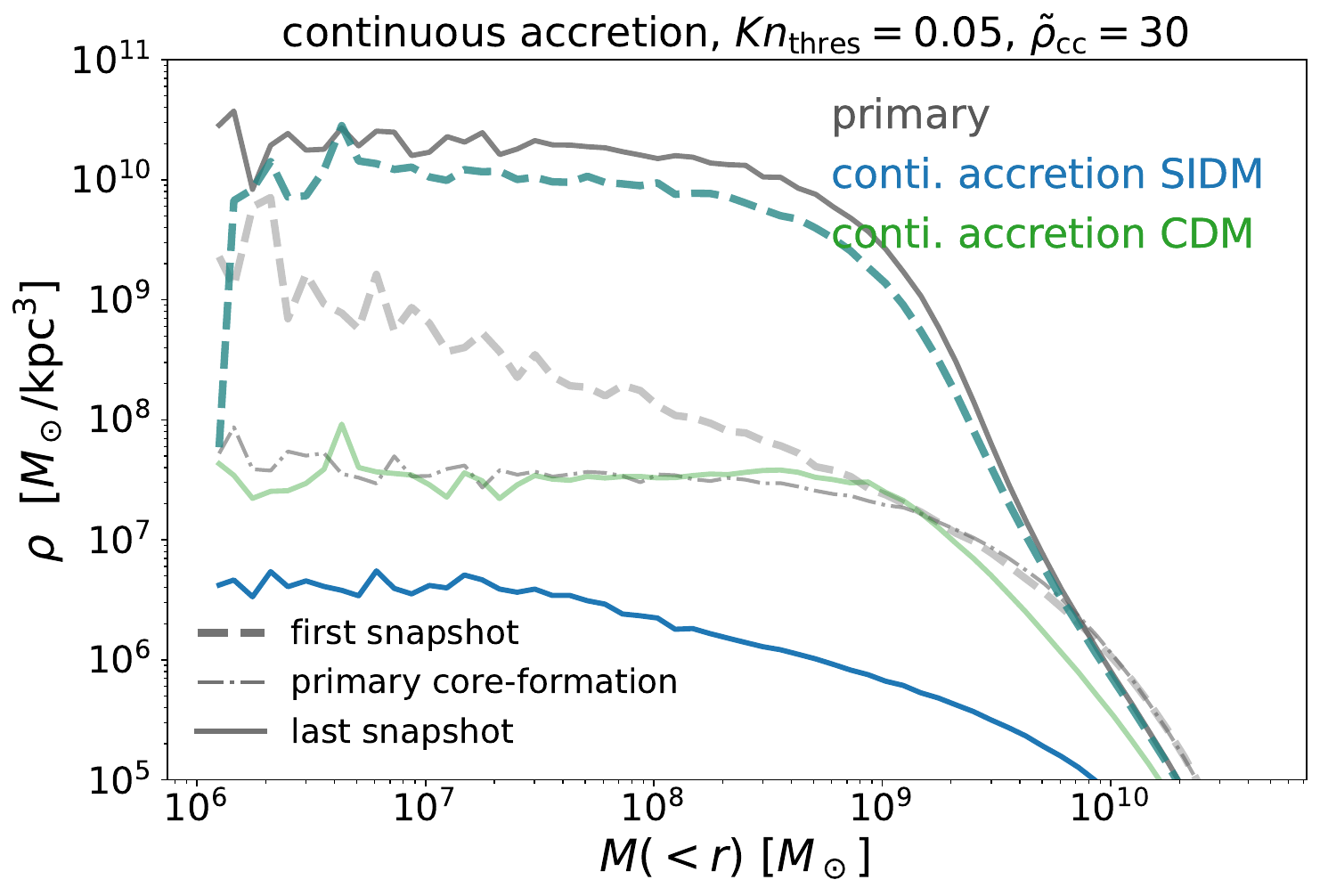}
        \caption{}
        \label{fig:density-rho-m-conti-rhocc30}    
    \end{subfigure}
    ~
    \begin{subfigure}[t]{0.45\textwidth}
        \centering
        \includegraphics[width=\textwidth, clip,trim=0.2cm 0cm 0.2cm 0cm]{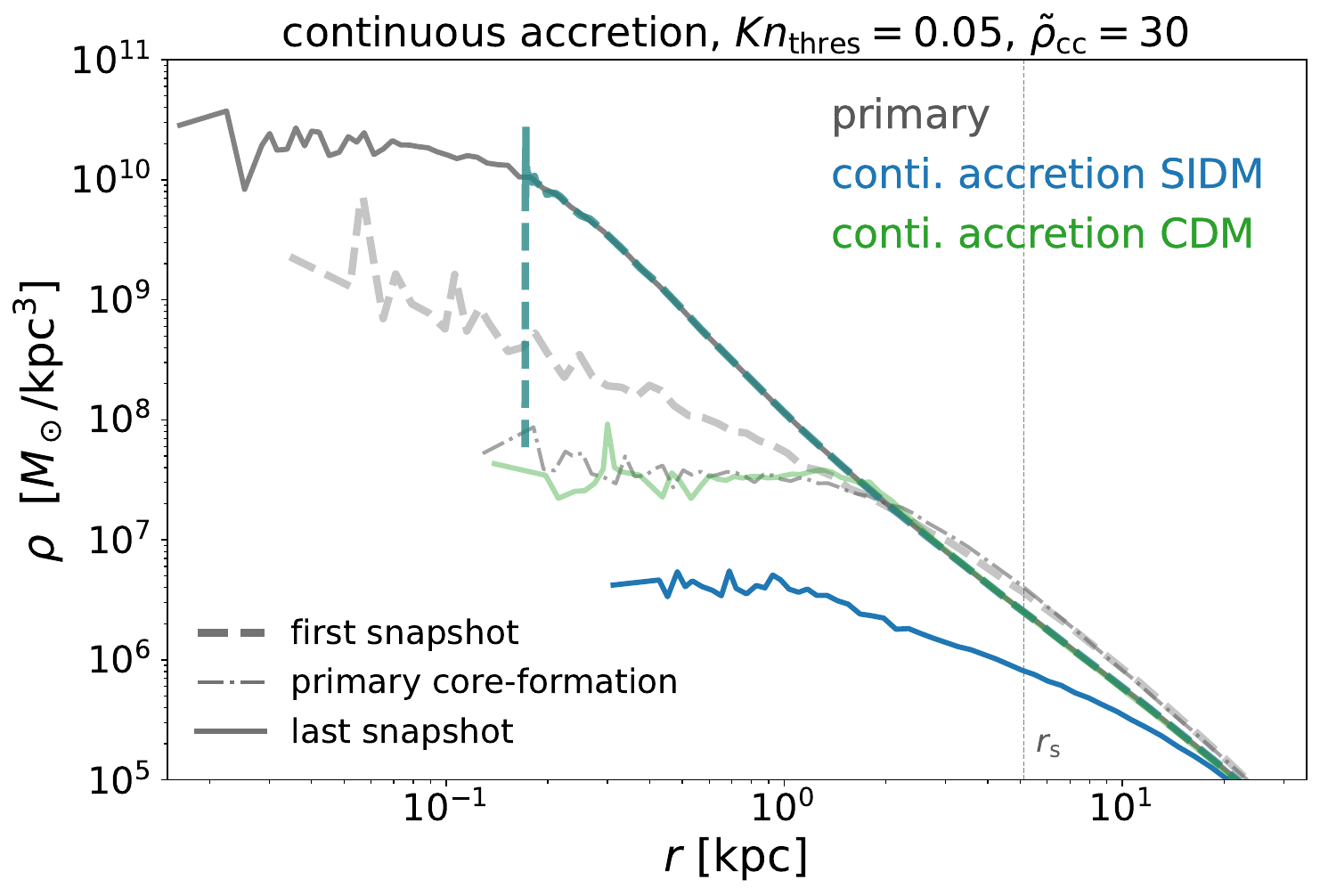}
        \caption{}
        \label{fig:density-rho-r-conti-rhocc30}    
    \end{subfigure}
    \caption{Similar to Fig. \ref{fig:density-profiles}, but showing the $Kn_{\rm thres}=0.05, \tilde{\rho}_{cc}=30$ parameter set.}
    \label{fig:density-profiles-rhocc30}
\end{figure*}

\begin{figure*}
    \centering
    \begin{subfigure}[t]{0.45\textwidth}
        \centering
        \includegraphics[width=\textwidth, clip,trim=0.2cm 0cm 0.2cm 0cm]{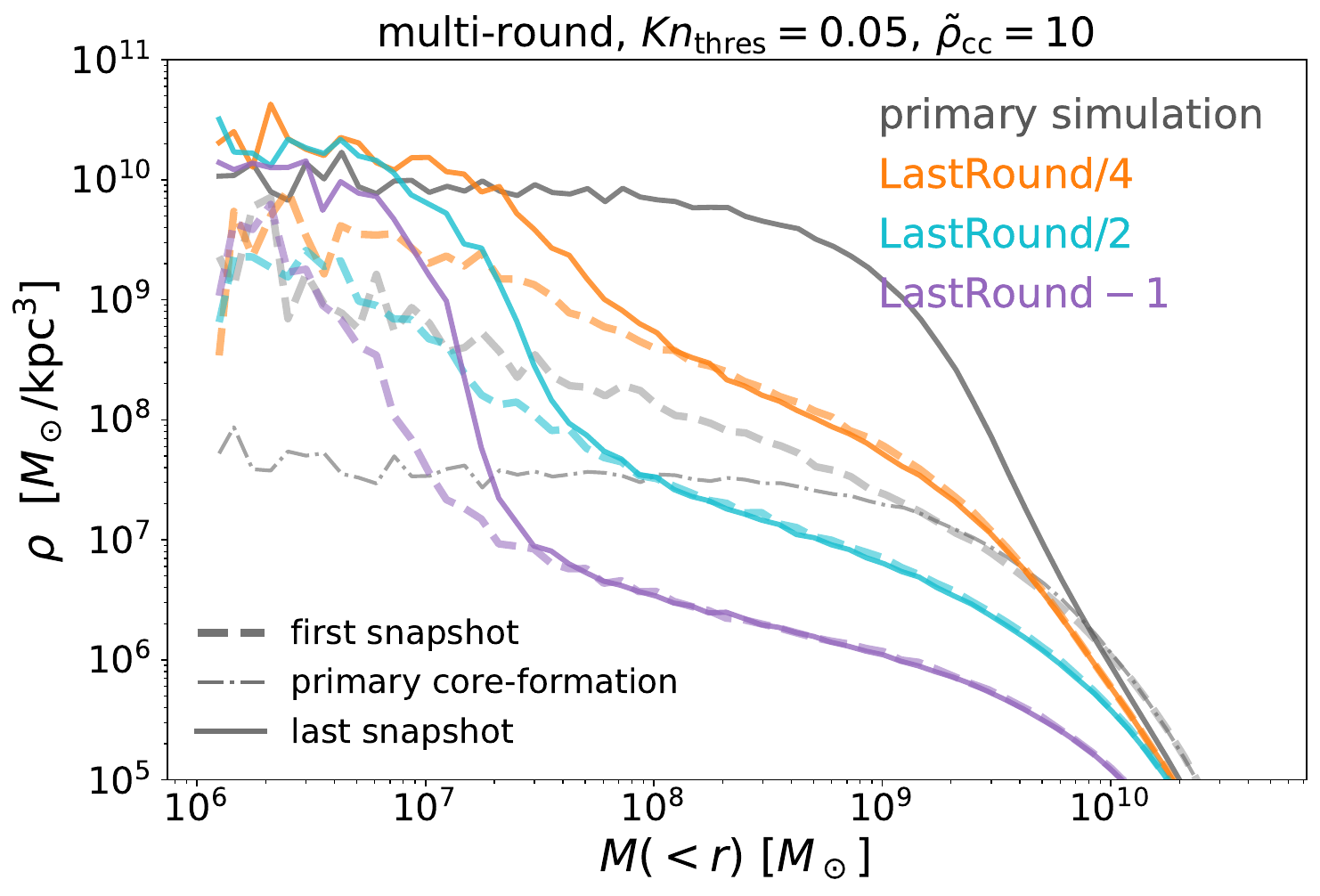}
        \caption{}
        \label{fig:density-rho-m-rhocc10}
    \end{subfigure}
    ~
    \begin{subfigure}[t]{0.45\textwidth}
        \centering
        \includegraphics[width=\textwidth, clip,trim=0.2cm 0cm 0.2cm 0cm]{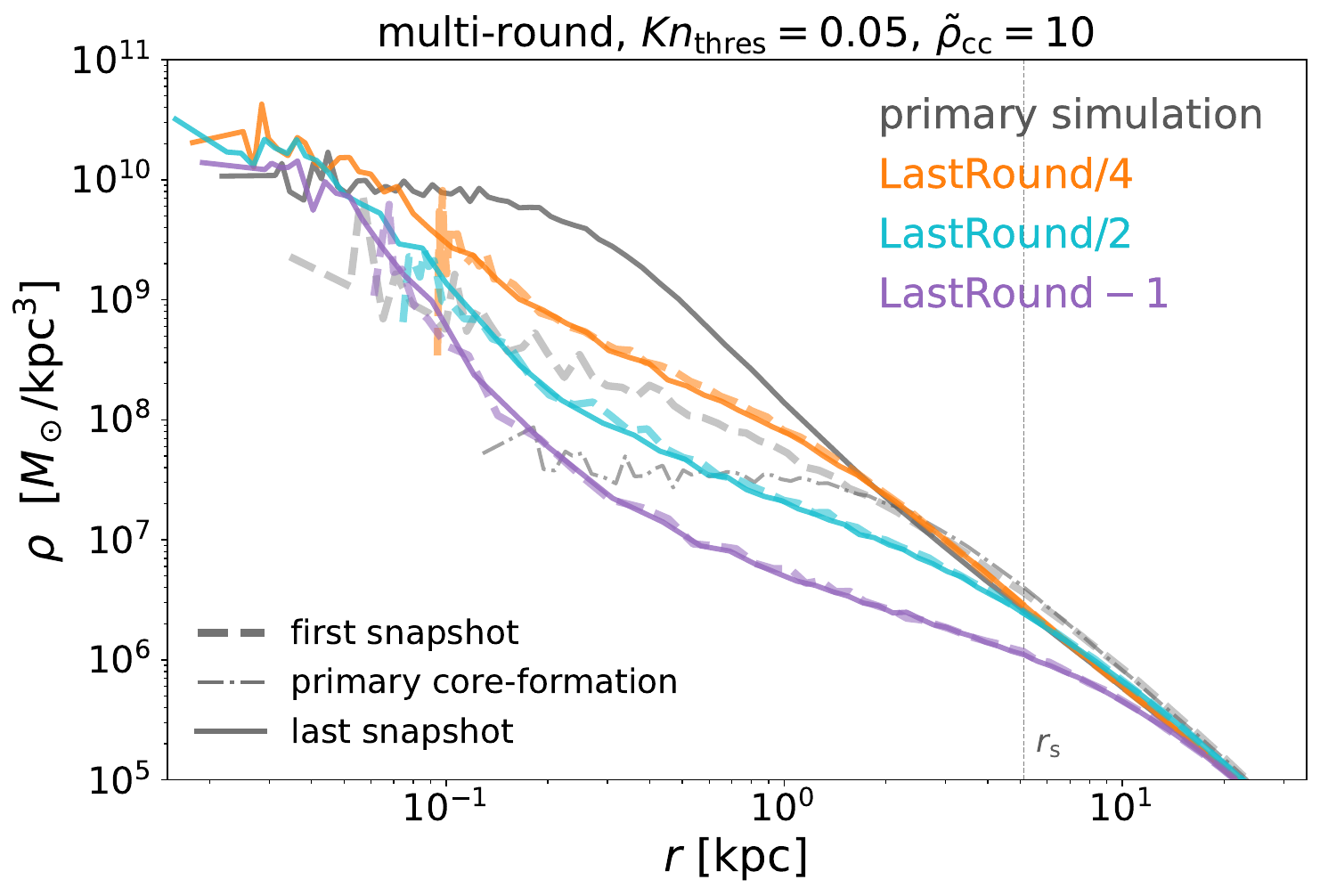}
        \caption{}
        \label{fig:density-rho-r-rhocc10}
    \end{subfigure}
    ~
    \begin{subfigure}[t]{0.45\textwidth}
        \centering
        \includegraphics[width=\textwidth, clip,trim=0.2cm 0cm 0.2cm 0cm]{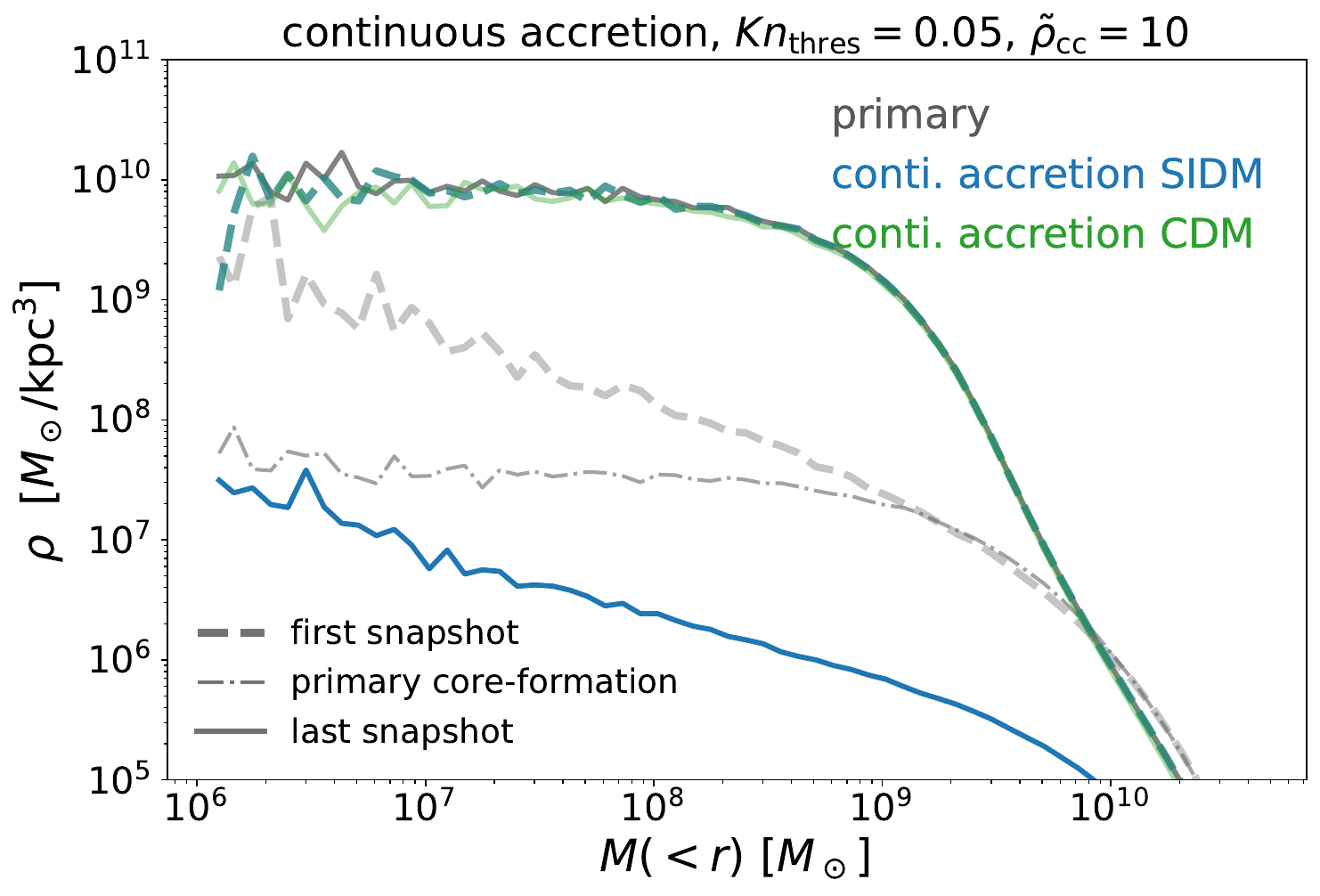}
        \caption{}
        \label{fig:density-rho-m-conti-rhocc10}    
    \end{subfigure}
    ~
    \begin{subfigure}[t]{0.45\textwidth}
        \centering
        \includegraphics[width=\textwidth, clip,trim=0.2cm 0cm 0.2cm 0cm]{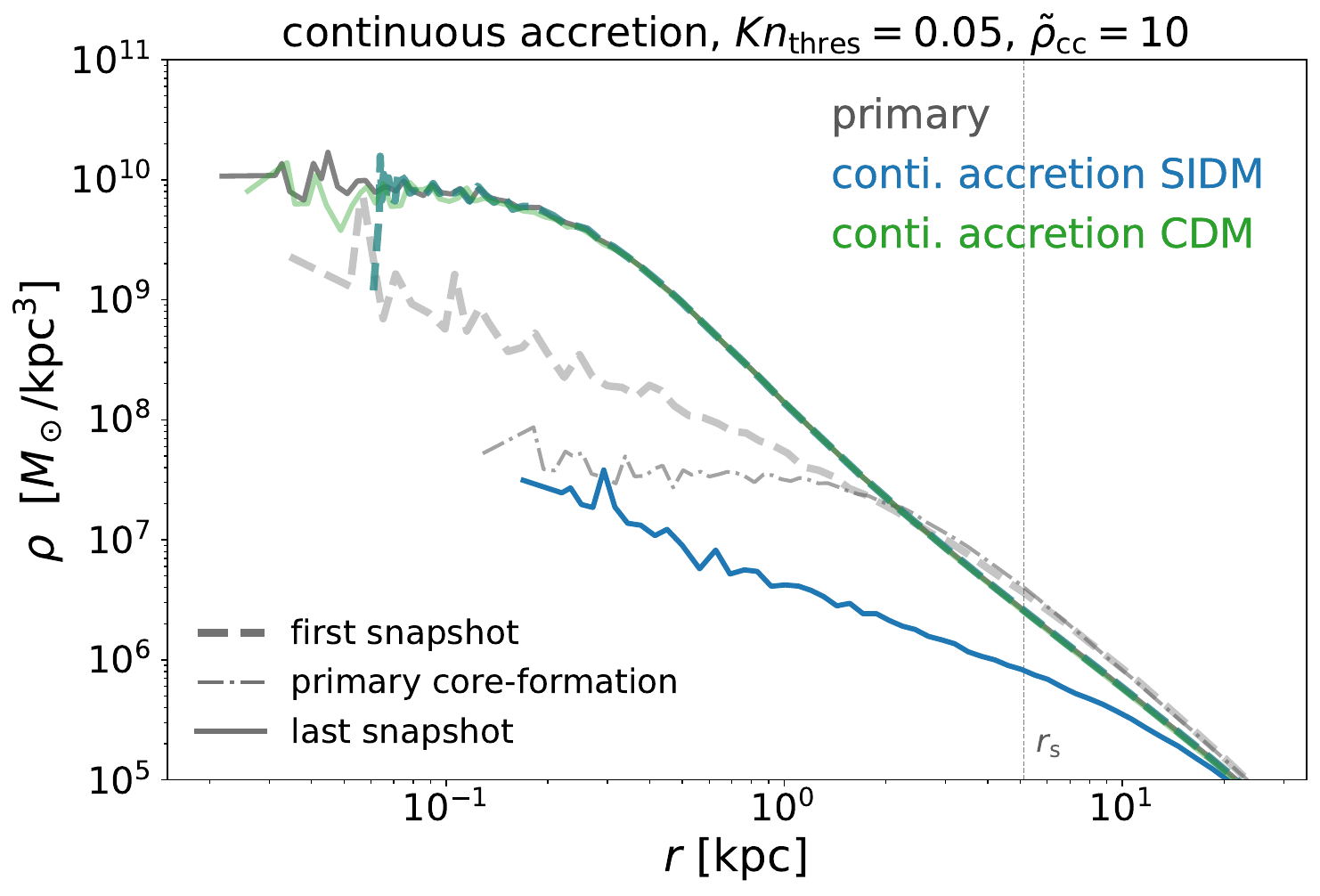}
        \caption{}
        \label{fig:density-rho-r-conti-rhocc10}    
    \end{subfigure}
    \caption{Similar to Fig. \ref{fig:density-profiles}, but showing the $Kn_{\rm thres}=0.05, \tilde{\rho}_{cc}=10$ parameter set.}
    \label{fig:density-profiles-rhocc10}
\end{figure*}

\begin{figure}
    \centering
        \includegraphics[width=0.45\textwidth, clip,trim=0.2cm 0cm 0.2cm 0cm]{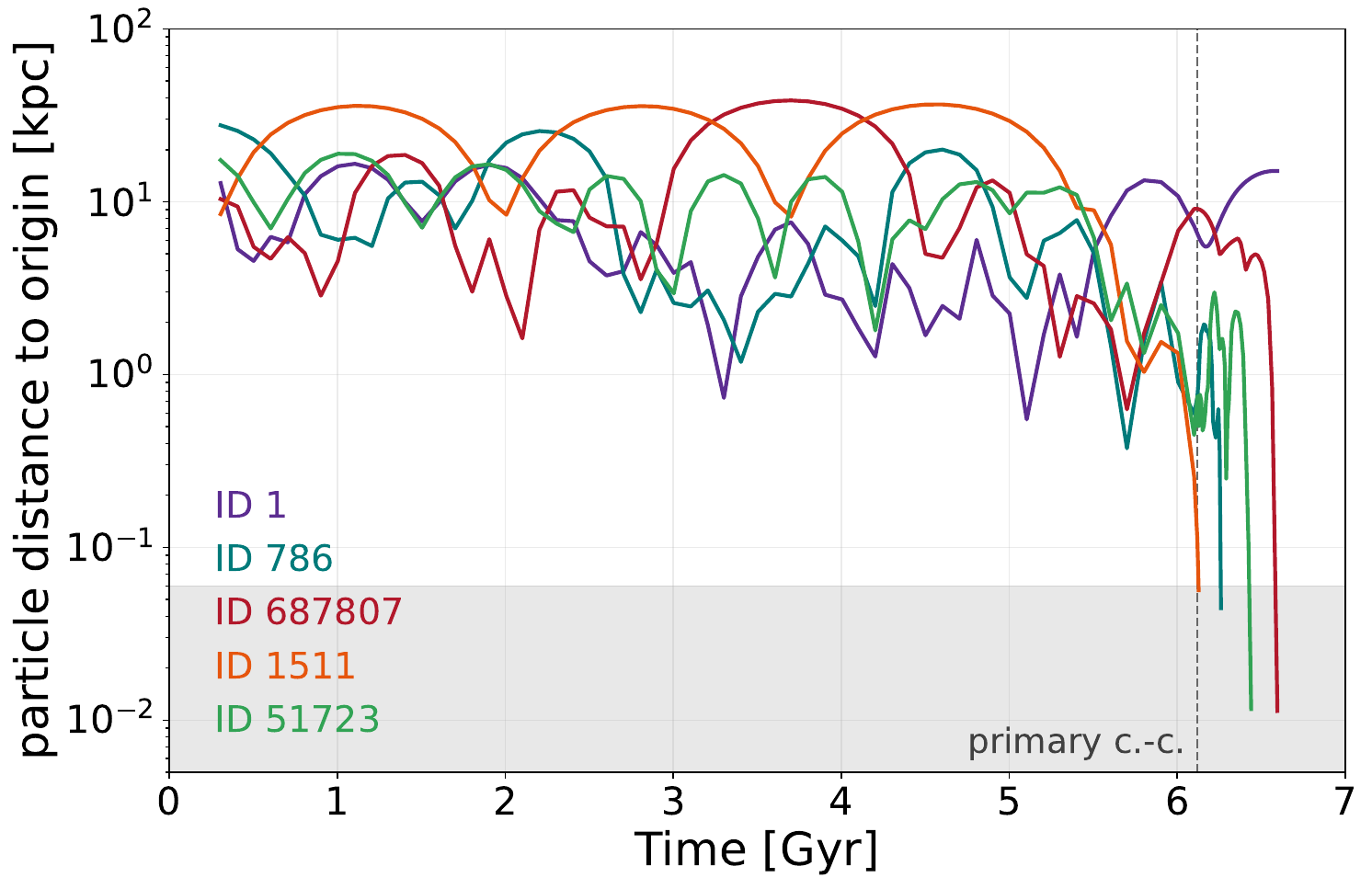}
    \caption{Tracing individual particles' orbital distances to halo center. This simple demonstration suggests that most, if not all particles fall into the deep core-collapse region and thus BH accretion boundary because of a last scattering that favors radial orbits.}
    \label{fig:particle-track}
\end{figure}

In this appendix, we provide supplementary figures to the main manuscript:

\begin{itemize}
    \item Figs. \ref{fig:density-profiles-rhocc30} and \ref{fig:density-profiles-rhocc10} extend the structural evolution analysis of Fig. \ref{fig:density-profiles} to other tested core-collapse thresholds $\tilde{\rho}_{cc}=10$ and 30. These results are visually converged, confirming that the severe gravothermal depletion of DM in the inner halo is a highly robust macroscopic consequence.

    \item Fig. \ref{fig:particle-track} traces the orbital trajectories of individual DM particles. This simple demonstration suggests that particles plunge into the deep SMFP core-collapse region --- and thus cross the BH accretion boundary --- because of a final scattering event that strongly biases them toward radial orbits.

\end{itemize}

%



\bibliography{apssamp}

\end{document}